\documentclass[aps,pra,showpacs,floatfix,superscriptaddress,11pt]{revtex4-1}
\usepackage{xr}

\usepackage{amsmath,amssymb,physics}
\usepackage[toc,page]{appendix}
\usepackage{verbatim,graphicx,mathtools}
\usepackage{txfonts}
\usepackage{color}
\usepackage[all]{xy}
\usepackage{color}
\usepackage[all]{xy}
\usepackage{subfigure}
\usepackage{xcolor}

\newcommand{\eref}[1]{(\ref{#1})}
\newcommand{\nn}{\nonumber}
\newcommand{\be}{\begin{eqnarray}}
\newcommand{\ee}{\end{eqnarray}}

\newcommand{\bmat}{\left ( \begin{array}{cc} }
	\newcommand{\emat}{\end{array} \right ) }

\newcounter{jvct}

\newcommand{\beq}{\begin{equation}}
\newcommand{\beqs}{\begin{equation*}}
\newcommand{\eeq}{\end{equation}}
\newcommand{\eeqs}{\end{equation*}}

\allowdisplaybreaks

\begin{document}

\begin{flushright}
KIAS-P19003
\end{flushright}

%%%%%%%%%%%%%%%%%%%%%%%%%%%%%%%%%%%%%%%%%%%%%%%%%%%%%%%%%%%%%%%%
\title{Quantum chaos transition in a two-site SYK model dual to an eternal traversable wormhole}
\author{Antonio M. Garc\'\i a-Garc\'\i a}
\email{amgg@sjtu.edu.cn}
\affiliation{Shanghai Center for Complex Physics, 
	School of Physics and Astronomy, Shanghai Jiao Tong
	University, Shanghai 200240, China}

\author{Tomoki Nosaka}
\email{nosaka@yukawa.kyoto-u.ac.jp}

\author{Dario Rosa}
\email{Dario85@kias.re.kr}
\affiliation{School of Physics, Korea Institute for Advanced Study,
	85 Hoegiro Dongdaemun-gu, Seoul 02455, Republic of Korea}

\author{Jacobus J. M. Verbaarschot}
\email{jacobus.verbaarschot@stonybrook.edu}
\affiliation{Department of Physics and Astronomy, Stony Brook University, Stony Brook, New York 11794, USA}

\begin{abstract} 
	It has been recently proposed by Maldacena and Qi that an eternal traversable wormhole in a two dimensional Anti de Sitter space (${\rm AdS}_2$) is the gravity dual of the low temperature limit of  two  Sachdev-Ye-Kitaev (SYK) models coupled by a relevant interaction (which we will refer to as spin operator). In this paper, we study spectral and eigenstate properties of this coupled SYK model. We have found that level statistics in the tail of the spectrum, and for a sufficiently weak coupling, shows substantial deviations from random matrix theory which  suggests that traversable wormholes are not quantum chaotic. By contrast, for sufficiently strong coupling, corresponding to the black hole phase, level statistics are well described by random matrix theory. This transition in level statistics coincides approximately with a previously reported  Hawking-Page transition for weak coupling. We have shown explicitly that this thermodynamic transition turns into a sharp crossover as the coupling increases. Likewise, this critical coupling also corresponds with the one at which the overlap between the ground state and the thermofield double state (TFD) is smallest. In the range of sizes we can reach by exact diagonalization, the ground state is well approximated by the TFD state only in the strong coupling limit. This is due to the fact that the ground state is close to the eigenstate of the spin operator corresponding to the lowest eigenvalue which is an exact TFD state at infinite temperature.  In this region, the spectral density is separated into blobs centered around the eigenvalues of the spin operator.  For weaker couplings, the exponential decay of coefficients in a tensor product basis, typical of the TFD, becomes power law. Finally, we have also found that the total Hamiltonian has an additional discrete symmetry which has not been reported previously.
 
\end{abstract}\maketitle

%\tableofcontents
\newpage
\section{{Introduction}}
The dynamics of quantum many-body systems is very rich and, in general,
highly dependent of the details of both initial conditions and the
Hamiltonian that governs its time evolution. However there are situations
where the time evolution has universal features. A paradigmatic example
is that of a quantum system whose classical counterpart is chaotic for
time scales of the order of the inverse of the  mean level spacing,
the so called Heisenberg time. 
In that region, where the discreteness of the spectrum is relevant,
and therefore quantum effects are always strong, quantum chaotic systems,
assuming localization effects are negligible, relax to a fully ergodic
state that only depends on the global symmetries of the system. 
Level statistics are a powerful tool to detect and classify universal
features in this region, because it is expected that the predictions of
random matrix theory will apply in this case. Indeed, agreement with random matrix theory results has been found in a broad variety
of systems: highly excited states of  nuclei \cite{wigner1951}, weakly
disordered systems with \cite{bertrand2016} and without
interactions \cite{efetov1983}, deterministic quantum chaotic
systems \cite{bohigas1984} or the low energy limit of QCD in a  box \cite{verbaarschot1993a,shuryak1993a}.
 
The field of quantum chaos  received an important boost after it was
claimed \cite{maldacena2015} that quantum black holes in the semiclassical
limit are quantum chaotic. More specifically, it was proposed
\cite{maldacena2015} that the growth of certain out-of-time-ordered
correlation functions in the semiclassical limit, and for intermediate times
of the order of the Ehrenfest time, is exponential as for quantum chaotic
systems \cite{larkin1969}. Moreover, the growth rate, given by the
Lyapunov exponent, obeys a universal bound which is saturated for
field theories with a gravity dual. Later this feature was also
observed \cite{kitaev2015,maldacena2016} in the strong coupling limit
of the Sachdev-Ye-Kitaev (SYK) model
\cite{bohigas1971,bohigas1971a,french1970,french1971,mon1975,benet2003,kota2014,sachdev1993,sachdev2010,kitaev2015,fu2018},
a zero dimensional fermionic model with infinite-range random interactions.
This simple quantum mechanical model has attracted a great deal  of attention because,
despite being quantum chaotic and strongly interacting,
it can be tackled analytically \cite{kitaev2015,maldacena2016,garcia2017,stanford2017,mertens2017,belokurov2017,bagrets2016,bagrets2017,jevicki2016,arefeva2018,Altland:2017eao}  and has all the features expected for a system with
a gravity dual \cite{kitaev2015,maldacena2016,jensen2016,jevicki2016}:
it does saturate the bound on chaos mentioned above, it also has a
finite zero temperature entropy \cite{sachdev1993}, a linear specific heat in
the low temperature limit \cite{kitaev2015,maldacena2016,garcia2017,cotler2016}
and a density of low energy excitations \cite{maldacena2016,cotler2016,garcia2017} that increases exponentially.
Level statistics of the SYK model was shown to be well described by random
matrix theory \cite{garcia2016,cotler2016,garcia2017,Altland:2017eao,garcia2018a} in the
spectral region where the SYK model is expected to have a gravity dual
which confirms that it is also quantum chaotic for long times. Based on the same pattern of conformal symmetry breaking, \cite{maldacena2016a,almheiri2015}, the gravity dual of the SYK model has been identified to be a Jackiw-Teitelboim \cite{jackiw1985,teitelboim1983,forste2017} background.

 A natural question to ask, assuming that the holographic duality applies for sufficiently long times, is whether these quantum chaotic features
 are exclusive to quantum black holes geometries, such as the Jackiw-Teitelboim background, or also apply to other backgrounds.
 Recently, Maldacena and Qi \cite{maldacena2018}, see also \cite{bak2018,kim2019}, constructed a near
 ${\rm AdS}_2$ background representing an eternal traversable wormhole.
 The quantum feasibility of a traversable wormhole was previously suggested
 \cite{gao2016} by showing that a double trace deformation could make
 the quantum matter stress tensor to have a negative average null
 energy without violating causality. An example of such construction for
 ${\rm AdS}_2$ geometries was worked out in \cite{maldacena2017} and for
 higher dimensions, and other backgrounds, in \cite{marlof2018,tumurtushaa2018,maldacena2018b,anabalon2018} Out of equilibrium features for intermediate times were investigated in Ref.~\cite{gao2018} and entanglement and other
 quantum information observables were studied in Ref.~\cite{bao2018}.

 Returning to eternal traversable wormholes, it was conjectured
 \cite{maldacena2018} that its ground state is well approximated by a thermofield double
 state (TFD) and that its field theory dual is a two-site SYK model coupled
 by a relevant interaction in the limit of weak coupling and low temperature.
 It was found that 
 for sufficiently high temperatures, or stronger coupling, the system
 undergoes a thermodynamic Hawking-Page transition from the wormhole phase
 to a two black hole phase. By increasing the strength of the coupling,
 the transition takes place at higher temperatures and finally, after
 reaching the critical coupling, this first order 
 transition terminates.
 Intriguingly,  the wormhole
 phase is still controlled by a generalized Schwarzian action when
 the residual SL($2$) symmetry of the standard SYK model is broken \cite{maldacena2018}.
 The associated Liouville quantum mechanics problem \cite{bagrets2016}
 has graviton bound states which are interpreted as excited states of
 the quantum wormhole background. 
 A non-random SYK model with the same large $N$ limit was investigated
 previously in Ref. \cite{ferrari2017} with similar conclusions.

 In this paper, we study different aspects of this coupled SYK model.
 Our motivation is twofold. First, we  aim to investigate the
 long time dynamics of this two-site coupled SYK model by level statistics
 in order to clarify whether quantum chaos and random matrix theory are
 generic in quantum gravity backgrounds or are restricted to black holes.
 Second, we  would like to gain a more detailed understanding
 of the phase diagram of the system. More specifically, we aim to
 determine the range of validity of the TFD as a ground state of
 the system and also the exact nature of the termination of the first order
 Hawking-Page transition at finite coupling. We have found that indeed
 these two main goals are closely related. We show that the first order
 transition suddenly becomes a sharp crossover above the critical coupling.
 Around the same coupling strength, 
 the number of non-zero coefficients, in a tensor product basis of the two independent SYK's, that contribute substantially to the ground state of the system increases sharply. This is typical of chaotic-integrable transitions. More specifically, except for strong coupling and in the range of sizes we can explore numerically, we have observed that the energy dependence of these coefficients is power law rather than the exponential decay typical of the TFD state.

Level statistics are also sensitive to the value of the coupling constant. In the tail of the spectrum, and couplings weaker than the critical one, corresponding to the wormhole phase,
 level statistics are not described by random matrix theory.
 However, well above the critical coupling, spectral correlations in the full
 spectrum are well described by random matrix theory. 

 The organization of the paper is as follows. First, we introduce
 the model with a special emphasis on the description of its symmetries
 and its ground state. In particular, we identify a global spin symmetry
 in the combined system that has an important effect on spectral statistics and thermodynamic properties. In Section III, we investigate the ground state of the system.
 For the number of Majoranas we can simulate by exact diagonalization, the ground state is well approximated by TFD only in the limit of strong coupling between the two SYK's. 
 Thermodynamic properties are investigated in Section IV. We have found that
 the first order transition at a finite coupling, reported in Ref.~\cite{maldacena2018}, is followed by a sharp crossover for stronger coupling. 
 At the critical coupling  the overlap between the ground state and
 TFD is smallest.
 In the strong coupling limit, the spectral density  develops well separated
 blobs centered around the spin eigenvalues that control the free energy.
 Section V is devoted to the analysis of level statistics.
 For weak coupling, and in the infrared region of the spectrum,
 corresponding to the traversable wormhole phase, we did not find
 agreement with the random matrix theory, thus suggesting that the
 gravitational bound states in this phase are not quantum chaotic.
 However, for sufficiently large coupling, corresponding to the black hole
 phase, level statistics are well described by random matrix theory.
 Finally, in Section VI, devoted to outlook and conclusions, we speculate
 that the termination of the observed transition at strong coupling is reminiscent of a Gross-Witten transition \cite{Gross:1980he}, induced by the gradual
 reduction of the effect of interactions by increasing the coupling between the two SYK models.   

\section{The model and its symmetries}

We study $N$ Majorana fermions in $(0 + 1)$ dimensions where the first $N/2$ fermions, $\psi_L$,
are labeled by Left ($L$), and the remaining $N/2$,  labeled by $R$, $\psi_R$, will be called the Right ($R$) fermions. In each of these two subspaces, the dynamics is governed by the SYK model,
\begin{align}
H_{L} \, &= \, \frac{1}{4!} \sum_{i,j,k,l=1}^{N/2} J_{ijkl} \, \psi_{L,i} \, \psi_{L,j} \, \psi_{L,k} \, \psi_{L,l} \,\nonumber \\
H_{R} \, &= \, \frac{1}{4!} \sum_{i,j,k,l=1}^{N/2} J_{ijkl} \, \psi_{R,i} \, \psi_{R,j} \, \psi_{R,k} \, \psi_{R,l}
\end{align}
where $\psi_{L,i},\psi_{R,i}$ are Majorana fermions
$\{ \psi_{A,i}, \psi_{B,j} \} = \delta_{AB}\delta_{ij}$ $(A,B=L,R)$ and $J_{ijkl}$ are Gaussian distributed random variables
with $\langle J_{ijkl}\rangle =0$ and standard deviation $\langle J_{ijkl}^2\rangle  = 4\sqrt{6}J/{N^{3/2}}$
\cite{kitaev2015,maldacena2016}. 

The two SYK model are coupled by the following interaction: 
\begin{equation} \label{Hint} 
H_{\rm int} = i k \sum_{j=1}^{N/2}    \psi_{L,j} \psi_{R,j}, 
\end{equation}
so that the total Hamiltonian is simply given by, 
\begin{equation} \label{Htotal} 
H_{\rm total} = H_{L} + \alpha H_{R} + H_{\rm  int}
\end{equation}

For the moment we set $\alpha = 1$ in which case the Hamiltonian has an extra global spin symmetry that we will discuss in detail shortly. For the study of level statistics we will find it computationally advantageous to
break this symmetry by setting $\alpha \neq 1$ to a slightly different value which does not affect qualitatively
 most of the key features of the model \cite{maldacena2018}, like for example the observed gap between the ground state and the first excited state.

The anti-commutation relations of $\psi_{L,i},\psi_{R,i}$ can be realized by introducing $N$ Gamma matrices $\Gamma_i$ as
\begin{align}
\label{eq:gammapsi}
\psi_{L,i}=\frac{1}{\sqrt{2}}\Gamma_i,\quad \psi_{R,i}
=\frac{1}{\sqrt{2}}\Gamma_{{N}/{2}+i}.
\end{align}
Since each term in the full Hamiltonian is an even power of the Gamma matrices, the full Hamiltonian preserves chirality, defined as the eigenvalue of $\Gamma_c=\Gamma_1\Gamma_2\cdots\Gamma_N$.
We have summarized the notation for the Gamma matrices
and related symmetries
in appendix \ref{Gammaapp1}.

\subsection{$S$ mod 4 symmetry}
Let us define the spin operator $S$ as
\begin{align}
S=i\sum_{j=1}^{{N}/{2}}\psi_{L,j}\psi_{R,j}=\frac{i}{2}\sum_{i=1}^{{N}/{2}}\Gamma_j\Gamma_{{N}/{2}+j},
\end{align}
that is, $H_{\rm int}=kS$.
Below we show that $S$ has the discrete spectrum $S=-N/4,-N/4+1,\cdots,N/4$ and that, if $\alpha =1$, the full Hamiltonian \eqref{Htotal} preserves $S$ mod 4 in addition to the chirality ($S$ mod 2).
We shall call $S$ the spin operator.

First, let us deduce the spectrum of $S$.
If we define $\Gamma^\pm_i$ as
\begin{align}
\Gamma^\pm_i=\Gamma_{{N}/{2}+i}\pm i\Gamma_i,
\end{align}
then $S$ can be written as
\begin{align}
S=\frac{1}{4}\sum_{i=1}^{{N}/{2}}\Gamma^+_i\Gamma^-_i-\frac{N}{4}.
\end{align}
The operators $\Gamma^\pm_i$ also satisfy the following (anti)commutation relations

\begin{align}
\{\Gamma^\pm_i,\Gamma^{\pm'}_j\}&=0\quad (i\neq j),\quad 
\Gamma^\pm_i\Gamma^\pm_i=0\quad \text{(with no contraction)}\nonumber \\
[S,\Gamma^\pm_i]&=\pm \Gamma^\pm_i,
\end{align}
from which we can derive the following  spectrum:
\begin{align}
\begin{tabular}{|c|c|c|}
\hline
eigenvalue of $S$&eigenstates                                    &degeneracy\\ \hline
$-\frac{N}{4}$   &$|S=-N/4\rangle$                               &$1$\\ \hline
$-\frac{N}{4}+1$ &$\Gamma^+_i|S=-N/4\rangle$                     &$N/2$\\ \hline
$-\frac{N}{4}+2$ &$\Gamma^+_{[i}\Gamma^+_{j]}|S=-N/4\rangle$&${N/2\choose 2}$\\ \hline
\vdots&\vdots&\vdots\\ \hline
$-\frac{N}{4}+p$ &$\Gamma^+_{[i_1}\Gamma^+_{i_2}\cdots\Gamma^+_{i_p]}|S=-N/4\rangle$&${N/2\choose p}$\\ \hline
\vdots&\vdots&\vdots\\ \hline
$\frac{N}{4}$ &$\Gamma^+_1\Gamma^+_2\cdots\Gamma^+_{N/2}|S=-N/4\rangle$&$1$\\ \hline
\end{tabular}
\label{spectrumofS}
\end{align}
where $|S=-N/4\rangle$ is the lowest spin state  satisfying
\begin{align}
\Gamma^-_i|S=-N/4\rangle=0.
\end{align}

Next we discuss the $S$ mod 4 symmetry.
If we define
\begin{align}
P_{ij}&=2(\psi_{L,i}\psi_{L,j}+\psi_{R,i}\psi_{R,j})=\Gamma_i\Gamma_j+\Gamma_{N/2+i}\Gamma_{N/2+j},\nonumber \\
Q_{ij}&=2(\psi_{L,i}\psi_{L,j}-\psi_{R,i}\psi_{R,j})=\Gamma_i\Gamma_j-\Gamma_{N/2+i}\Gamma_{N/2+j},
\end{align}
the full Hamiltonian \eqref{Htotal} with $\alpha=1$ can be written as
\begin{align}
H=\frac{1}{8}\sum_{i<j<k<l}^{N/2}J_{ijkl}(P_{ij}P_{kl}+Q_{ij}Q_{kl})+kS.
\end{align}
The $P_{ij}$ and $Q_{ij}$ operators satisfy the following commutation relations with $S$
\begin{align}
[S,P_{ij}]=0,\quad
[S,[S,Q_{ij}]]-4Q_{ij}=0.
\label{[S,Q]}
\end{align}
That is, $P_{ij}$ preserves the spin $S$.
To interpret the commutation relation for $Q_{ij}$ it is convenient to act Eq.~\eqref{[S,Q]} on an eigenstate of $S$,
  with the spin eigenvalue $s$, which we shall call $|\phi\rangle$. One easily derives
\begin{align}
(S^2-2sS+s^2-4)Q_{ij}|\phi\rangle=0.
\end{align}
This implies that an expansion of  $Q_{ij}|\phi\rangle$ contains only spin eigenstates with $S=s\pm 2$.
Hence $S$ mod 4 is a symmetry of $Q_{ij}Q_{k\ell}$.
Putting the above computations together we find that the full Hamiltonian Eq.~\eqref{Htotal} preserves also
the $S$~mod~4 symmetry.

It is instructive to express the exponent of the spin operator as a product 
\be
e^{i\theta S} =\prod_{i=1}^{N/2} \left(\cos \frac{\theta}{2} -\sin \frac{\theta}{2} \Gamma_i\Gamma_{ N/2+i} \right ).
\label{productform}
\ee
For $\theta = \pi $ we recover the chirality operator, while $\theta = \pi/2$ gives the $S$ mod 4 symmetry,
\be
   [e^{\pi iS/2},H]=0.
   \ee
This also shows that the chirality symmetry operator is the square of the $S$ mod 4 operator.
In appendix \ref{Gammaapp13}, we give an alternative proof of the $S$ mod 4 symmetry based on the product form Eq.~\eref{productform}.

This $S$ mod 4 symmetry plays an important role when we analyze the level correlations.
For $\alpha=1$, the full Hamiltonian splits into four blocks labeled by $S$ mod 4, regardless of the value of the random coupling $J_{ijkl}$. When analyzing level correlations, unitary symmetries
have to be taken into account exactly and we have to focus on a single block.
We revisit this issue in section \ref{sec_Levelstatistics}.
For $\alpha \neq 1$, the full Hamiltonian only preserves chirality and there is no $S$ mod 4 symmetry.

\section{Comparison of the TFD state with ground state}
\label{sec_TFD}
Here we consider the case with identical L and R systems, i.e. $\alpha=1$.
For $k=0$ (and $N=0$ mod 4 so that $N/2$ is even), the system we are considering is $H=H_{SYK}\otimes 1+1\otimes H_{SYK}$.
That is, the system consists of two decoupled SYK systems with $N/2$ fermions,
the L and R systems, which are identical to each other.
It was already pointed out in \cite{maldacena2018} that when the interaction between L-system and R-system is turned on the ground state has a large overlap
with the Thermo Field Double (TFD) state of $H_{SYK}\otimes 1+1\otimes H_{SYK}$.
In this section, we shall study this similarity in more detail.

Because the left and right systems  are identical, the left and right $\Gamma$-matrices
can be represented as
\be
\Gamma_k^L& =& \gamma_k \otimes 1 ,\nn\\
\Gamma^R_k& =&\gamma_c \otimes  U \gamma_k  U^{-1},
\label{Tensorbasis}
\ee
where $U$ is a unitary matrix (we use the notation that the lower case
gamma matrices are of size $2^{N/4}\times 2^{N/4}$ and upper case gamma matrices
are of size $2^{N/4}\times 2^{N/4}$).

The factor $\gamma_c$ ensures that the left $\Gamma$ matrices anti-commute with the right $\Gamma$ matrices. Below, we will call this basis the tensor basis.
If the eigenvalues and eigenstates of $H_L$ are given by
\be
H_L |n\rangle_L = E_n  |n\rangle_L,
\ee
then
\be
H_R |n\rangle_R = E_n U|n\rangle_R, \;\text{with} \; |n\rangle_R = U|n \rangle_L.
\ee
Let us define the TFD state,
\be
|{\rm TFD}\rangle = \sum_n e^{-2\beta  \, E_n+i\phi_n} |n\rangle_L e^{i\frac \pi 4 \gamma_c} CK|n\rangle_R,
\label{tfd-gen}
\ee
where we have included a phase factor, $\exp i\phi_n$ and $CK$ is the charge conjugation operator.
We will next show that 
\be
\label{eq:smallestspin}
|I\rangle \equiv|S=- N/4\rangle=\sum_n |n\rangle_L\otimes e^{i\frac \pi 4 \gamma_c}CK|n\rangle_R
\ee
is the ground state of the spin operator. To this end we calculate the expectation value
\be
\langle I|S|I\rangle &=&2^{-N/4}\sum_{n,m}  \Bigl(\langle n|_L\otimes \langle  e^{i\frac \pi 4 \gamma_c}  CK n |_R\Bigr)
S
\Bigl(|m\rangle_L\otimes e^{i\frac \pi 4 \gamma_c}C K|m\rangle_R\Bigr)\nn\\
&=& 2^{-N/4}\frac i2\sum_{n,m}\sum_k \langle n|_L \gamma_k \gamma_c |m\rangle_L \langle  e^{i\frac \pi 4 \gamma_c} CK n |_R
\gamma_k   e^{i\frac \pi 4 \gamma_c}CK|m\rangle_R\nn\\
&=& 2^{-N/4}i\sum_{n,m}\sum_k\chi_m \langle n|_L \gamma_k  |m\rangle_L
\frac i2\chi_m\langle m|_R \gamma_k |n\rangle_R,\label{ground1}
\ee
where $\chi_m$ is the chirality of the state $|m\rangle$,
\be
\gamma_c|m\rangle= \chi_m|m\rangle.
\ee
Choosing $|m\rangle_R =|m\rangle_L $ (i.e. $U=1$  in Eq. \eref{Tensorbasis})
 the sum over $m$ can be eliminated by completeness.
\be
\text{r.h.s. of Eq. \eqref{ground1}}&=& -2^{-N/4}\frac 12\sum_{n}\sum_k \langle n|_R \gamma_k^2   |n\rangle_R\nn\\
&=& -\frac N4,\label{ground2}
\ee
which is the desired result. Therefore $|I\rangle$  is the state with the lowest spin. A similar prove can be used to show that $\| \Gamma_i^- |S=-N/4\rangle\|^2=0$ (no sum over $i$). This shows that
the  state $|I\rangle$ is annihilated  by all lowering operators ${\Gamma^-}_i$.
The details of the ground state which  depend on $N/2 \mod 8$ are worked out in Appendix \ref{Gammaapp14} for $N/2 \mod 8 =4$, in Appendix \ref{sec:tfdN20} for $N/2 \mod 8 =2$ and
in Appendix \ref{app:mod0} for $N/2 \mod 8 =0$.

At finite temperature,  we  define the TFD state  $|\text{TFD}\rangle_\beta$ as
\begin{align}
|TFD\rangle_\beta=\frac{1}{\sqrt{Z(N,\beta)}}e^{-\beta(H_{SYK}\otimes 1+1\otimes H_{SYK})}|S  =-N/4\rangle,
\label{TFD}
\end{align}
where $Z(N,\beta)$ is a normalization factor  such that $\langle \text{TFD}|_\beta\text{TFD}\rangle_\beta=1$.
As confirmed by numerical calculations for $ N$ up to 32, the complex phase factor of the $ |n\rangle \otimes |n\rangle$ components of the
ground state at finite coupling $k$, does not depend on $k$. This is a nontrivial result which we only understand perturbatively.

\subsection{$|\text{TFD}\rangle_\beta$ versus ground state of coupled SYK}

In \cite{maldacena2018} it has been argued that the ground state $|\text{gs}\rangle$ of the coupled system at finite coupling $k$ is close to the TFD state $|\text{TFD}\rangle_\beta$ by studying the overlap $\langle \text{TFD}|_\beta\text{gs}\rangle$.
Here the inverse temperature $\beta$ is a function of $k$ determined by maximizing the overlap.
However, it should be stressed that a large value of this overlap does not necessarily imply that the coefficients $(\langle n|\otimes \langle n|)|\text{gs}\rangle$, taken as a function of the energy levels of the single SYK system, behave like $e^{-2\beta E_n}$.
The reason is that for small coupling most of the strength of the wave function can be localized in one or few components of the wave function.
In this section we will show numerically that, at finite $N \leq 28$, this is actually the case and that for sufficiently small values of the coupling $k$ the coefficients display large deviations from the exponential behavior typical of the TFD state.

Before moving to a detailed analysis, let us first consider the results in some limiting cases.
In the limit $k\rightarrow\infty$, since $H$ is dominated by $H_{int}=kS$, the ground state $|gs\rangle$ coincides with $|\text{TFD}\rangle_{\beta=0}$, see Eqs. \eref{ground1}, \eref{ground2} and \eqref{TFD}.
In the limit $k\rightarrow 0$, on the other hand, the tensor product states $|m_\pm\rangle_L \otimes |n_{\pm'}\rangle_R$ are the exact eigenstates of $H$ with the eigenvalues $E_m+E_n$, where the $E_n$ are the eigenvalues of a single SYK, $H_{SYK}|n_\chi\rangle_{L,R}=E_n|n_\chi\rangle_{L,R}$, and the index $\chi=\pm$ denotes the chirality of the state.
The ground state is
$|1_\chi\rangle_L\otimes |1_{\chi'}\rangle_R$ which may be degenerate depending on the value of $N$.
Hence the ground state in the $k\rightarrow 0$ limit is given by
the  $|\text{TFD}\rangle$ state for  $\beta=\infty$. Note that  this does not necessarily imply that the ground state is close to a 
  TFD state for small but nonzero $k$.

\newcommand{\gs}{|{\rm gs}\rangle}

  Let us compare the ground state of the Hamiltonian  Eq.~\eqref{Htotal}   with the TFD state at finite temperature $|\text{TFD}\rangle_\beta$ for varying $k$.
 It can be expanded in the complete set $|m\rangle \otimes|n\rangle$
  \be
  \label{eq:gsexpansion} 
  \gs = \sum_n c_n |n\rangle\otimes|n\rangle + \sum_{m\ne n} b_{mn} |m\rangle\otimes |n\rangle.
  \ee
  Numerically, the phases of the $c_n$ do not depend on the coupling constant and are thus
  given by the phases of the ground state  of $S$ (perturbed by a $\epsilon(H_L+H_R)$ with $\epsilon$
  a small constant). This is not the case for the phases of the coefficients $b_{mn}$.
  We can understand this fact  to first order in perturbation theory in $1/k$ with $S$ as the zeroth
  order Hamiltonian, 
  \be
  \gs = |I \rangle + \sum_m|S_m\rangle \frac{\langle S_m | (H_L+H_R)|I\rangle}{S_0 - S_m}
  \ee
  with $|S_m\rangle $ an eigenstate of $S$ with eigenvalue $S_m$ and $S_0$ the spin of the ground state. Inserting a complete set of eigenstates
  of $H_L +H_R$, $\sum_{m,n}|m\rangle\otimes|n\rangle\langle m|\otimes \langle n|$, and noticing that $\Bigl(\langle m|\otimes \langle n|\Bigr)(H_L+H_R)|I\rangle=0$ for $m\neq n$, the matrix element can be written as
  \be
  \langle S_m | (H_L+H_R)|I\rangle =  \sum_n \langle S_m|\Bigl(|n\rangle\otimes |n\rangle\Bigr)\Bigl(\langle n|\otimes \langle n|\Bigr) (H_L+H_R)|I\rangle.
  \ee
  Only those states with the same $S$ mod 4 spin as the ground state contribute to the sum over $n$. It turns
  out that the complex phase of the matrix elements $ \langle S_m|\Bigl(|n\rangle \otimes |n\rangle\Bigr)$ does not depend on
  $m$ apart from an overall phase which cancels with the contribution to the matrix element
\be
  \langle S_m|\Bigl(|n\rangle\otimes |n\rangle\Bigr)\Bigl(\langle n|\otimes \langle n |\Bigr)(H_L+H_R)|I\rangle 
  = 2\langle S_m|\Bigl(|n\rangle\otimes |n\rangle\Bigr) E_n\Bigl(\langle n|\otimes \langle n|\Bigr) |I\rangle,
  \ee
  which is therefore real. The complex phase of the first order correction is therefore due to the phase
  of $\Bigl(\langle n |\otimes \langle n|\Big)|S_m\rangle$ which is the same as the phase of $\Bigl(\langle n |\otimes \langle n|\Bigr)|I\rangle$ (up to a minus sign).
  The same argument cannot be made for third and higher order perturbative
  corrections because off-diagonal  states, $|m\rangle\otimes |n \rangle$ (with
  $m\ne n$) contribute to the intermediate state sums.
 We conclude that at first and second order in  perturbation theory
  the complex phase of the ground state does not depend on $k$, up to a value of $k$ for which the perturbative
  contributions are larger than the $|n\rangle\otimes |n\rangle$ components of the ground state. At that point one or more
  components of the ground state change sign.

 In the numerical studies below  we will only consider the cases $N = 20$ and $N=28$, where $|\text{TFD}\rangle_\beta$ is defined as, see Eq. \eqref{TFD},
\begin{align}
|\text{TFD}\rangle_\beta=\frac{1}{\sqrt{Z(\beta)}}\sum_{n=1}^{2^{N/4-1}}e^{-2\beta E_n}(|n_+\rangle_L\otimes |n_-\rangle_R+|n_-\rangle_L\otimes |n_+\rangle_R) \ ,
\label{TFDexplicit}
\end{align}
and $Z(\beta)=2\sum_{n=1}^{2^{N/4-1}}e^{-4\beta E_n}$ is the normalization factor.

\begin{figure}
\centering
\resizebox{0.8\textwidth}{!}{\includegraphics{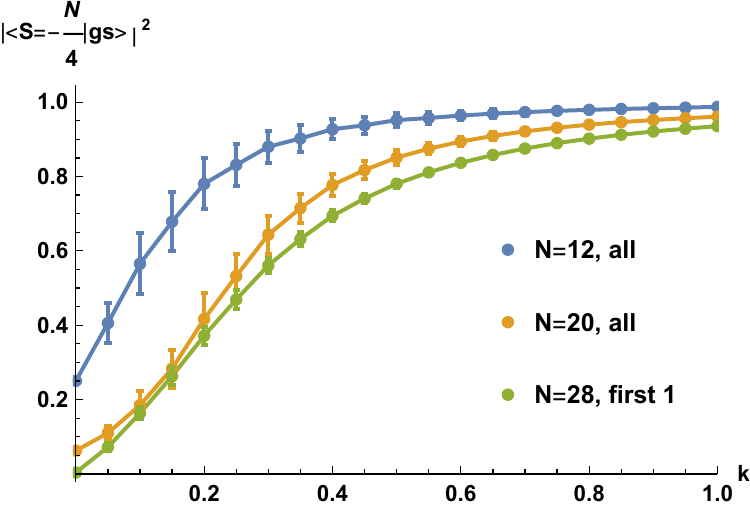}}
\vspace{-4mm}
\caption{Overlap between the ground state of the spin operator and the ground state of the coupled SYK model as a function of the coupling $k$.
Here the results for $N=12$ and $N=20$ are obtained by diagonalizing the full Hamiltonian and then extracting the eigenvector with the lowest energy eigenvalue,
while for $N=28$ we have obtained only the ground state eigenvector by applying
the so called Arnoldi method to the Hamiltonian shifted by a constant matrix $H-\text{diag}(100,100,\cdots,100)$.
Here the shift is required technically so that the eigenvalue of the ground state has the largest absolute value among the full spectrum.
%   Here the results for $N=12$ and $N=20$ are obtained by diagonalizing
%   $H_\text{total}$ while the results for $N=28$ are obtained by diagonalizing $H_\text{total}-\text{diag}(100,100,\cdots,100)$ in Mathematica.
Only for large $k$ is the ground state of the spin operator a good description of the ground state of the coupled SYK model.
}
\label{gsov}
\end{figure}

Since $k$ is the parameter of the Hamiltonian that interpolates between a ground state  $|\text{TFD}\rangle_{\beta=\infty}$ and $|\text{TFD}\rangle_{\beta=0}$ which also corresponds to the ground state of the spin operator,
we expect that the effective inverse temperature $\beta(k)$ monotonically decreases as $k$ increases.
Furthermore, we have computed $| \langle S=-N/4|gs\rangle|^2$ and found that this overlap is almost 1 already around $k\sim 1$ (see Fig.~\ref{gsov}).
Hence we expect the effective temperature changes substantially only in the regime $0<k<1$.

\begin{figure}
\includegraphics[width=8cm]{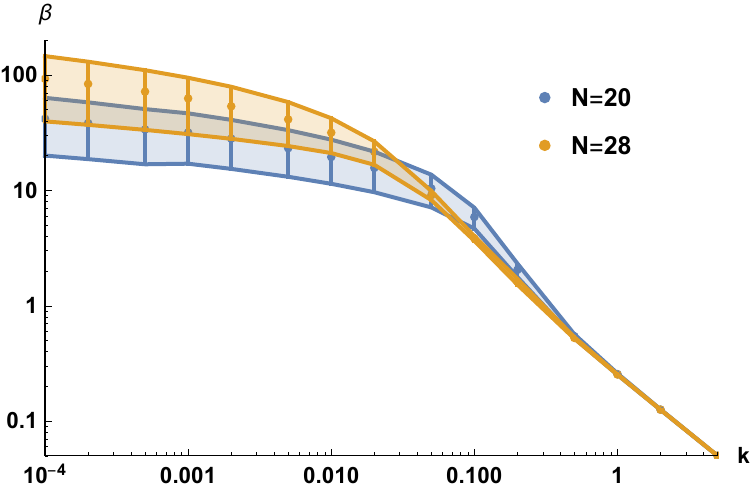}
\includegraphics[width=8cm]{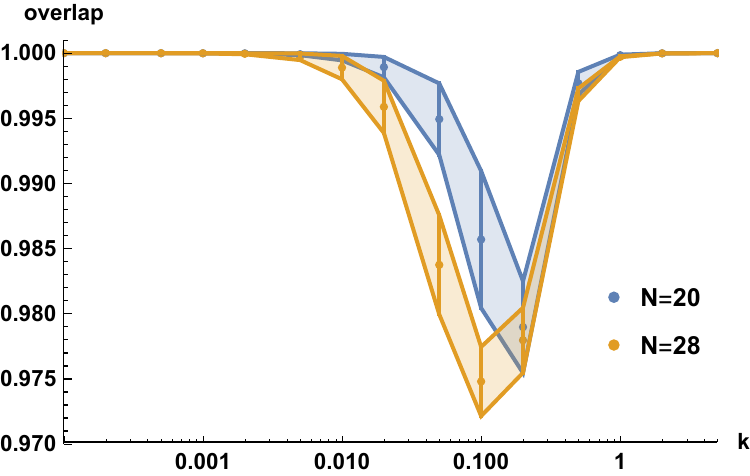}
\caption{
Left: inverse temperature $\beta$ as a function of the coupling $k$.
Right: overlap $\langle \text{TFD}|_\beta \text{gs}\rangle$ as a function of $k$.
}
\label{betakandoverlapTFD}
\end{figure}

Below, to be self-contained, we first redo the comparison of the ground state and the TFD state by maximization of their overlap.
This computation was already performed in \cite{maldacena2018}.
We have checked numerically, for $N=20$ and $N=28$, that the maximized overlap
with respect to $\beta$,
between the TFD state and the ground state, is close to $1$.
For this purpose, we have considered the ansatz Eq.~\eqref{TFDexplicit} for the ground state,
in which the $E_n$ are the energy levels of a single SYK model and $\beta$ is the fitting parameter  determined by maximizing the overlap.
We  normalize this state as $\langle \text{TFD}|_\beta\text{TFD}\rangle_\beta=1$ by choosing $Z(\beta)$ as
\begin{align}
Z(\beta)&=2\sum_{n=1}^{2^{N/4-1}}e^{-4\beta E_n} \ ,
\end{align}
which therefore has  not been considered as an independent fitting parameter.
We proceed as follows:
\begin{enumerate}
\item Generate  an ensemble of $2^{N/4}\times 2^{N/4}$ matrices $H_{SYK}=\frac{1}{2}\sum_{i<j<k<\ell}J_{ijk\ell}\gamma_i\gamma_j\gamma_k\gamma_\ell$ with Gaussian random $J_{ijk\ell}$,
  and compute their eigenvalues $\{E_n\}$ and eigenvectors $\{|n_\pm\rangle\}$ through exact numerical diagonalization.
Then construct the tensor product $|n_\pm\rangle_L\otimes |n_\mp\rangle_R$.

\item For the same ensemble of random couplings, generate $2^{N/2}\times 2^{N/2}$ matrices representing the Hamiltonian Eq.~\eqref{Htotal} with $\alpha=1$
using the tensor basis \eqref{Tensorbasis} and compute the ground state $|\text{gs}\rangle$ through the exact diagonalization.

\item Compute $c^\pm_n=(\langle n_\pm|_L\otimes \langle n_\mp|_R) |{\rm gs}\rangle$.
We have observed that $c^+_n=c^-_n$ and $c^+_n>0$ for all $n= 1,2,\cdots 2^{N/4}/2$, up to an overall phase provided that we define the phase of $|n_\pm\rangle_{L,R}$ such that Eq.~\eqref{TFDexplicit} holds without any extra phase (see appendix \ref{Gammaapp14} for detail).
Hence we shall denote $c^+_n=c^-_n$ simply by $c_n$.
\item For each ensemble realization determine $\beta$ by maximizing the overlap $\langle\text{TFD}|_\beta \text{gs}\rangle$.

\begin{align}
\langle \text{TFD}|_\beta \text{gs}\rangle=\frac{2}{\sqrt{Z(\beta)}}\sum_{n=1}^{2^{N/4-1}}c_ne^{-2\beta E_n}
\label{chi2}
\end{align}

\item Take average of both $\beta$ and $\langle \text{TFD}|_\beta \text{gs}\rangle$ over the ensemble.
\end{enumerate}
Following this procedure we obtained both the effective inverse temperature $\beta(k)$ and the overlap $\langle \text{TFD}|_\beta \text{gs}\rangle$.
The results are displayed in Fig.~\ref{betakandoverlapTFD}.
As we can see, the inverse temperature is indeed growing for small $k$ and it tends to zero at strong coupling as expected.
In agreement with \cite{maldacena2018}, we confirm that $\langle \text{TFD}|_\beta\text{gs}\rangle$ is pretty close to $1$ with relatively large deviations around $k \sim 0.1$.
More quantitatively, for small $k$, the inverse effective temperature depends on $k$ roughly as $1/k^{1/6}$
while at $k\approx 0.1$, there is a transition to a $1/k$ dependence.

\begin{figure}[t!]
\includegraphics[width=5cm]{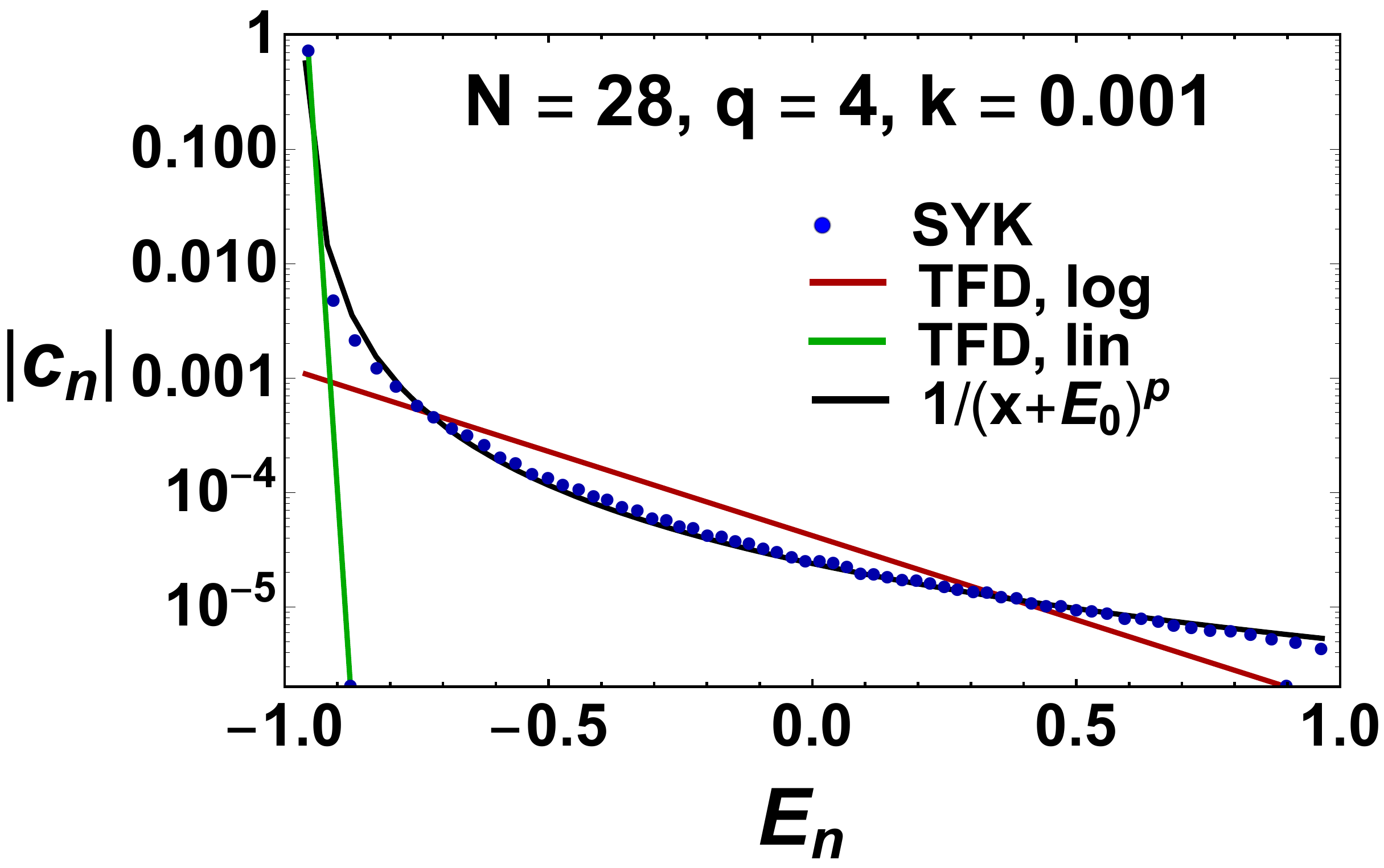}
\includegraphics[width=5cm]{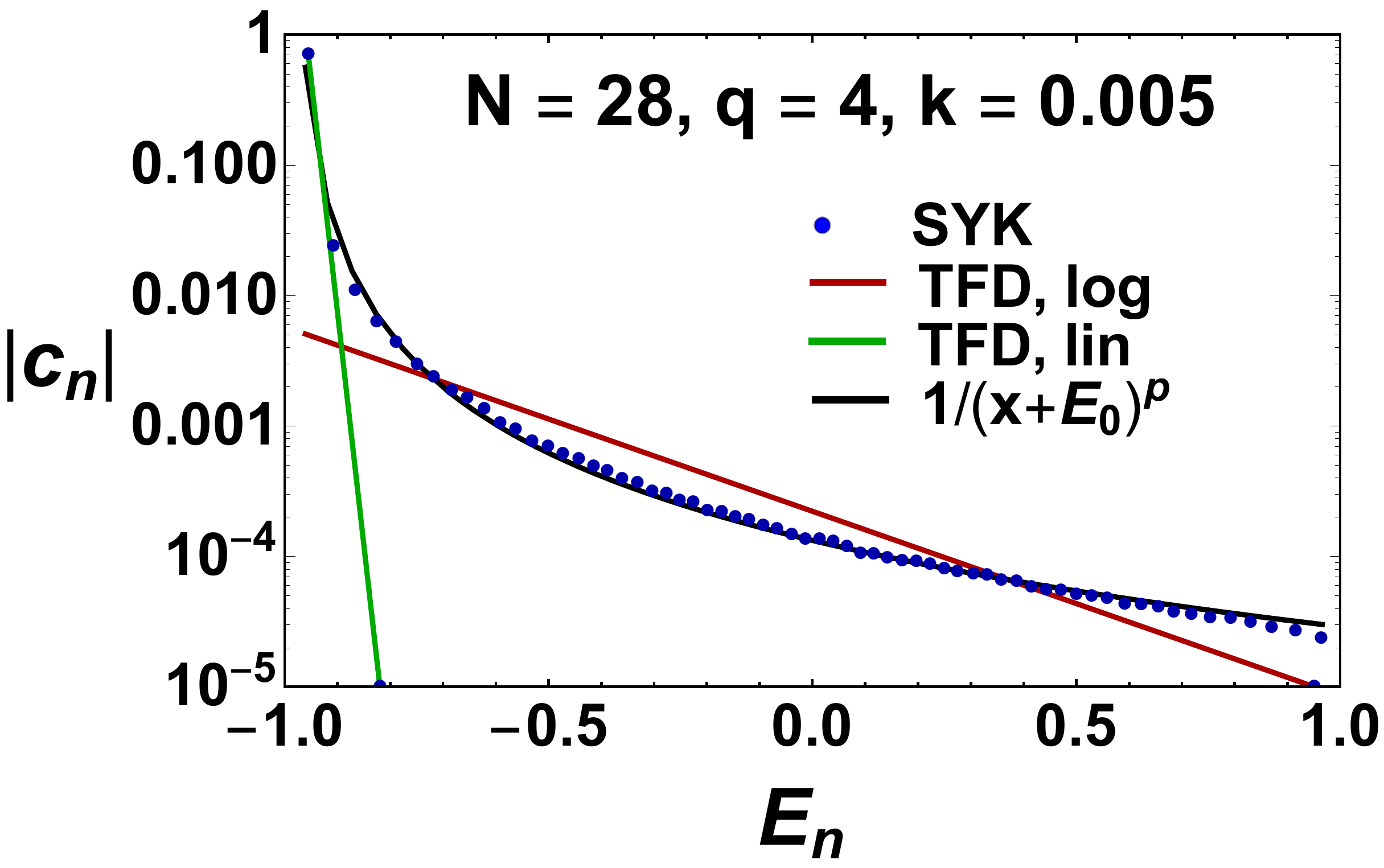}
\includegraphics[width=5cm]{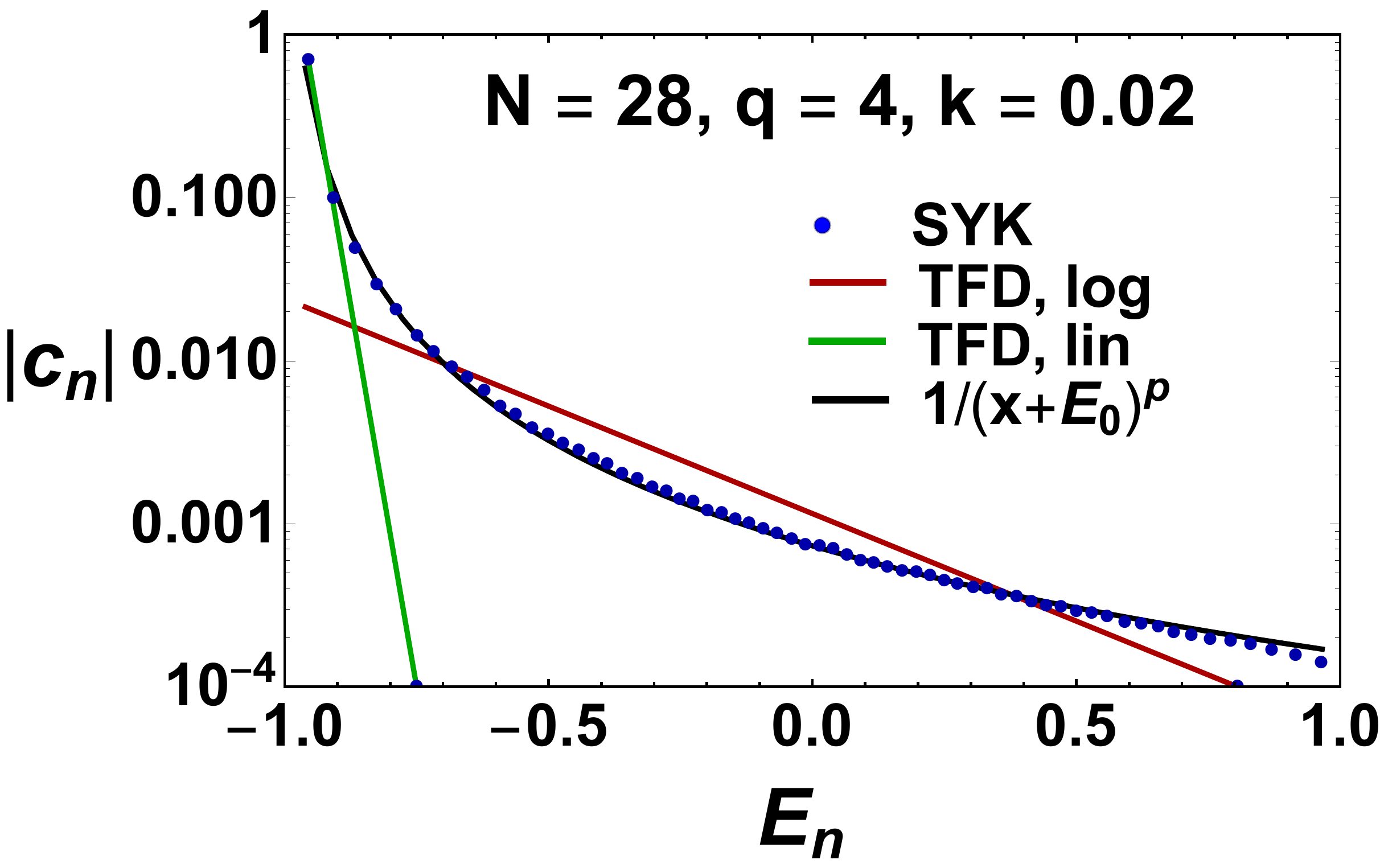}\\
\includegraphics[width=5cm]{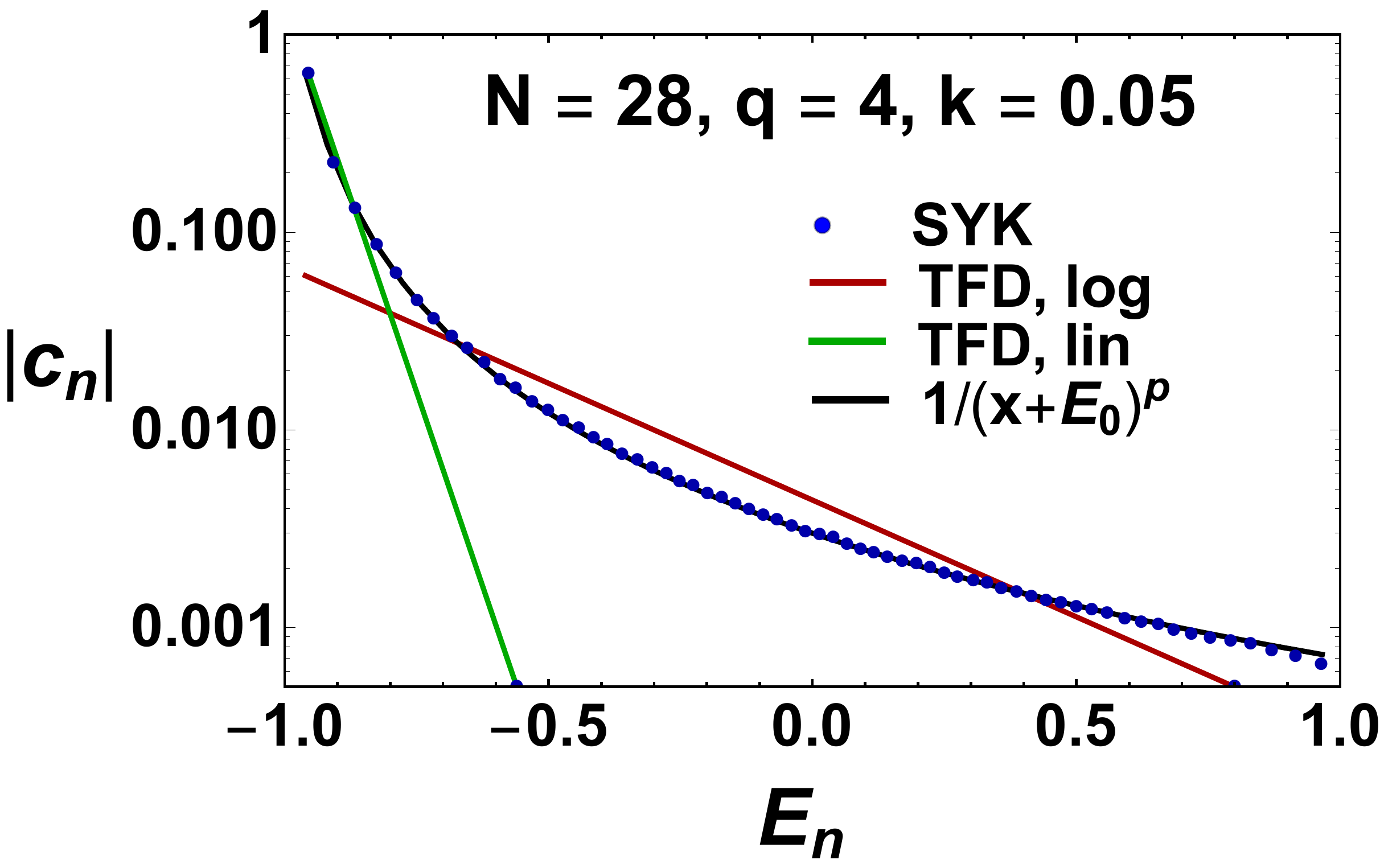}
\includegraphics[width=5cm]{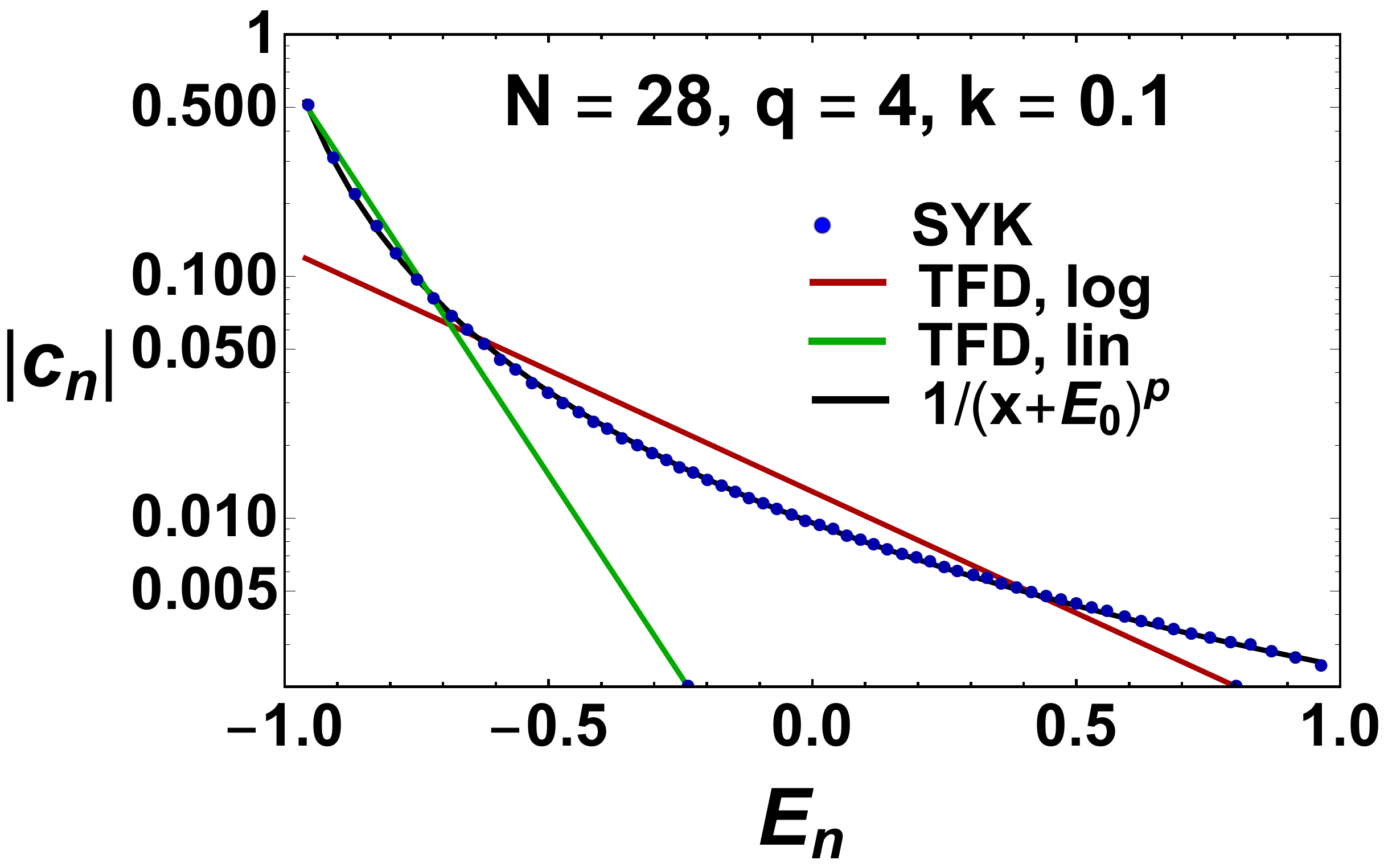}
\includegraphics[width=5cm]{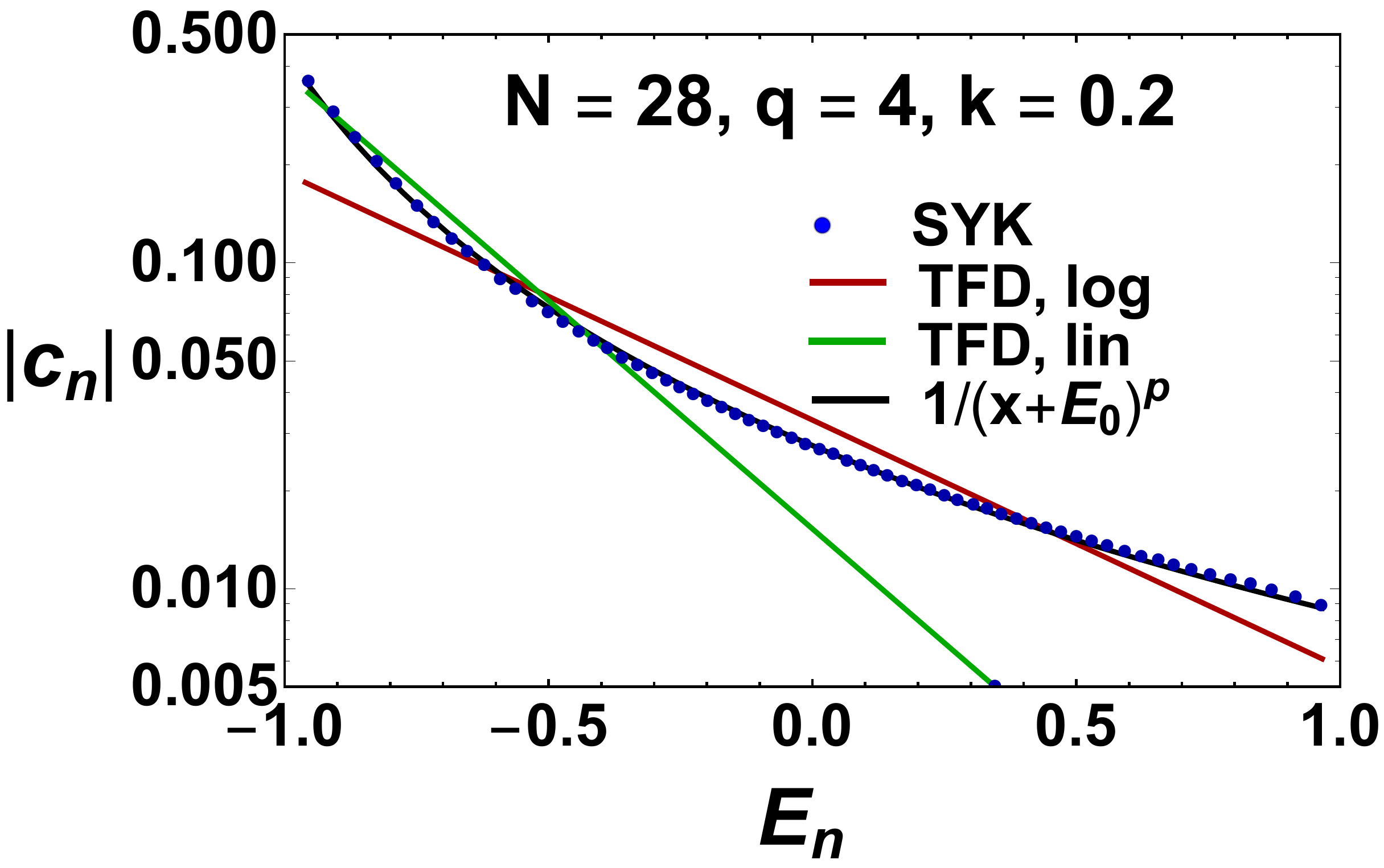}\\
\includegraphics[width=5cm]{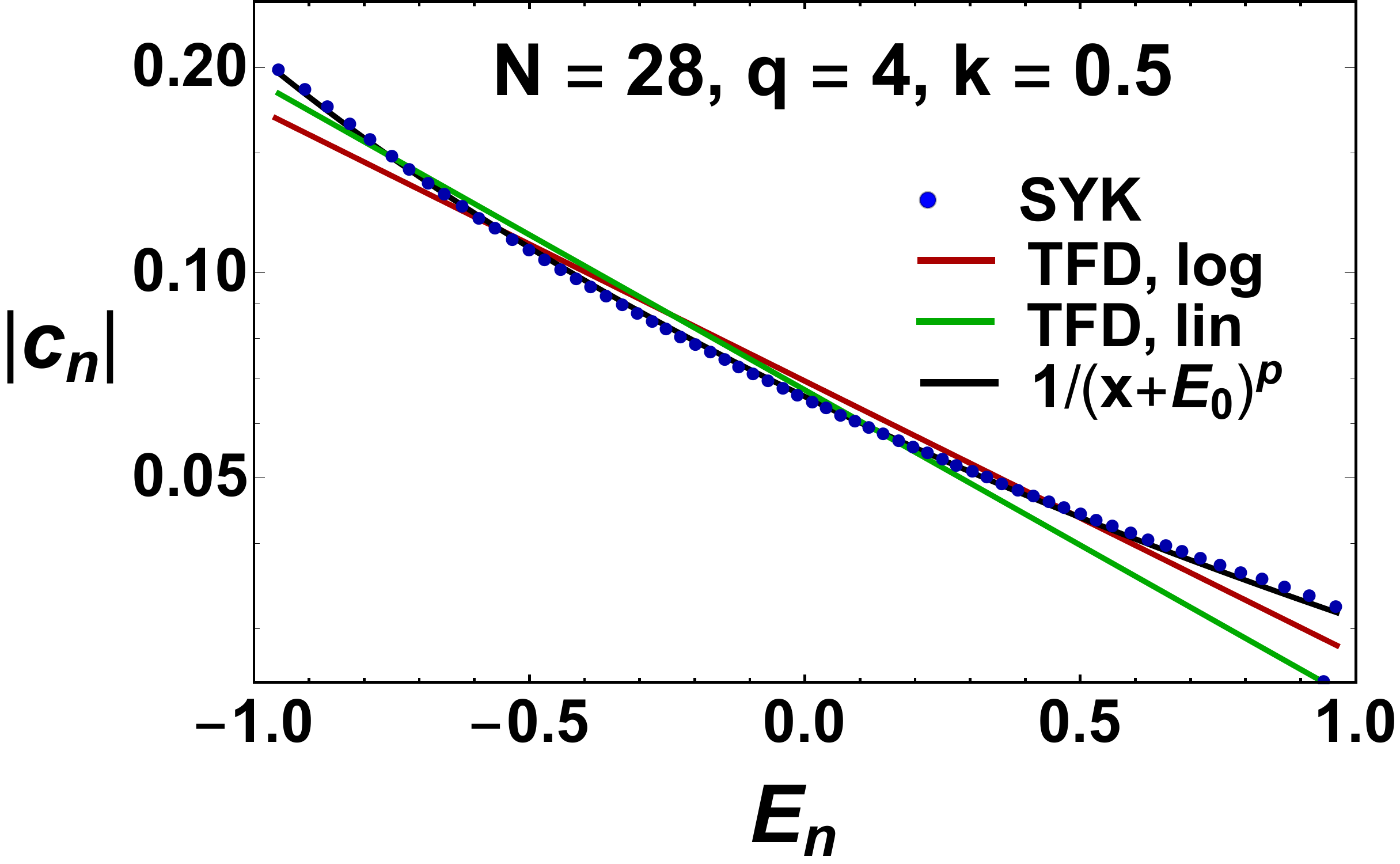}
\includegraphics[width=5cm]{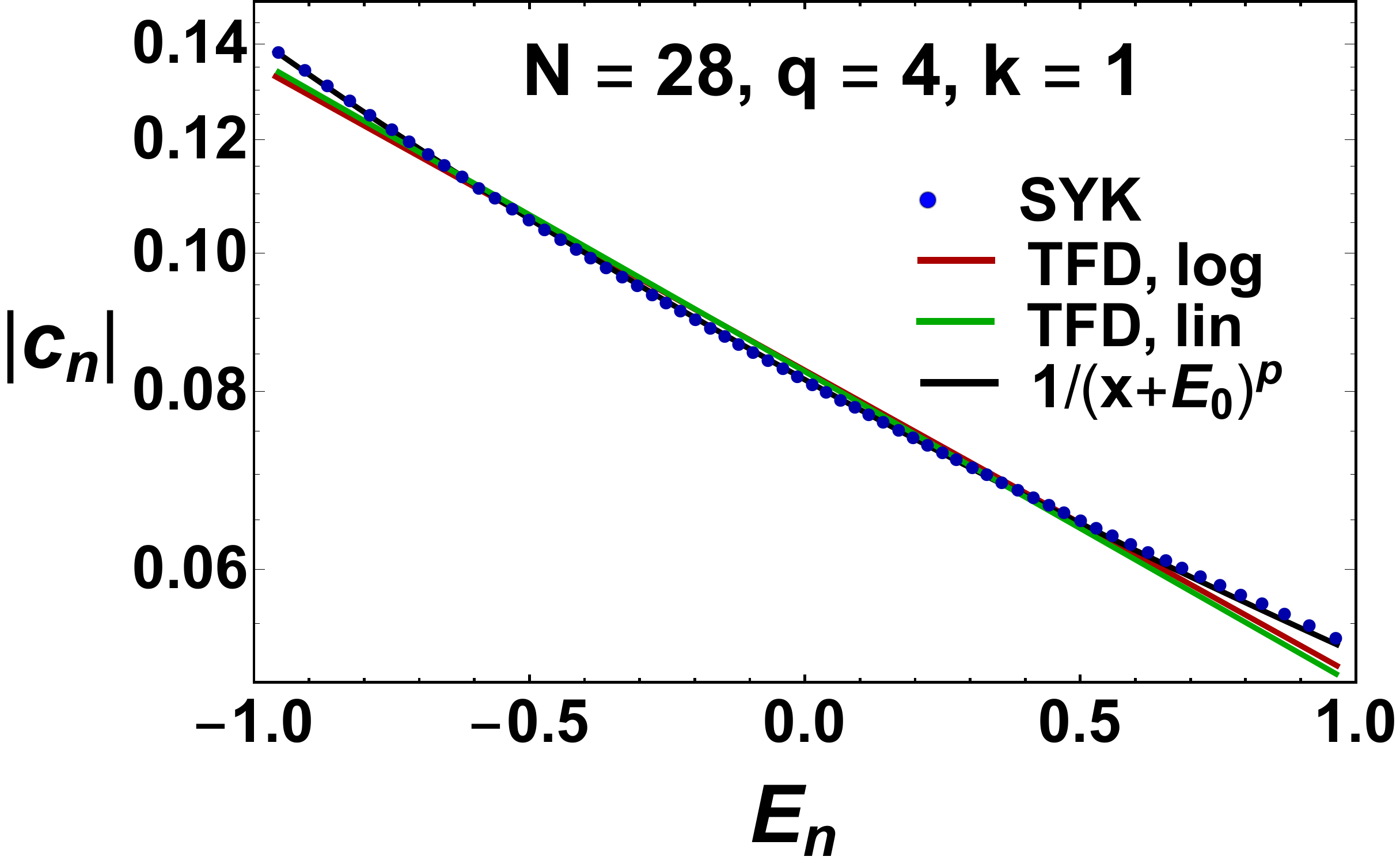}
\includegraphics[width=5cm]{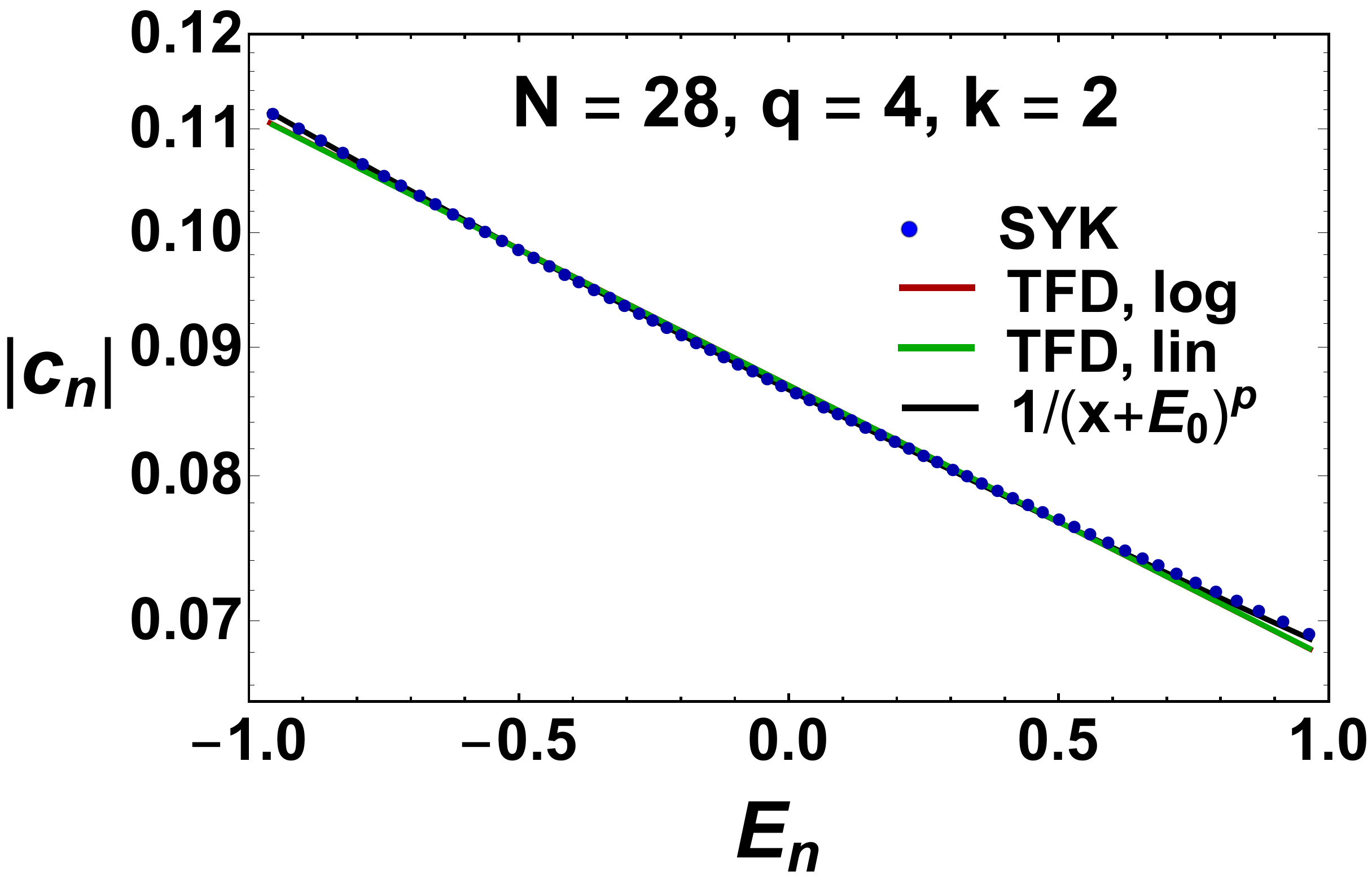}
  \caption{
    The agreement between the average ground state coefficients $c_n$ (black points)
    and the TFD
    obtained by fitting the $|c_n|$ configuration by configuration (green curve), or by
    fitting to the logarithm of the ensemble average of the $|c_n|$ (red and black curves). The 
    power law ansatz (black curve) gives  much better fit especially for small $k$.
    Note that the green curve has one fitting parameter ($\beta$), while the
    red (exponential) and black (power law)
    curves are two-paramter fits, $c$ and $\beta$ with $c$ the normalization constant, and $c$ and $e_0$, repectively. The value of the power in the latter case is
    $p=2.178$.
  }
\label{TFDvspower}
\end{figure}

As was already mentioned, the large overlap between the ground state and the TFD, does not necessarily imply that the coefficients $c_n$ have an exponential behavior in energy. 
Indeed, it could be simply that one or a few coefficients in \eqref{eq:gsexpansion}
are much larger than the others. In this case, the overlap $\langle \text{TFD}|_\beta\text{gs}\rangle$  would be very close to be $1$, even though all the other coefficients could behave in a completely different way.

Hence, to check whether the coefficients of the ground state have an exponential profile, we have studied their average  on a logarithmic scale versus the ensemble average of each of the $E_n$
(for the case $N = 28$ only). Results, depicted in  Fig.~\ref{TFDvspower}, clearly show that an exponential form is not a good fit to the
$|c_n|$ coefficients. For comparison, we have also included the coefficients of the
TFD state with $\beta$ determined by maximizing the overlap (green curves) as described before, and fitting
the logarithm of the ensemble average of $|c_n|$ to the TFD form also leaving the normalization constant as a free parameter (red curve).
None of these fitting to the TFD state 
give good results. By contrast, a power law ansatz (black line),
\begin{align}
\label{eq:poweransatzfirst}
c_n \sim \frac{c}{(E_n - e_0)^p},
\end{align}
provides an excellent fitting. In order to compare the two parameter fits, we
fix $p$ at  $p=2.178$.  
The fitted  value of $e_0$ is somewhat smaller than the
ground state energy $E_1$ for small $k$ and decreasing for increasing values of $k$. The value of $c$ is
in principle determined by the normalization, but leaving it as a free parameter gives
a better fit. For small coupling $k< 0.5$ the value of $c$ is quite
different from the one obtained from the normalization. The reason is that most of the strength of the
wave function is in $c_1$ which does not contribute much to the $\chi^2$ when you fit the logarithm of the
$|c_n|$. One could also take $p$ as a $k$-dependent fitting parameter which gives
slightly better fits. In any case, except in the regime of strong coupling, power law fits provide a much better description of the numerical results than exponential fits related to the TFD state.
For $k \geq 1$ the TFD ansatz is a good description of the ground state though.
This is consistent with the expectation that at large coupling the system is dominated by the spin operator and the TFD state (with $\beta \to 0$) is the ground state of the Hamiltonian Eq.~(\ref{Htotal}).

For $k>0.2$ the energy dependence of the coefficients can also be described by
the sum of two exponentials with a $\chi^2$ that is comparable to the power law fit,
but it has one more parameter. A Gaussian dependence, $\exp(-\beta x -\alpha x^2)$,
first introduced in the context of a nuclear many body system \cite{Verbaarschot:1979kts},
also gives a reasonable fit in this parameter region. 

Once again, we stress that the deviation from the TFD form
is not in contradiction with the good overlap between the TFD and the ground state, since the dominant coefficients are well reproduced by the TFD at least for small $k$, as it can be seen from Fig.~\ref{TFDvspower}.

These numerical results suggest that the coefficients of the ground state are not well
approximated by the TFD ansatz, and that instead they are very close to a power
law ansatz. However, we should note that $N=28$ may not
be close 
to  the large
$N$ limit of this system where analytical arguments
\cite{maldacena2018}, show that the ground state
is well  approximated by a TFD state.
This indicates that the convergence to the large $N$ limit is slow and most likely
  nonuniform.
Interestingly, it has been observed in \cite{Cottrell2018}, that the Hamiltonian Eq.~\eqref{Htotal} is just an approximation of a Hamiltonian whose exact ground state is the TFD.

\begin{figure}
\centerline{\includegraphics[width=7cm]{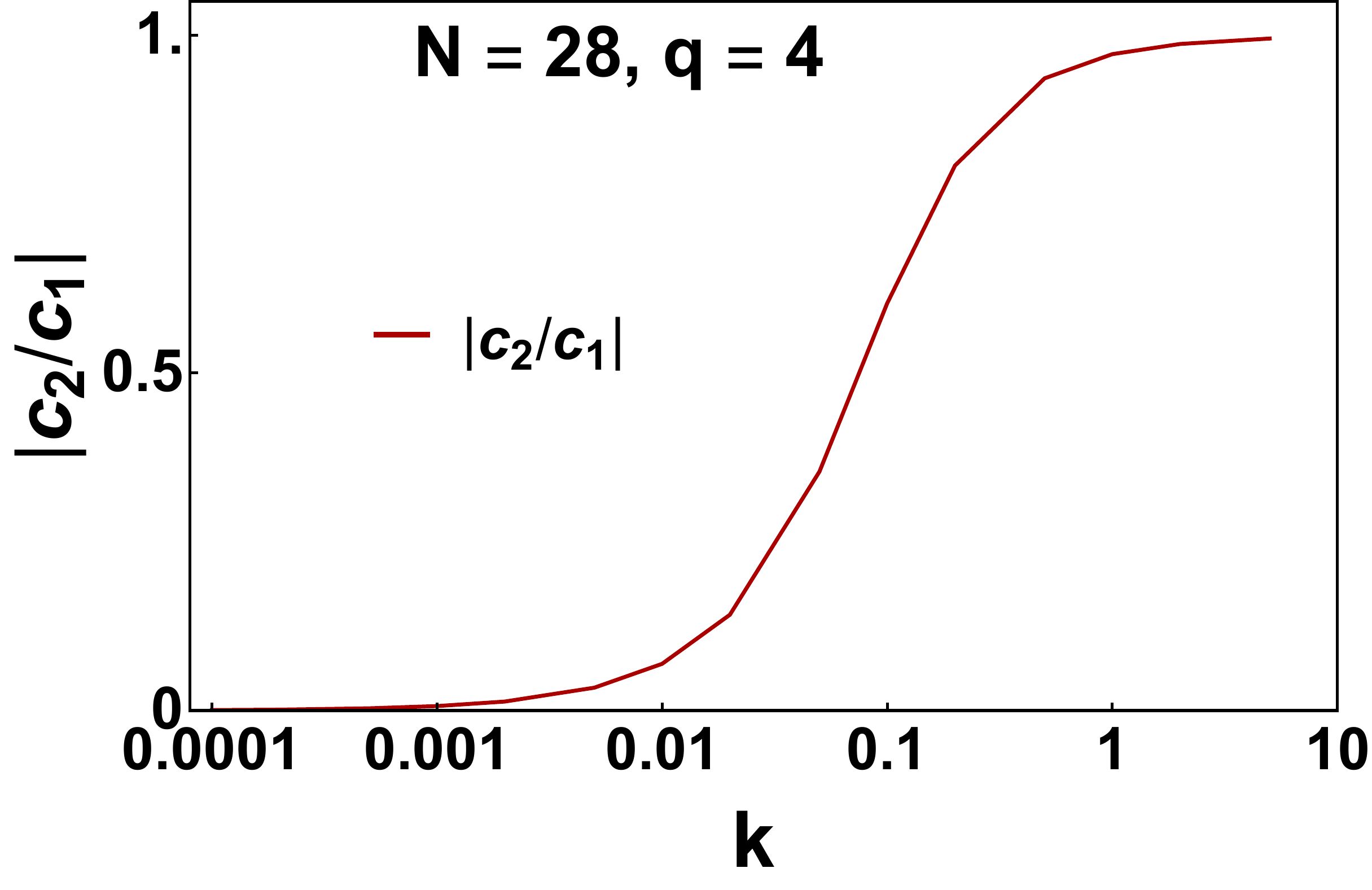}
\includegraphics[width=7cm]{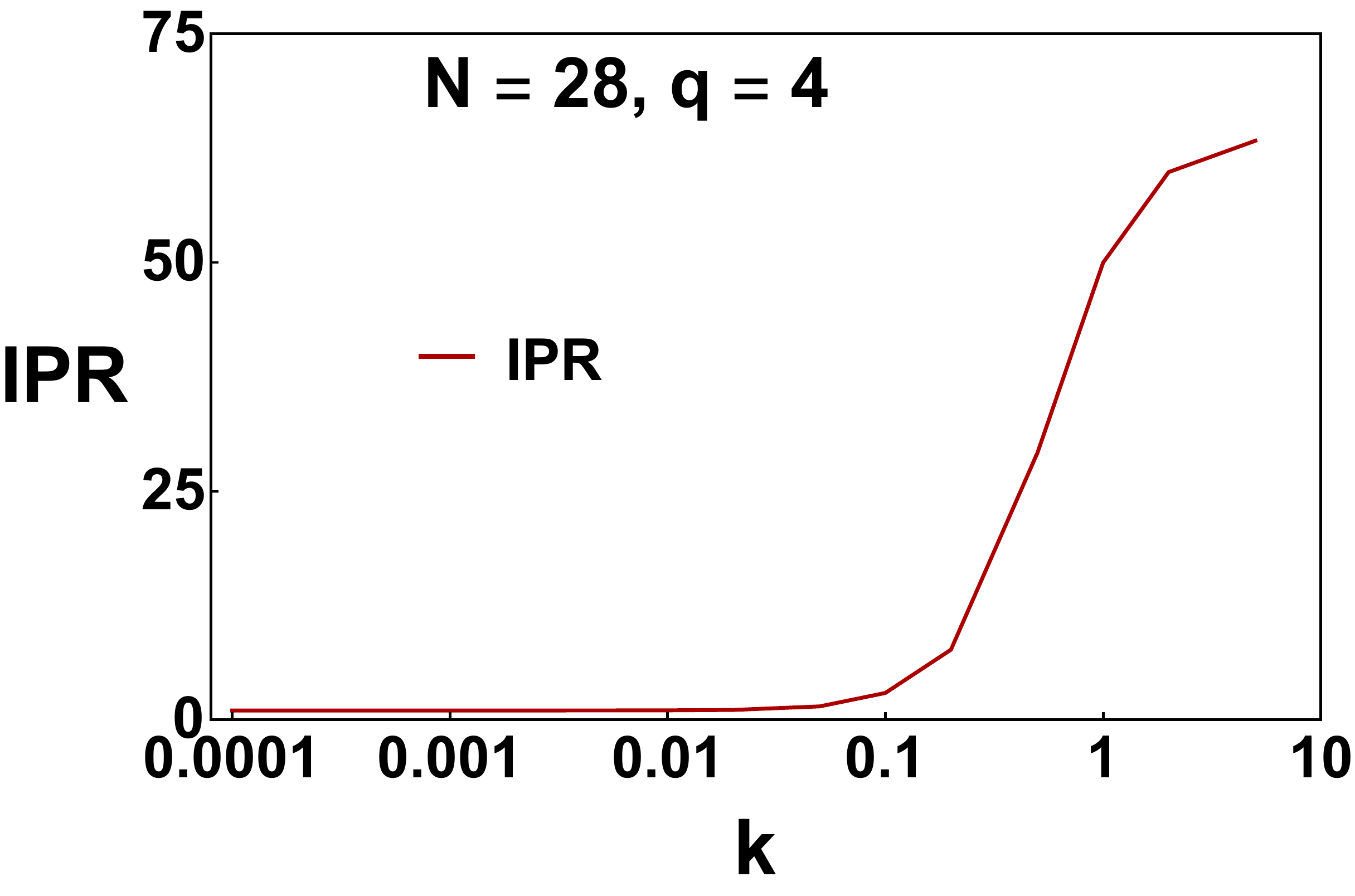}}
\caption{The ratio of $|c_2/c_1|$ (left) and the inverse participation ratio (right) versus $k$. 
}
\label{collect}
\end{figure}

In Fig. \ref{collect} we show the ratio $|c_2/c_1|$ (left) and the
inverse participation ratio (IPR) (right)  defined as
\be
{\rm IPR} = \frac{(\sum_n|c_n|^2)^2 }{\sum_n|c_n|^4}.
\ee
 We observe a crossover at $k \approx 0.1$ where $\beta(k) \approx E_2- E_1$.
Around the same value of $k$, the inverse participation ratio
increases dramatically. This is a feature typical in metal-insulator and
integrable-chaotic transitions.

The results of this section suggest that the ground state of the model undergoes a qualitative change
around $k\sim 0.1-0.2$.
We shall see in next section that precisely in this region the first order transition mentioned in the introduction turns into a sharp crossover.
We turn now to the study of low energy excitations by the analysis of thermodynamic properties and spectral correlations. 

\section{Thermodynamic properties}
\label{sec_Therm}
In the first subsection, we study thermodynamic properties of Hamiltonian Eq.~(\ref{Htotal}) for $\alpha = 1$ and $N \leq 34$ by exact diagonalization techniques. In the second subsection, we repeat this analysis in the large $N$ limit by solving the Schwinger-Dyson equations to study
the Hawking-Page phase transition, first reported in Ref.~\cite{maldacena2018}, in more detail.

\subsection{Numerical Study}

In order to investigate the thermodynamic properties of the Hamiltonian Eq.~(\ref{Htotal}),
we compute its spectrum by exact diagonalization techniques for up to $N = 34$ Majoranas.
We are mostly interested in the low temperature limit where, according to the results
of Ref.~\cite{maldacena2018}, a  first order Hawking-Page transition occurs at a finite value
of the coupling $k$. The transition, according to Ref. \cite{maldacena2018}, seems to terminate
for $k \geq k_c$ though it is not clear whether it becomes second order or just a crossover. 
 
\begin{figure}[t!]
 	\centering
 	\resizebox{0.325\textwidth}{!}{\includegraphics{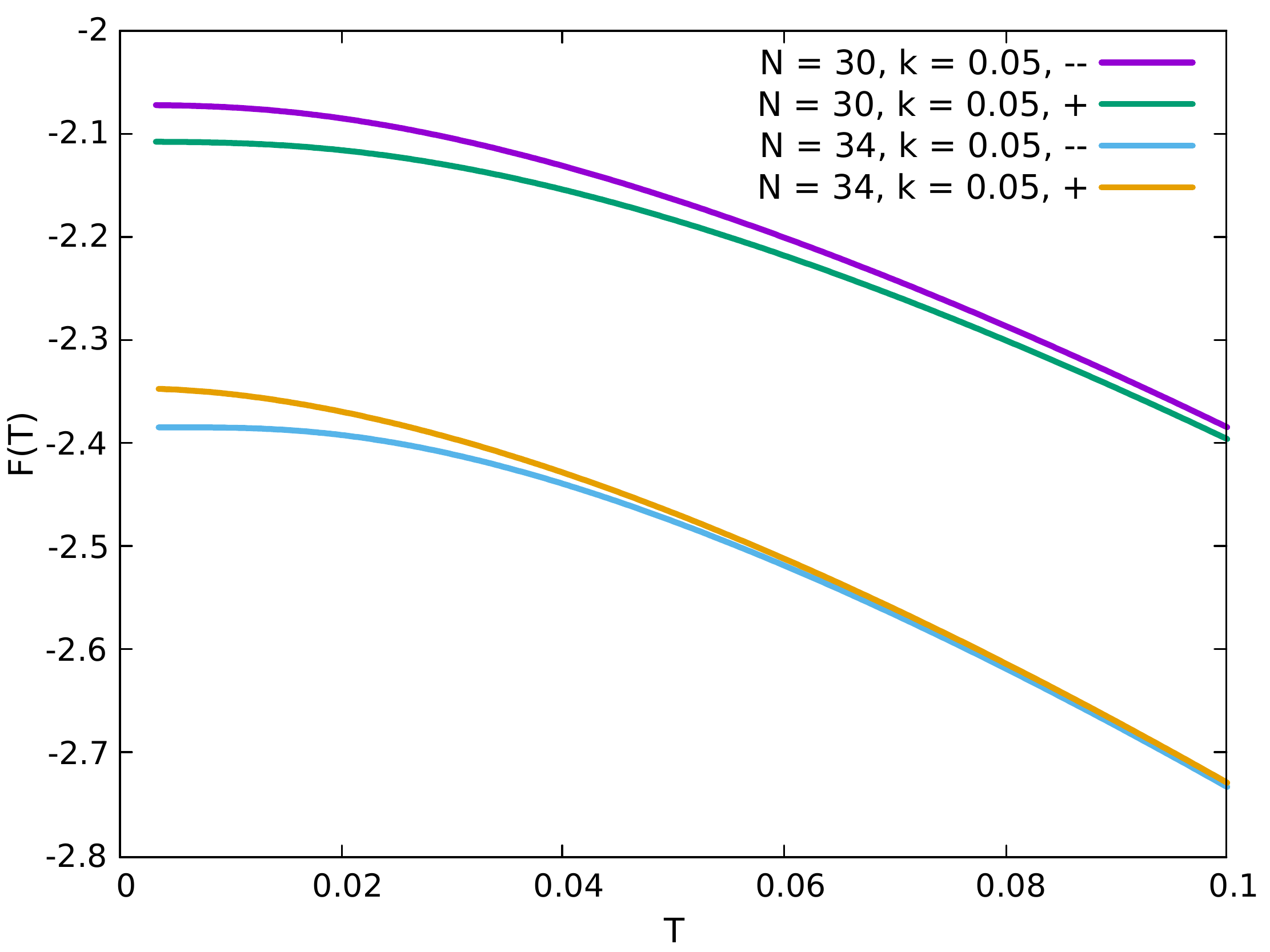}}  
 	\resizebox{0.325\textwidth}{!}{\includegraphics{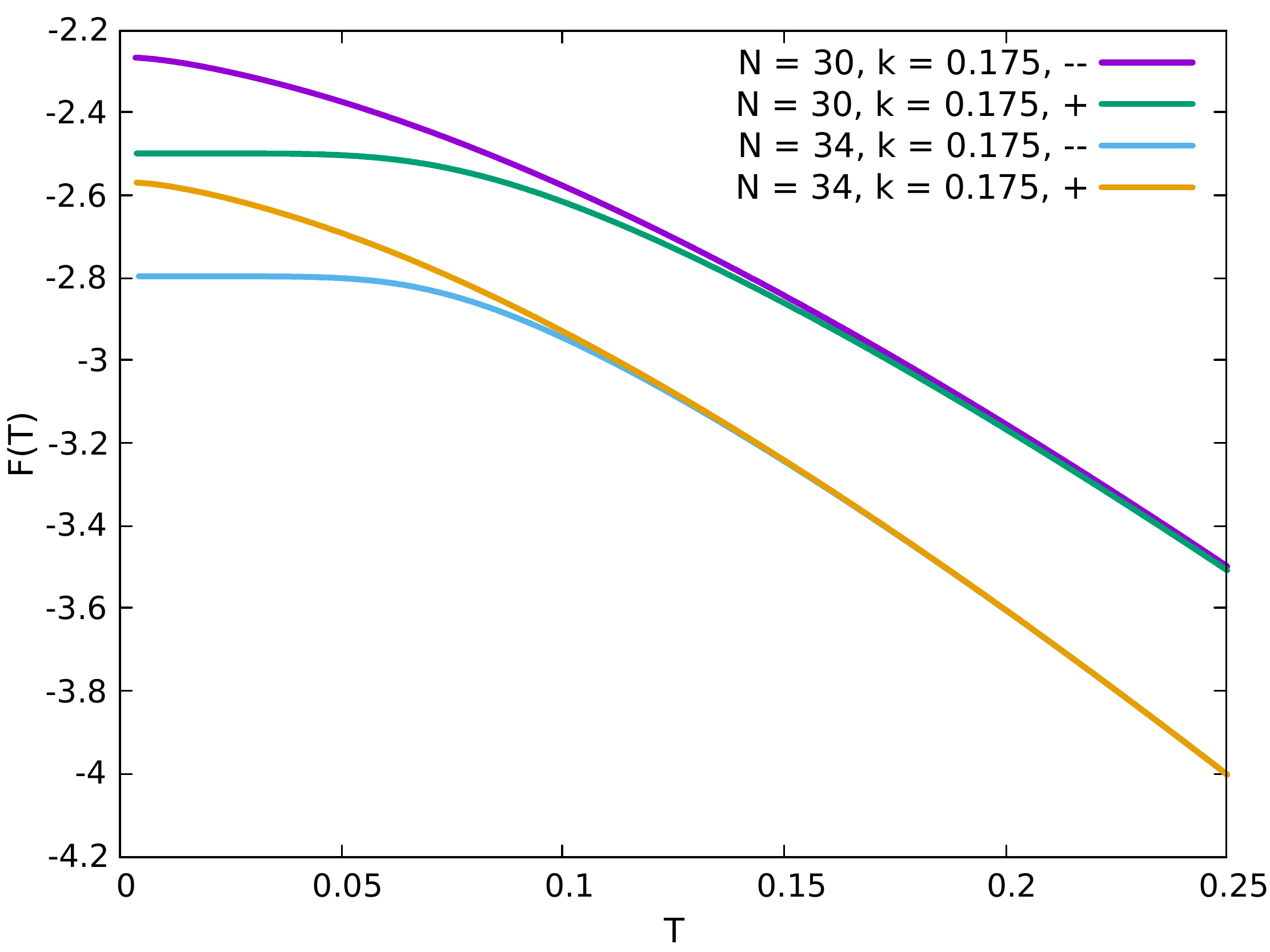}}
 	\resizebox{0.325\textwidth}{!}{\includegraphics{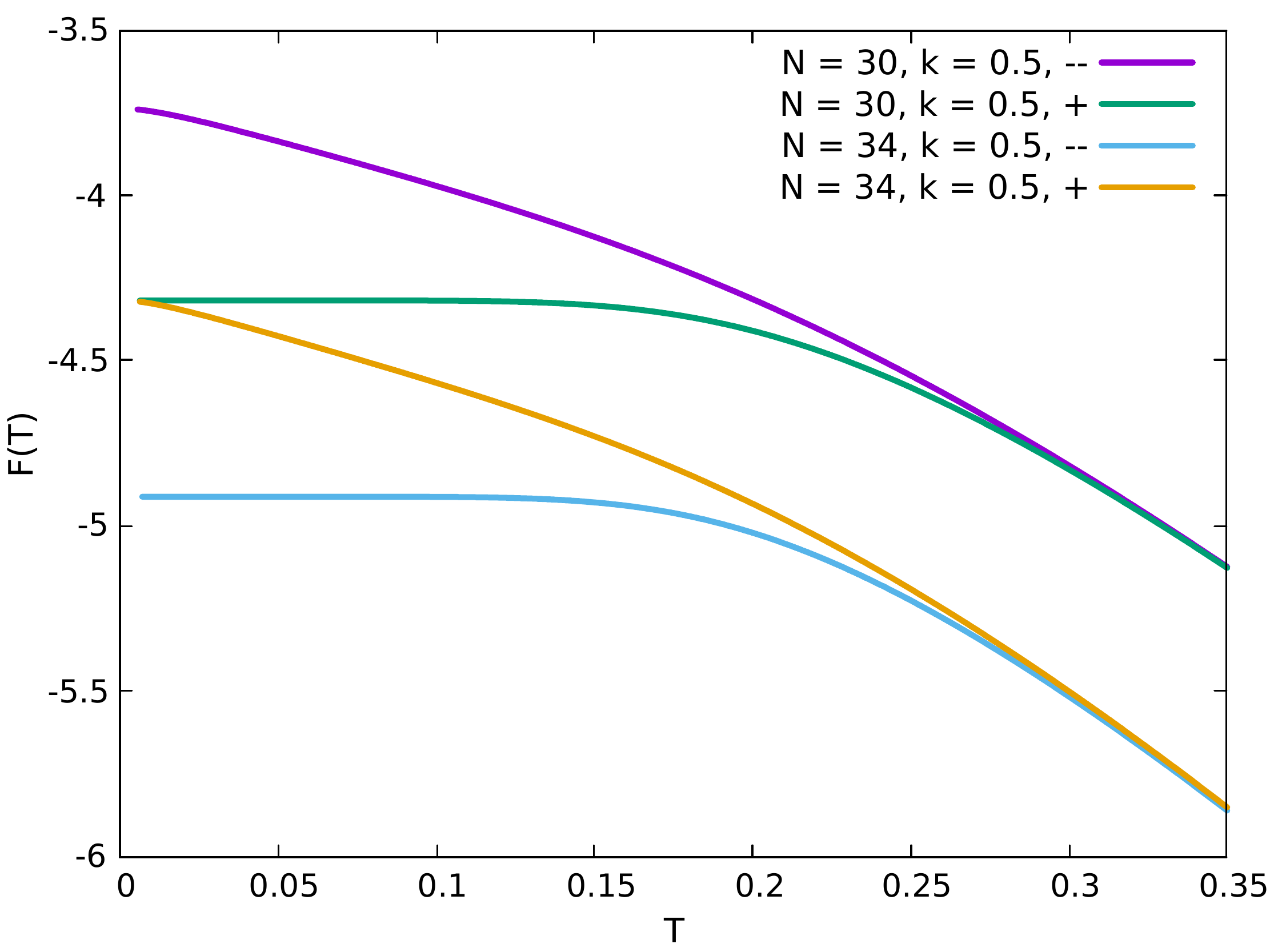}}  
 	\vspace{-4mm}
 	\caption{Free energy as a function of temperature for $N = 30$ and $34$ Majoranas with positive and negative parity and different values of $k$. The observed temperature independent part of the free energy has its physical origin in the gap between the ground state and the first excited state.
      } 
 	\label{fekN}
 \end{figure}

We study the free energy as a function of $k$ and $N$. For any $N \gg 1$ we have found,
see Fig.~\ref{fekN}, that for sufficiently large $k$ and low temperature, the free energy,
computed in the chiral sector containing the ground state, is constant which signals the existence
of a gap in the spectrum, one of the distinctive features of the AdS graviton gas. 
Interestingly, unlike the standard SYK model, the chirality of the ground state depends on
$N$, namely, for $N = 26$ it is positive but for $N = 28$ and $N=30$ it is negative. We do not have a clear understanding of this feature.

 \begin{figure}[t!]
 	\centering
 	\resizebox{0.32\textwidth}{!}{\includegraphics{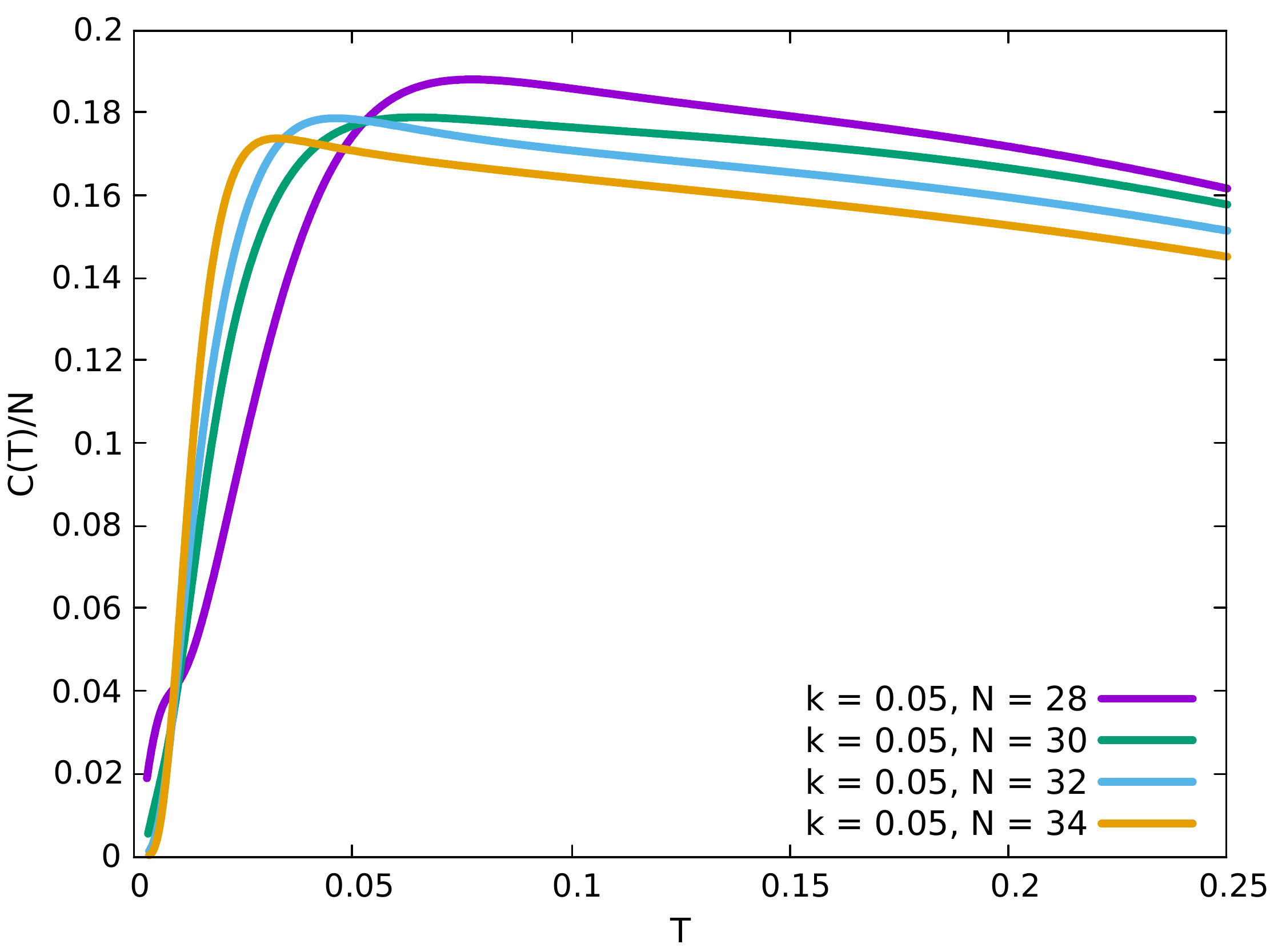}}  
 	\resizebox{0.32\textwidth}{!}{\includegraphics{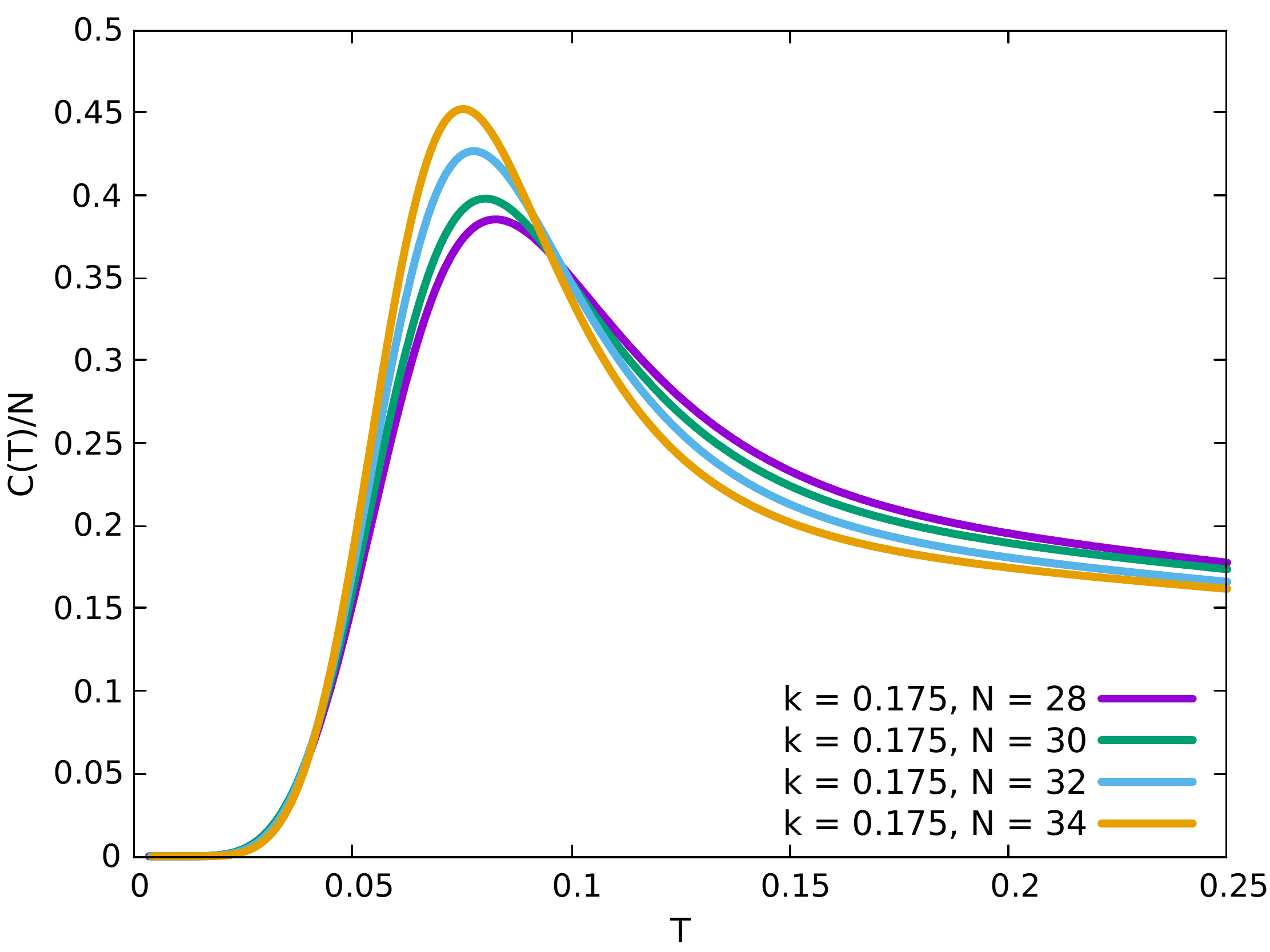}}
  \resizebox{0.32\textwidth}{!}{\includegraphics{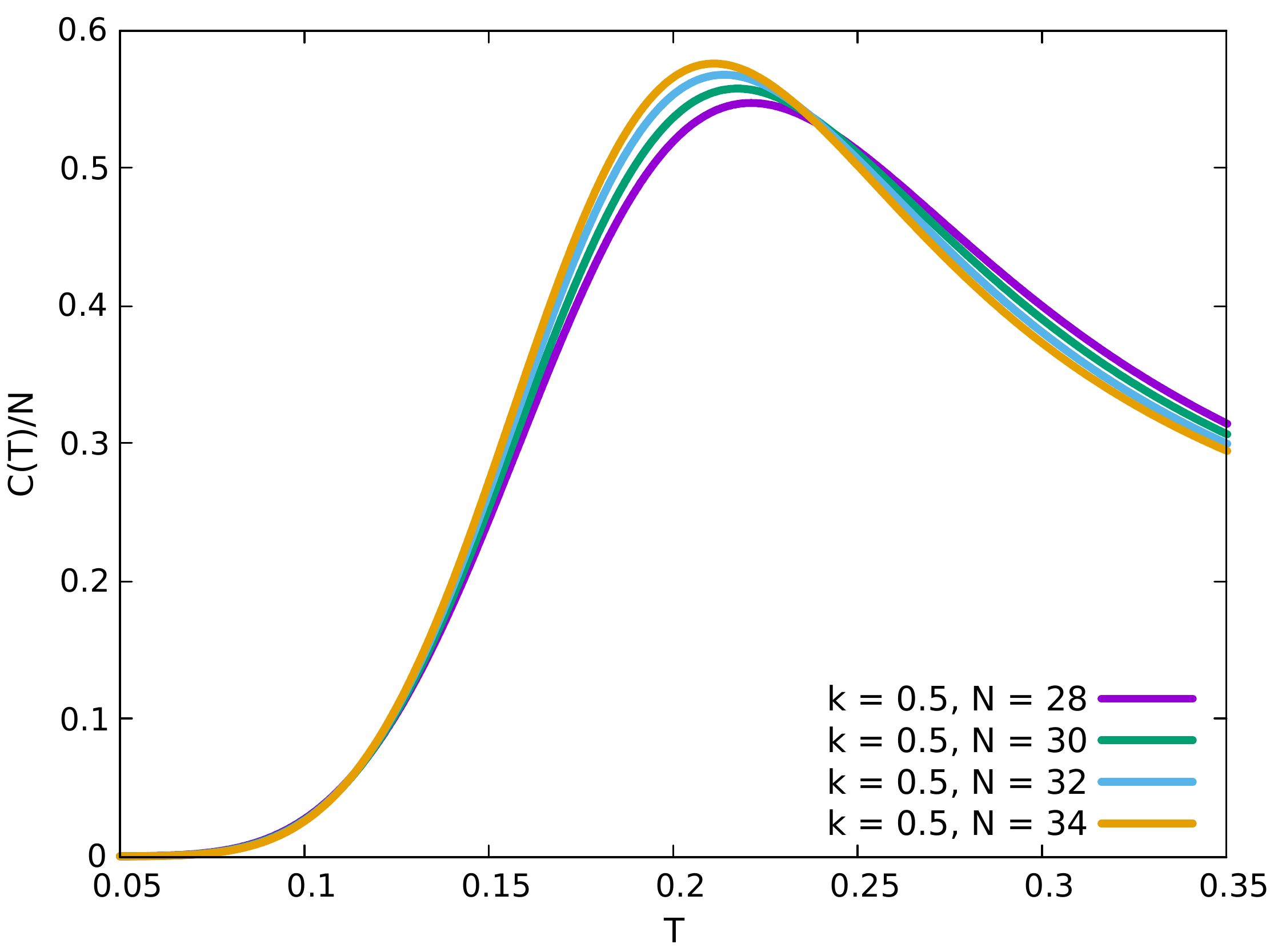}}  
 		\vspace{-4mm}
 		\caption{Specific heat as a function of temperature for different values of $N$. For      
 	          $k_c =0.175$ (central) we observe that as $N$ increases, the observed peak becomes increasingly narrow
                  which suggests the existence of the phase transition. For larger $k$ (right) the peak is broad
                  which suggests a crossover. For small $k$ (left), finite size effects seems to be stronger and we cannot reach any firm conclusion.
                } 
 		\label{shkN}
 	\end{figure}
 At higher temperature, the free energy, corresponding to the chiral block that includes the ground state, starts to decrease and becomes increasingly close to the one corresponding to the other chiral block. The temperature at which the
 two becomes close gives a rough estimation of the critical temperature of 
 the Hawking-Page transition from the graviton gas to the black hole background.
 A more accurate estimation of the critical temperature can be  obtained from the specific heat. For a first order
 phase transition, the specific heat diverges at the transition, for a second order transition it shows
 a finite jump and for higher order phase transitions, the specific heat is smooth.
 
 In Fig.~\ref{shkN}, we show results for the size dependence of the specific heat for different
 values of $k$. We note that thermodynamic phase transitions only occur in the $N \to \infty$ limit
 so information about the size dependence is necessary for a correct understanding of the transition.
 For large couplings, the specific heat has a broad maximum around the value of the gap, with a weak
 size dependence. Although the range of sizes is rather limited, this is a strong indication of a crossover,
 not a transition.
 For very weak coupling $k = 0.05$, the maximum is almost unnoticeable for $N = 28$ which suggests
 that larger couplings or larger sizes are necessary for the wormhole phase to occur. Even for a
 larger $N = 34$, there is only a slight indication of a gap in the spectrum and no clear signature
 of the transition. Larger values of $N$ are necessary to reach any firm conclusion. For a coupling strength of
 $k \approx0.2$ around the one for which we have observed a qualitative change in the ground state,
 there is a sharp jump of the specific heat for temperatures of the order of the gap. The jump in the specific heat is noticeably sharper as $N$ increases which strongly suggests a phase transition takes place.
 \begin{figure}[t!]
 	\centering
 	\resizebox{0.72\textwidth}{!}{\includegraphics{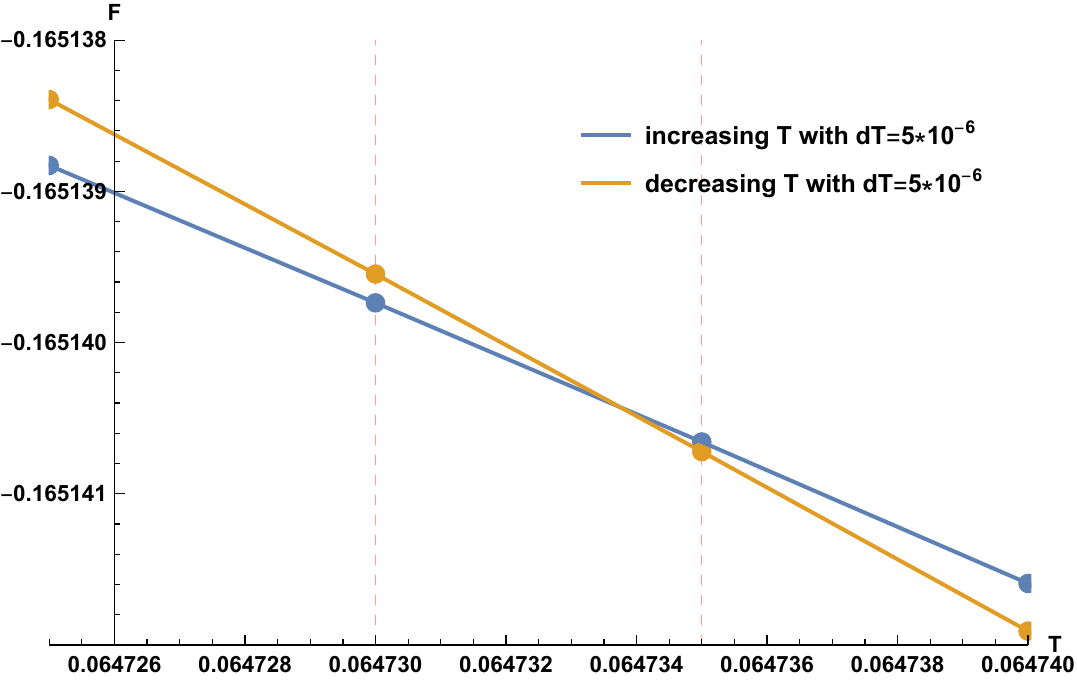}}  
 		\vspace{-4mm}
 		\caption{
Free energy $F$  of the model Eq.~(\ref{Htotal}) versus temperature $T$.  
The two free energies, computed by solving the Schwinger-Dyson equations by raising and lowering the temperature, with a temperature step of ${dT} = 5 \cdot 10^{-6}$,
                  intersect at an angle. The full free energy then develops a kink, which signals a first order phase transition around the critical point $k_c = 0.175$.}
                %\jvc{Lines are too thin, Title too small. Put legend in the figure. Title should move to the caption which I already did.}} 
 		\label{shSD}
 	\end{figure}
 
\subsection{Solution of Schwinger-Dyson Equations}
 
 The above analysis, though illustrative, is not conclusive because of the limited size we can study by
 exact diagonalization. 
 In order to gain a more quantitative understanding, we have computed the free energy and specific heat
 in the large $N$ limit by solving the Schwinger-Dyson equations numerically. The details of this calculation have been already 
 explained in great detail in \cite{maldacena2018}, to which we refer. The
 purpose of this section is  to perform a more precise
 analysis around the critical coupling $k_c$. 

 \begin{figure}[b!]
 	\centering
 	\resizebox{0.72\textwidth}{!}{\includegraphics{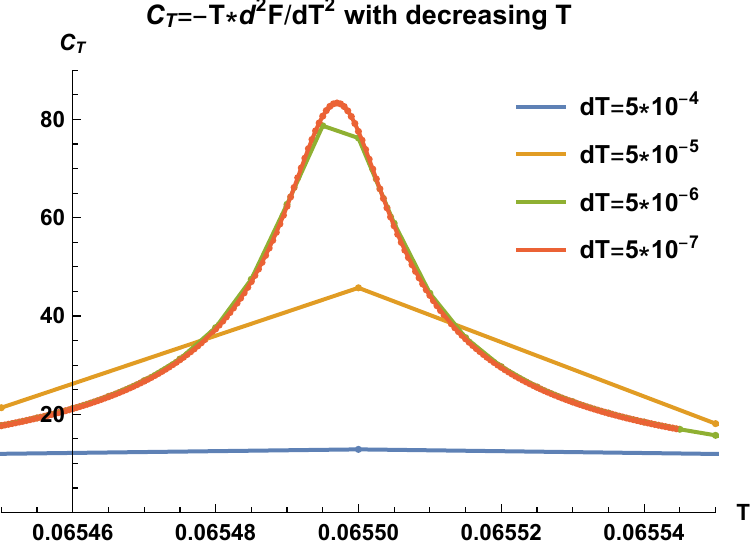}} 
 		\vspace{-4mm}
 		\caption{Specific heat of the Hamiltonian Eq.~(\ref{Htotal}) versus temperature.
                  The first order phase transition suddenly turns to a sharp but smooth crossover slightly above the critical value, $k_c = 0.177$. By taking a very small temperature step to the very small value of $dT=5 10^{-7}$, the peak in the specific heat turns out to be finite and smooth.}
 		\label{shSDhigh}
 	\end{figure}

 We have found that up to $k = k_c \approx 0.175$
 the system undergoes a first order phase transition. In order to reach this conclusion we have carried out a
 careful analysis of the dependence of the derivative of the free energy on the temperature step: the first
 order phase transition is clearly visible by noticing that, up to $k = k_c$, the free energy shows a hysteresis
 curve as function of the temperature, see Fig.~\ref{shSD}. 

The critical value of $k_c \sim 0.175$ is  consistent with the one, obtained in the previous section, at which the ground state changes qualitatively.
%agreement between the ground state and the TFD state becomes much worse.
 We have also observed, see Fig.~\ref{shSDhigh}, that for slightly larger values of $k \geq 0.177$, this first order phase transition turns out to be a sharp but smooth crossover, which becomes broader with increasing $k$. To reach this conclusion, it has been necessary to study how the peak in the specific heat behaves when decreasing the temperature step: as shown in  Fig.~\ref{shSDhigh}, by further decreasing the temperature step to the very small value of $dT = 5 \cdot 10^{-7}$, the peak turns out to be a sharp, but smooth, crossover.
 We therefore conclude that the window  for a hypothetical second order phase transition must be very narrow, $0.175 \leq k \leq 0.177$.  We note that our results are in perfect agreement with those of \cite{ferrari2017}, where  a detailed analysis of the Schwinger-Dyson equations associated with a very similar model without disorder was carried out.
  
\begin{figure}
 	\centering
 	\resizebox{0.9\textwidth}{!}{\includegraphics{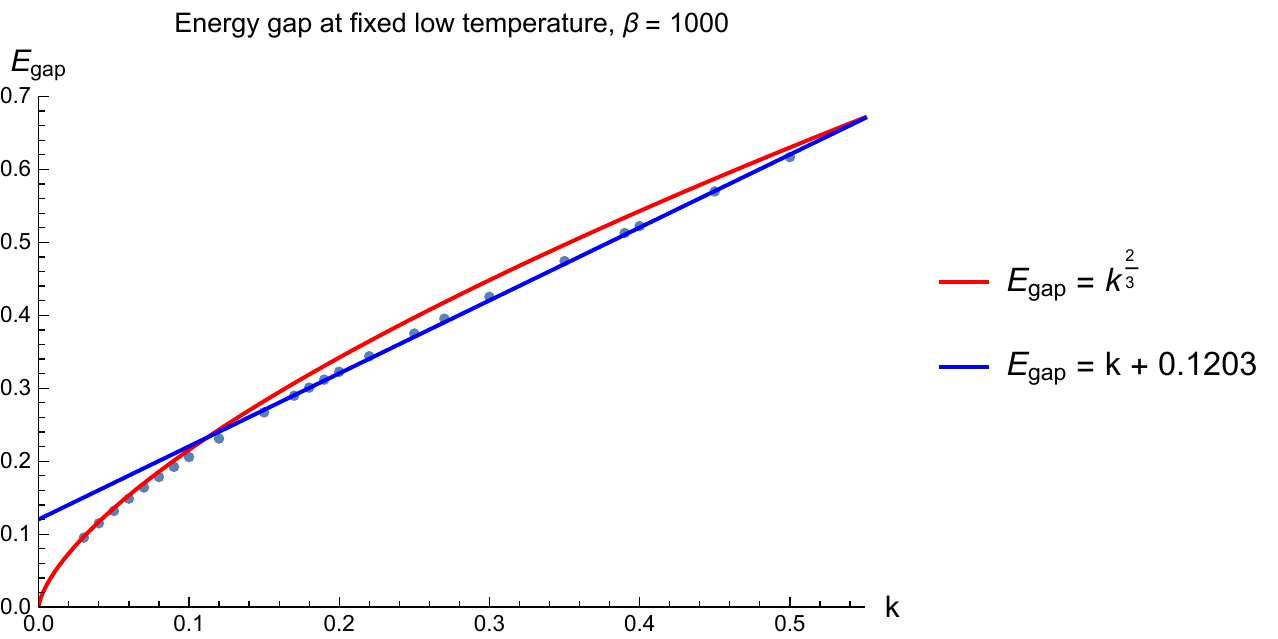}}
 		\vspace{-4mm}
 		\caption{The energy gap, computed at fixed low temperature, as a function of the coupling $k$.
We see a crossover from the power law dependence $E_{\text{gap}}\sim k^{2/3}$ to the linear dependence $E_{\text{gap}}\sim k$. This change of behavior was first reported in Ref.~\cite{maldacena2018}.} 
 		\label{gapek}
 	\end{figure}
We have also computed, see Fig.~\ref{gapek}, the energy gap $E_g$ between the ground state and the first excited state as function $k$ for a fixed low temperature by a careful fitting of the exponential decay of the relevant
Green's function. The same computation has been already performed and explained in \cite{maldacena2018}. Here, we just reproduce the result for completeness.
 The dependence of the gap on the coupling $k$ is a further confirmation of the existence
of a transition. 
Around $k \sim k_c$ we observe a change of behavior from $E_g \propto k^{2/3}$
for $k \ll k_c$ to $E_g \propto k$ for $k \gg k_c$. The latter linear behavior is expected if the gap is due to direct hopping induced by the coupling term in the Hamiltonian which is one-body (two Majoranas) and therefore non-interacting.

\begin{figure}[t!]
	\centering
\resizebox{1\textwidth}{!}{\includegraphics{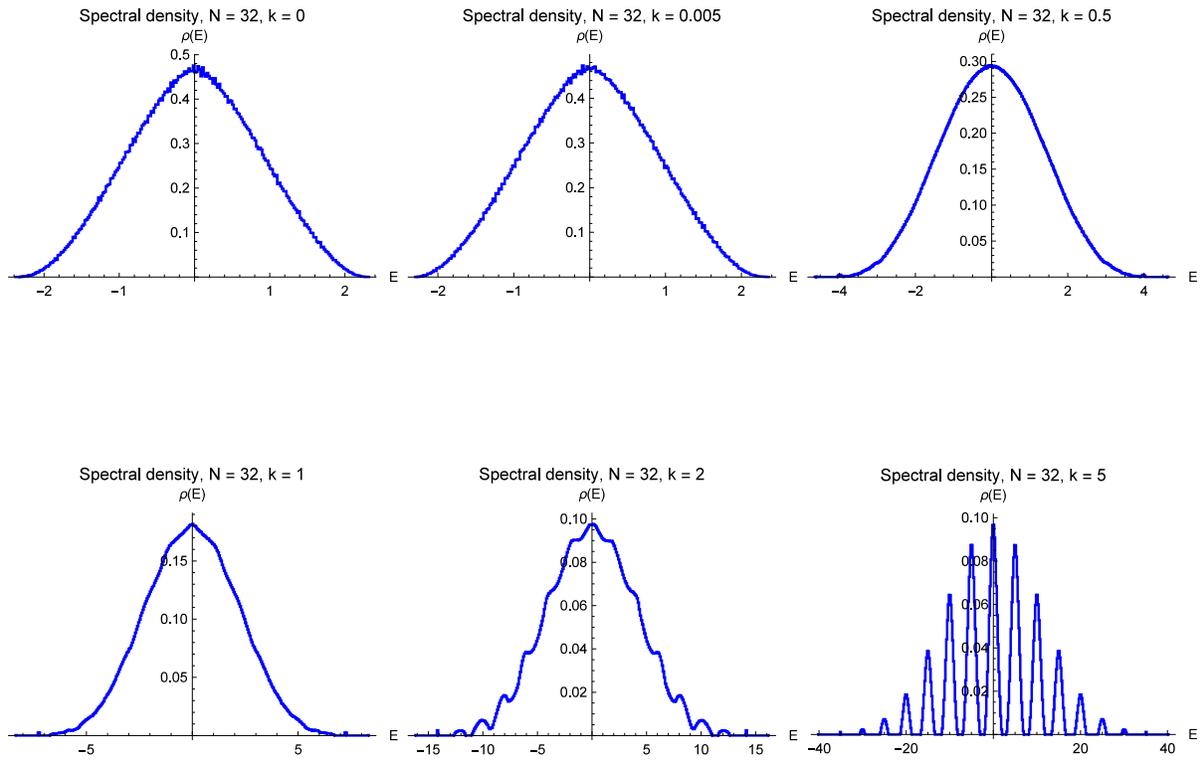}}  
	\vspace{-4mm}
	\caption{Density of states as a function of the coupling $k$. For sufficiently large $k$ the spectral density
          splits into separate blobs centered around the eigenvalues of the spin operator. In this region
          of parameters, no Hawking-Page transition occurs, and the interactions becomes increasingly weak
          with respect to the hopping between the left and right SYK model.} 
	\label{figdenblob}
\end{figure}
Physically, by increasing the coupling $k$, we are effectively reducing the many-body interaction inside each SYK model in favor of a direct coupling between the two SYK models which is a
diagonalizable one-body (two Majoranas) interaction.
This is vaguely reminiscent of a Gross-Witten transition \cite{gross1980}, or a BEC-BCS crossover \cite{bourdel2004}, where
the reduction of the interaction strength induces a higher order transition or just a crossover.  
 
 For larger $k > 1$, we have found, see Fig.~\ref{figdenblob}, that the spectral density starts to split into different regions. There is a simple explanation of this phenomenon.
% As was mentioned earlier, the coupled system has a spin symmetry.
 The interaction term in the Hamiltonian, which becomes dominant in the large $k$ limit,
 is proportional to the spin operator which has a discrete spectrum with large degeneracies
(see (\ref{spectrumofS})). 
%  (see Table (\ref{spectrumofS})). 
Therefore we
expect that, in this region, the spectral properties are largely controlled by this coupling term and
the interactions inside each of the SYK models 
spread out the degenerate states.
Since the eigenvalues of the coupling term range from $-kN/4$ to $kN/4$, we expect the full spectrum of
the model to cluster around these few eigenvalues which leads to the appearance of the above mentioned
blobs in the spectral density. This is exactly what we observe in the large $k$ limit  (see Fig. \ref{figdenblob}).

\begin{figure}
	\centering
\resizebox{0.32\textwidth}{!}{\includegraphics{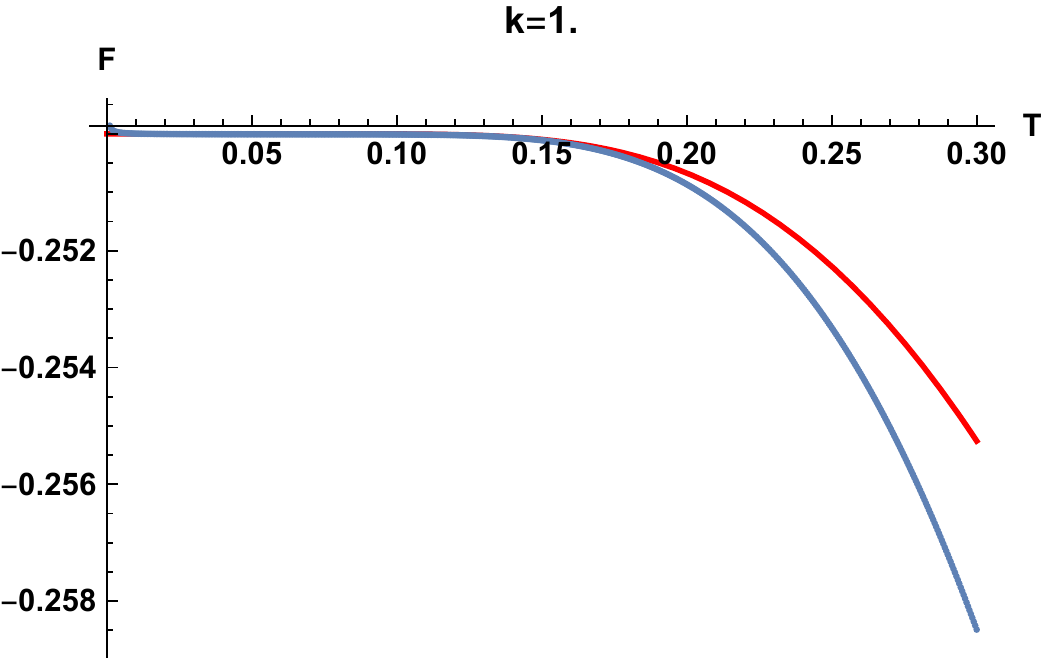}}  
\resizebox{0.32\textwidth}{!}{\includegraphics{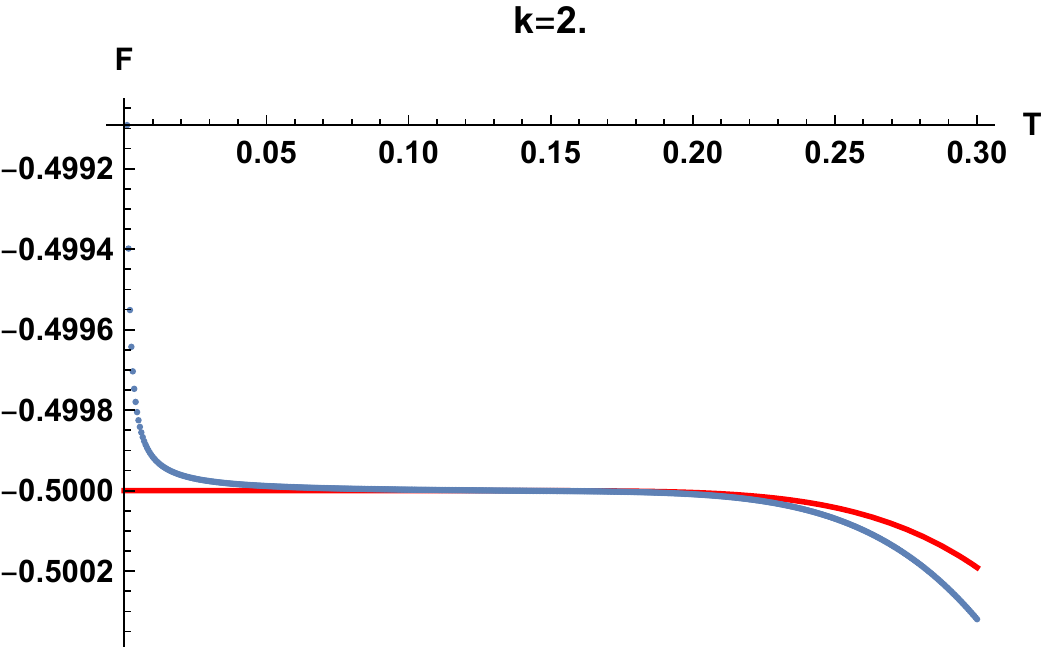}}  
 \resizebox{0.32\textwidth}{!}{\includegraphics{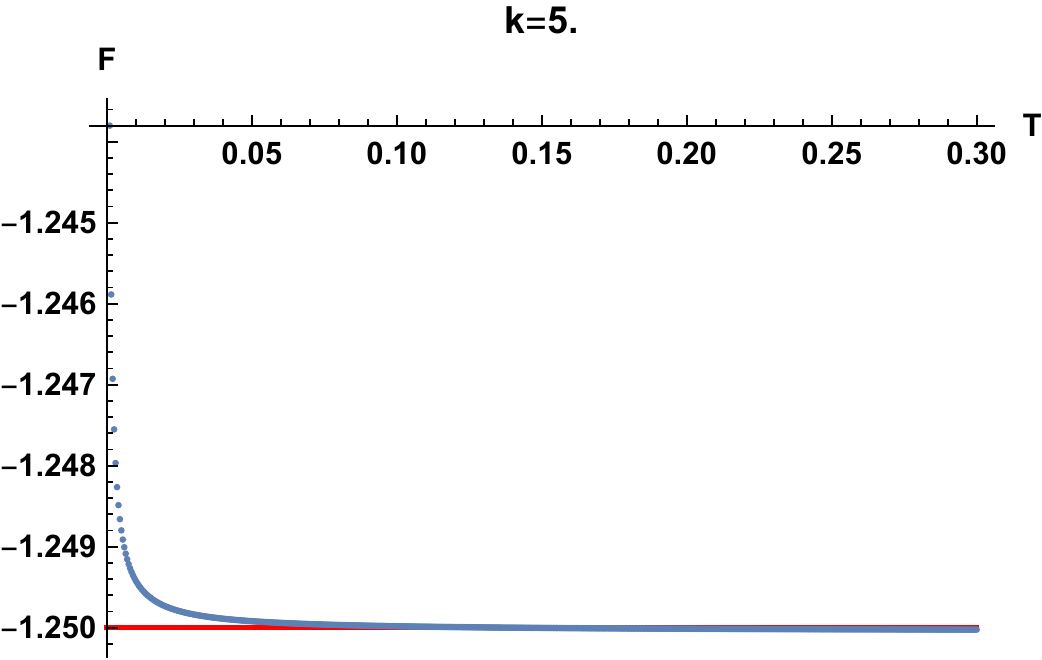}}  
	\vspace{-4mm}
	\caption{Comparison of the free energy, numerically obtained from the Schwinger-Dyson equation (light blue line), with the analytic expression Eq. \eqref{Fanalytic} for large $k$ (red line).
          Here, the free energy is normalized such that 
 the plateau value coincides with $-k/4$.}
%\jvc{Why are you doing that? I do not see the reason. Can you make the red curve as thick as the blue curve?}}
        
	\label{SD_vs_deltablobapprox}
\end{figure}

The free energy can be computed analytically in the limit $k\rightarrow\infty$ as follows.
If we ignore the SYK terms, we are left with the discrete spectrum just given by Eq. \eqref{spectrumofS}, for which the free energy is given by
\begin{align}
F=-\frac{1}{\beta}\log \sum_{p=0}^{N/2}{\frac{N}{2}\choose p} e^{-\beta k(-N/4+p)}
=-\frac{k}{4}-\frac{1}{2\beta}\log(1+e^{-\beta k}).
\label{Fanalytic}
\end{align}
In Fig.~\ref{SD_vs_deltablobapprox}, we compare the numerical free energy for $k=1,\;2$ and $5$ with this analytical approximation and find that the agreement becomes excellent around $k \sim 5$.

All these results indicate that the physical reason for the termination of the Hawking-Page transition is just the gradual reduction of the  interaction strength.
This is a strong indication that a gravity interpretation is restricted to low temperatures and
weak coupling where the first order transition takes place.

\section{Level statistics}
\label{sec_Levelstatistics}
We now turn to the study of level statistics, with the main aim to characterize the dynamics in the
traversable wormhole phase observed in the low temperature limit of the free energy.
For that purpose, we obtain the exact spectrum of the model by exact diagonalization techniques for up to $N = 34$ Majoranas.  We are especially interested in the spectral correlations of the smallest eigenvalues above the ground state, which are likely to be closely related to gravitational modes of the wormhole phase.

For a meaningful analysis of spectral correlations it is in general necessary to unfold the spectrum so that
the mean level spacing is the same
across the spectrum. For that purpose, we employ  the
splines method that fits locally consecutive subsets of many
$(> 10)$ eigenvalues with low order polynomials.
Results are insensitive to the degree of the fitting polynomials.

We aim to study the evolution for long times, of the order of the Heisenberg time, in order
to  find the range of applicability of random matrix results, a feature of quantum chaotic systems, rather than finite size deviations from it.
Hence, 
we will focus on short range spectral correlators, such as the level spacing distribution, $P(s)$,
and the adjacent gap ratio. The former is defined as the probability to find two consecutive eigenvalues
$E_{i}, E_{i+1}$ at a distance $s = (E_{i+1}-E_{i})/\Delta$ (with $\Delta$ the average local level spacing).
For a fully quantum chaotic system it is given by Wigner-Dyson statistics \cite{mehta2004} which is well
approximated by the so called Wigner surmise which depends on the universality class \cite{guhr1998}.
For the Gaussian Orthogonal Ensemble (GOE), corresponding to systems with time reversal invariance, it is given
by:
$
P_\mathrm{W,GOE}(s) \approx \frac{\pi}{2}s\exp(-\pi s^2/4).
$
For an insulator, or a generic integrable system, it is given by Poisson statistics, 
$P_\mathrm{P}(s) = e^{-s}$.  
The adjacent gap ratio  is defined as \cite{luitz2015,oganesyan2007,bertrand2016}, 
\begin{equation}
r_i = \frac{\min(\delta_i, \delta_{i+1})}{\max(\delta_i, \delta_{i+1})} 
\label{eq:agr}
\end{equation}
for the ordered spectrum $E_{i-1} < E_i < E_{i+1}$  where $\delta_i = E_i - E_{i-1}$.
For a Poisson distribution it is equal to $\left\langle r \right\rangle_\mathrm{P} \approx 0.386$ while for a random matrix ensemble it depends on the symmetry class, with
$\langle r \rangle\approx 0.530$ for the GOE \cite{atas2016}. The advantage of $\langle r \rangle$ over $P(s)$, is that it does not require to unfold the spectrum. 
The adjacent gap ratio was also studied in \cite{numasawa2019} for another model related to the traversable wormhole \cite{Kourkoulou:2017zaj}.

\begin{figure}[t!]
	\centering
	\resizebox{0.8\textwidth}{!}{\includegraphics{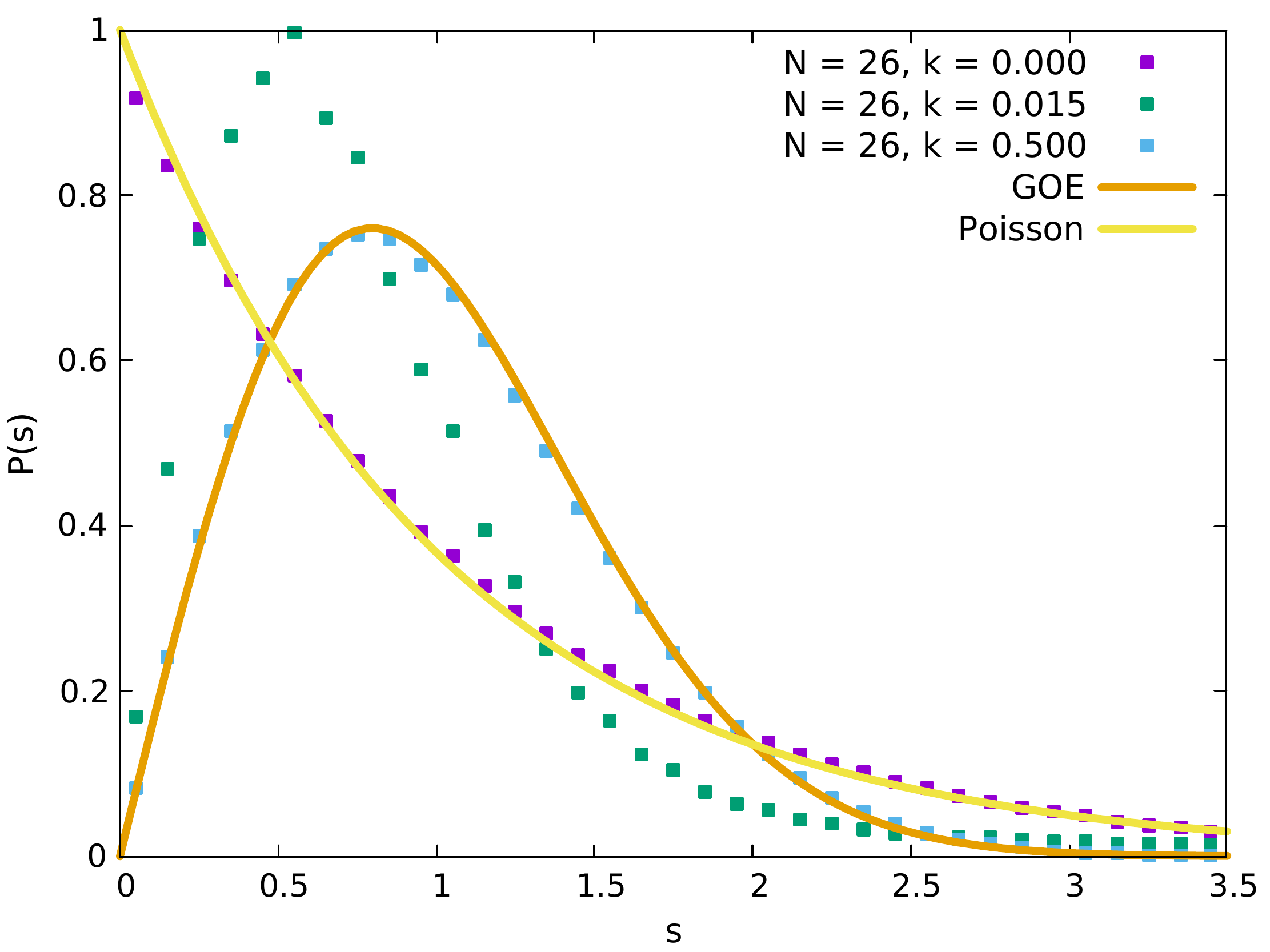}}  
	\vspace{-4mm}
	\caption{Nearest neighbor spacing distribution $P(s)$ for $N = 26$ and different $k$'s in the bulk of the spectrum. As $k$ increases, we observe a crossover from Poisson statistics to the $GOE$ prediction typical of a quantum chaotic system.} 
	\label{psbulk}
\end{figure}

We note that for a correct analysis of spectral correlations it is necessary to consider only eigenvalues with the same (good) quantum number.
This means that we have to consider eigenvalues only of a given spin and parity sector. Due to the block structure of the Hamiltonian, it is straightforward to select
eigenvalues of the same parity. For the spin symmetry, we could not find a numerically inexpensive method
to go beyond comparatively small sizes $N = 28$. We avoid this problem by
introducing an asymmetry in the couplings ($\alpha = 1.15$ in Eq.~(\ref{Htotal})) between left and right SYK models.
It has been argued \cite{maldacena2018} that the main features of the model, including the existence
of a gapped phase, though with a smaller gap, for low temperature which is a signature of the wormhole phase,
are robust to a small interaction strength asymmetry which breaks the spin symmetry. 
We start our analysis
of level correlations with the case of very small coupling.

\begin{figure}
	\centering
	\resizebox{0.49\textwidth}{!}{\includegraphics{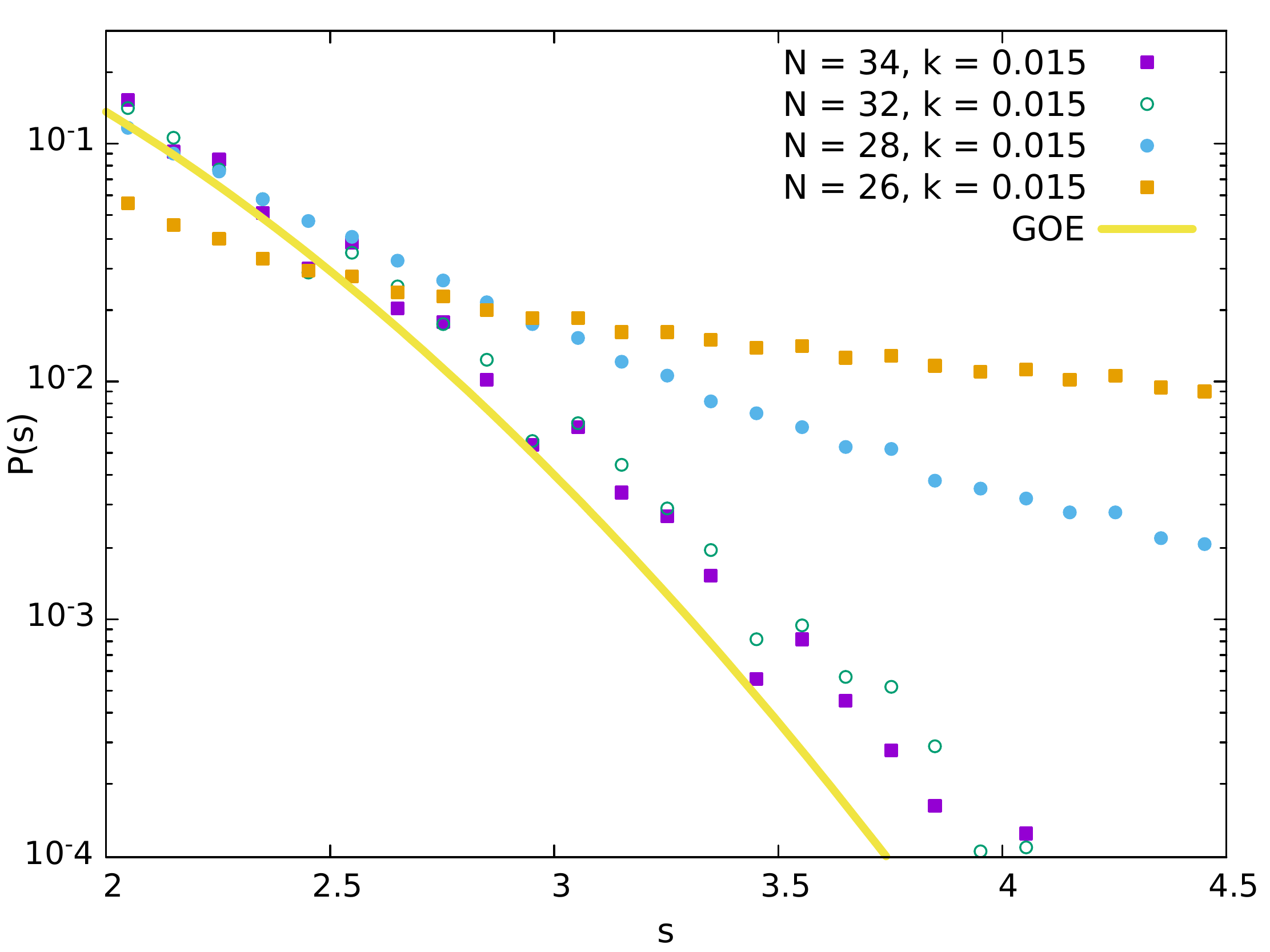}}
	\resizebox{0.49\textwidth}{!}{\includegraphics{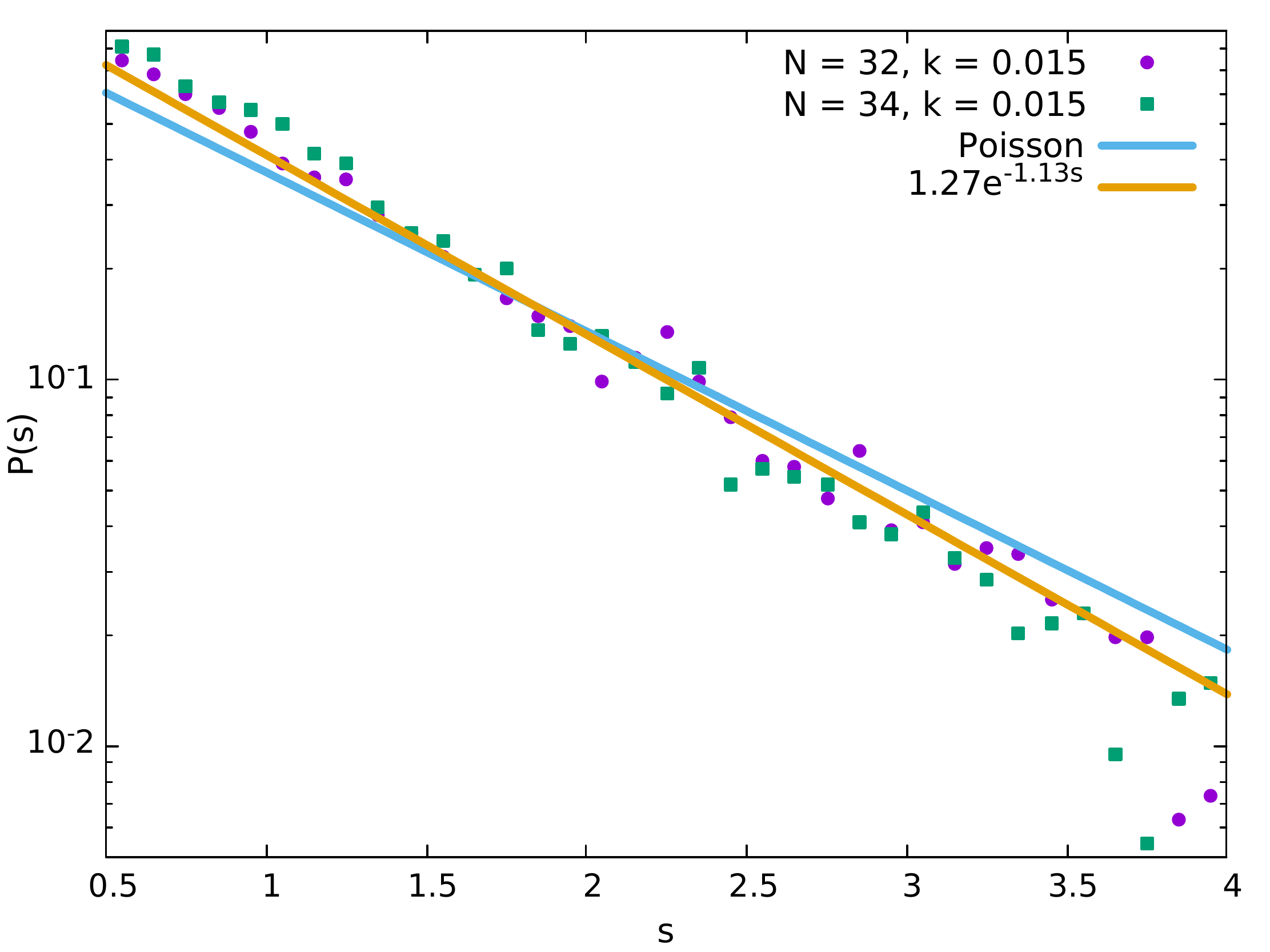}}    
	\vspace{-4mm}
	\caption{Left: The nearest neighbor spacing distribution, $P(s)$, for small $k$ and  different
          values of $N$ in the bulk of the spectrum. The agreement with the random matrix prediction becomes
          better as $N$ increases. This is a strong indication that, in the bulk of the spectrum,
          Poisson statistics for $k = 0$ is not robust to  a small coupling between the left and right SYK models,
          namely, the system is always quantum chaotic in the high temperature limit. Right: The nearest spacing
          distribution,
          $P(s)$, for small $k$ and $N = 32$ or $N=34$ for approximately the lowest $0.1\%$ of the eigenvalues.
          Agreement with Poisson statistics is very good. This is a strong indication that the infrared limit
          of the spectrum for sufficiently small $k$ is similar to that for $k = 0$ corresponding to two black holes.} 
	\label{pssmallk}
\end{figure}
\subsection{Small $k$}

For $k = 0$, the eigenvalues of the full Hamiltonian for a given chirality
have degeneracies that depend on the value of $N$. For example, we find
a two fold or four fold  degeneracy  $N=20$, and
a two fold degeneracy for  $N=24$ and $N=32$.
Once the degeneracy is removed, we have found, see Fig.~\ref{psbulk}, spectral correlations in the bulk are well described by Poisson statistics. This is the expected behavior for a system which is defined as the tensor product of two many-body quantum chaotic systems whose spectral correlations are described by random matrix theory.
For very small $k \ll 2^{-N/2}$,
the spectral degeneracy is barely lifted. As a consequence, the distance between eigenvalues that were
degenerate for $k = 0$ is much smaller than that between neighboring eigenvalues which makes  the analysis of level statistics difficult. For $k \lessapprox 2^{-N/4}$,
the degeneracy is already lifted,
so the statistical analysis does not require any artificial pruning of the spectrum. 
In this region, the level statistics in the tail and in the bulk of the spectrum are qualitatively different.

In Fig.~\ref{pssmallk}, we plot the nearest neigbor spacing distribution in the bulk of the spectrum for small $k$
and different values of $N$. Level repulsion is observed in all cases, but for $N = 26$ and $N=28$ the decay
for large distances is clearly exponential with an exponent close to the one corresponding to a Poisson
distribution. However for $N = 32$ and $N= 34$  we find good agreement with random matrix correlations.
Although a careful finite size scaling analysis would be necessary to reach a final conclusion, the latter
is a strong suggestion that in the bulk of the spectrum, corresponding to high temperatures,
even a small coupling $k$, makes the system quantum chaotic. However, see Fig.~\ref{pssmallk}, it seems
that  the lowest eigenvalues in the same region of small $k$, for $N = 32$ and $N= 34$ are correlated to Poisson statistics
like for $k = 0$. Rather than a metal-insulator transition at a certain value of $k$, this behavior
 suggests that for sufficiently weak coupling the two SYK's, each of them quantum chaotic,
are effectively disconnected systems for low energies (tail) while for sufficiently high energies (bulk)
the behavior is similar to a single quantum chaotic system whose levels  are correlated according
to random matrix theory. It is tempting to speculate that this indicates a transition from two black holes to
one black hole for sufficiently high temperatures. However, the fact that we do not have a clear geometrical understanding on how this transition can occur, together with the limited range of $N$ values which we could study numerically, prevents us to reach any firm conclusion. 
 
\begin{figure}[t!]
	\centering
	\resizebox{0.49\textwidth}{!}{\includegraphics{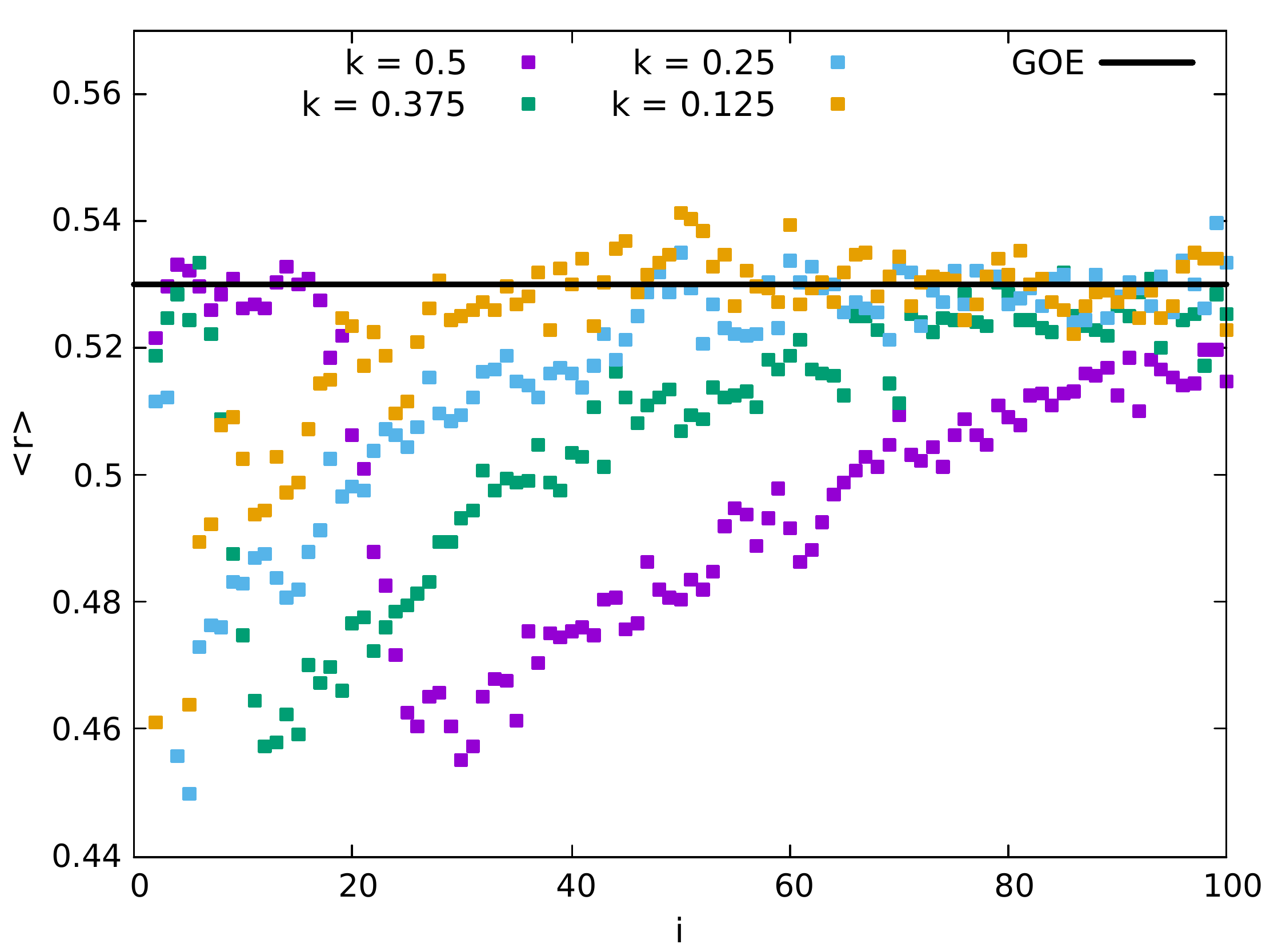}}  
	\resizebox{0.49\textwidth}{!}{\includegraphics{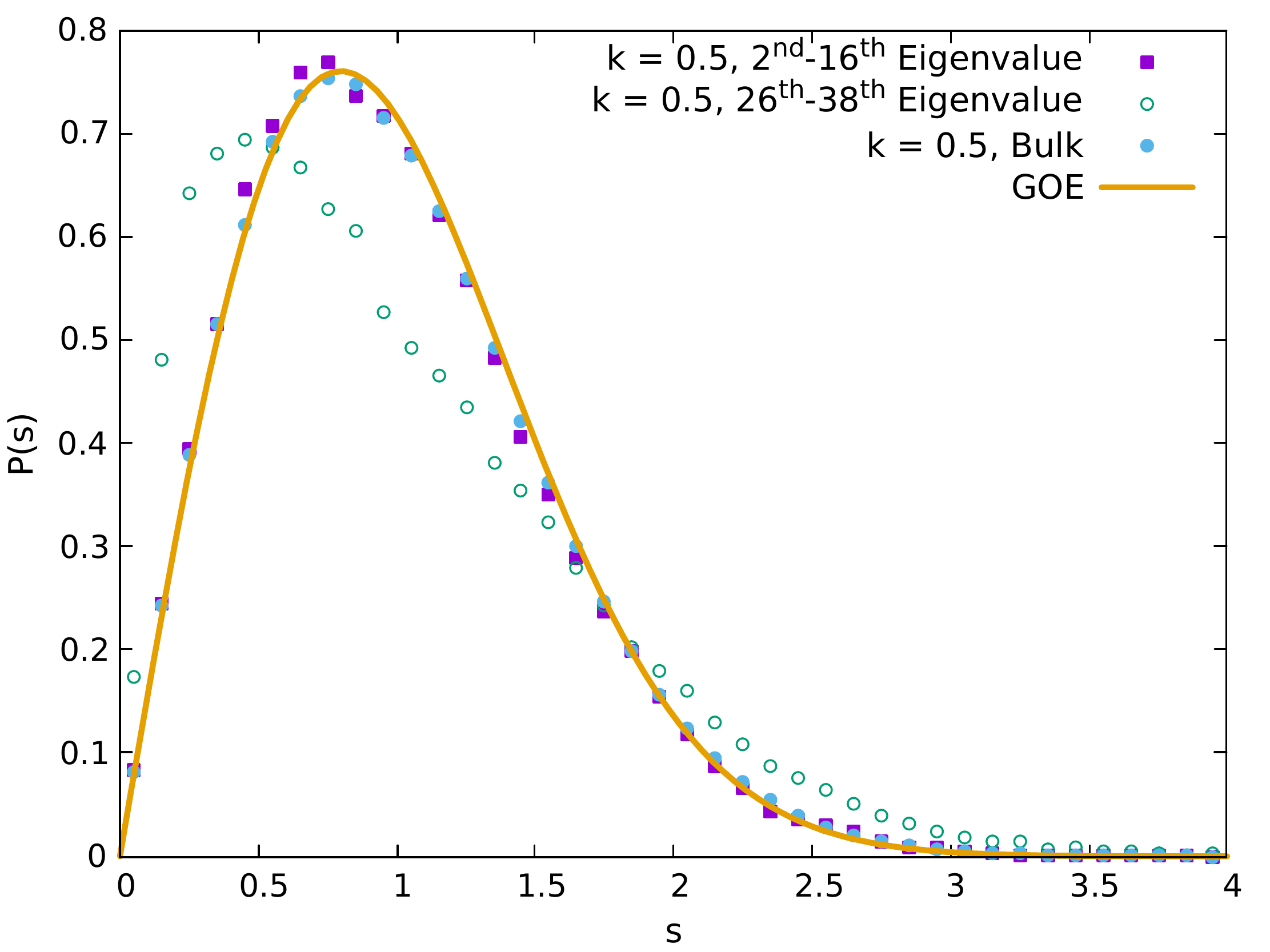}}  
	%	\resizebox{0.44\textwidth}{!}{\includegraphics{figden.pdf}}  
	\vspace{-4mm}
	\caption{Left: Adjacent gap ratio $\langle r \rangle$, Eq.~(\ref{eq:agr}), for $N = 26$ and
          different values of $k$ for the infrared part of the spectrum. We observe that,
          as $k$ increases, some of the lowest eigenvalues become correlated according to the random matrix theory. Except for $k \approx 0$, the bulk of the spectrum is correlated according to the random matrix prediction. Right: Results for the level spacing distribution $P(s)$ fully confirm the agreement with
          the Wigner surmise typical of random matrix spectral correlations.} 
	\label{fig2}
\end{figure}
Summarizing, our results show that, while in the bulk of the spectrum a very small value of $k$ is sufficient to
induce the transition from  Poisson statistics to random matrix theory spectral correlations, in the tail of the spectrum, Poisson statistics is robust, and a larger coupling is necessary to induce the transition to level statistics described by random matrix theory. We investigate this case next. 

\subsection{Critical $k$}

As $k$ increases, the bulk of the spectrum is  still correlated according to random matrix theory with
no quantitative change from the region of small $k$ investigated previously. For this reason, we will focus exclusively on the spectral correlations in the tail of the spectrum.

In Fig.~\ref{fig2} we depict results for the average adjacent gap ratio $ \langle r \rangle$
with different values of $k$ for the chiral block that includes the ground state. In the tail of the spectrum
we do observe clear deviations from the GOE prediction that become gradually smaller as $k$ increases.
Interestingly, for $k \approx 0.14$, the ratio for the lowest eigenvalues becomes suddenly very close to
the GOE value while the subsequent eigenvalues still deviate substantially from it. A further increase
of the coupling $k$ leads to more of the lowest eigenvalues becoming correlated according to the GOE
while the subsequent ones still show deviations.
Finally, for $k \sim 1$, all eigenvalues in the spectrum are GOE correlated.
We stress that the value of $k$ for which the lowest eigenvalue becomes GOE correlated is very similar
to the one at which the termination of the Hawking-Page transition, between the traversable wormhole and the black hole phase, takes place. This is a strong suggestion that deviations from random matrix theory in the tail, with values close to Poisson statistics,  may be related to the wormhole phase and
that the sudden appearance of the GOE correlated eigenvalues deep in the infrared may signal
an instability of the wormhole phase.
This instability might be expected close to the transition to a black hole phase.

\begin{figure}[t!]
	\centering
	\resizebox{0.49\textwidth}{!}{\includegraphics{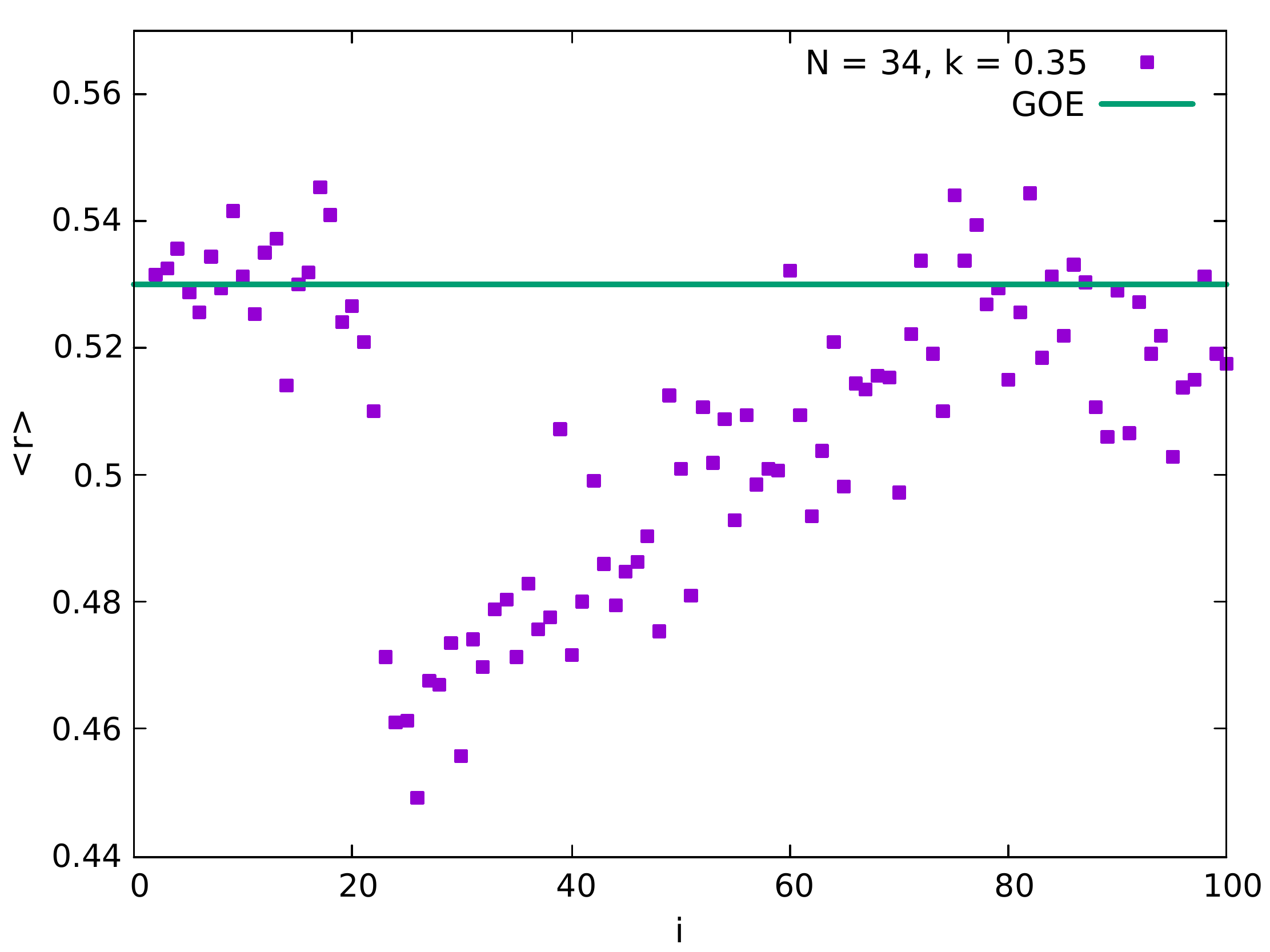}}  
	\resizebox{0.49\textwidth}{!}{\includegraphics{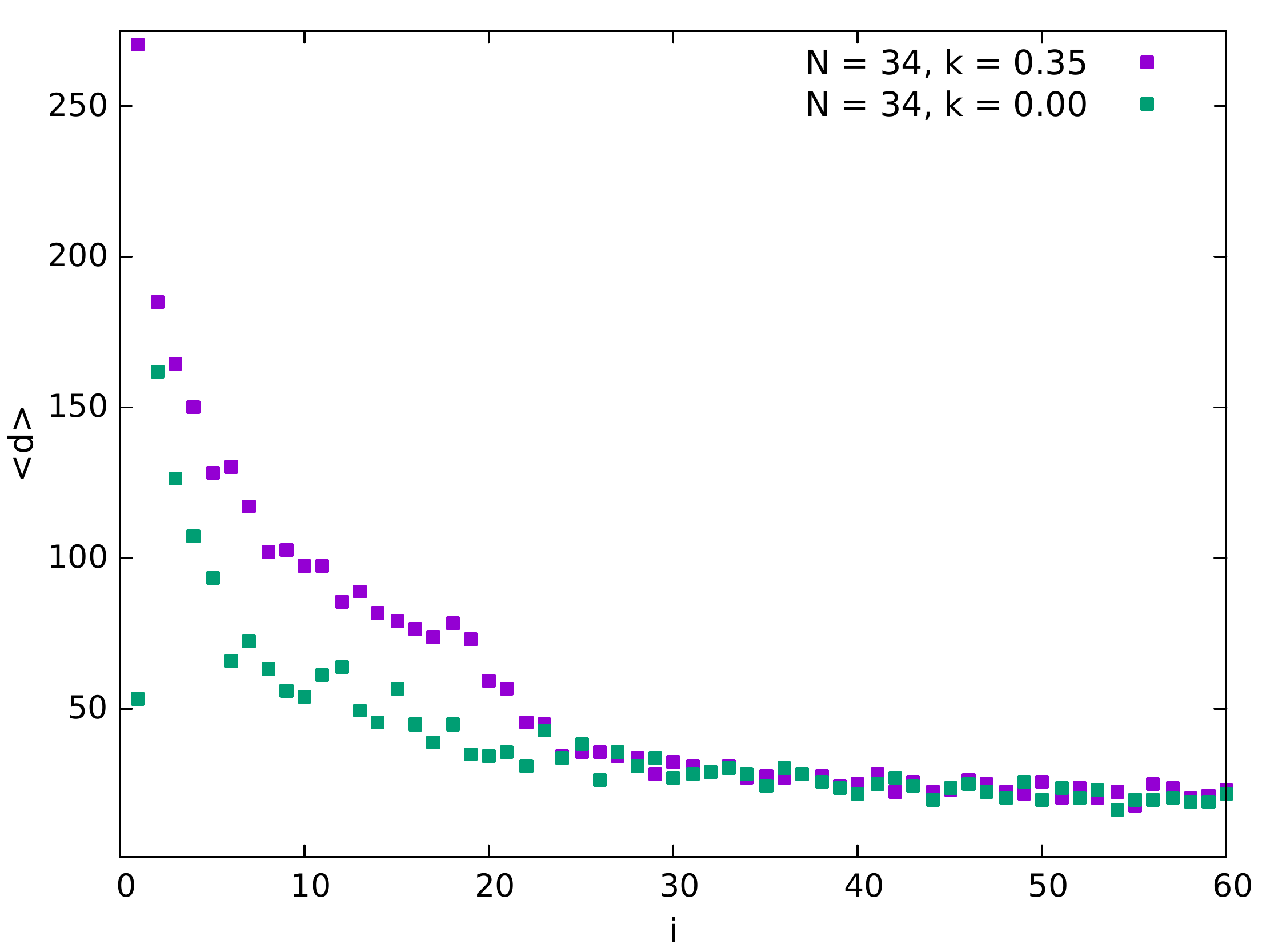}}  
	\vspace{-4mm}
	\caption{Left:  Adjacent gap ratio $\langle r \rangle$, Eq.~(\ref{eq:agr}), for $N = 34$ and $k=0.35$.
          More low-lying eigenvalues are correlated  according to random matrix
          theory for increasing $N$.
	  Right: Local average spacing $\langle d_i \rangle = \langle E_{i+1} - E_i \rangle$ for $N = 34$, $k = 0.35$ rescaled by the total average mean level spacing. 
          Roughly speaking, only the local spacing involving the lowest $25$ eigenvalues is $k$ dependent. This is also the number of eigenvalues which are correlated according to random matrix figure in the left figure.}
	\label{fig2a}
\end{figure}

As a further check that random matrix theory describes the level statistics of the lowest eigenvalues
for $k > 0.14$, we compare the level spacing distribution $P(s)$ of these eigenvalues with the
Wigner surmise. Results, depicted in Fig.~\ref{fig2} confirm that for $k \sim 0.5$, the level spacing
distribution for the lowest eigenvalues is in excellent agreement with the Wigner surmise  $P_\mathrm{W,GOE}(s)$. 
 
The coupling between the left and right site is a relevant perturbation which
we  expect to become dominant in the tail of the spectrum.
Interestingly, in Ref.~\cite{maldacena2018} it
was reported that, assuming $N$ fixed and $k$ sufficiently large, some of the lowest energy excitations
of the effective Hamiltonian
are of gravitational origin and not related to the breaking of conformal symmetry. Physically this
corresponds to excited states of the wormhole geometry. 
It is tempting to speculate that there is a connection between these excitations of the wormhole geometry  and the deviations from random matrix theory in the tail of the spectrum
 below the critical coupling. 
If this picture applies, these gravitational modes related to the wormhole phase are not quantum chaotic and the sudden transition to random matrix theory in level statistics is a dynamical aspect of the thermodynamic Hawking-Page transition.
That would imply that random matrix theory, and therefore quantum ergodicity, is only a signature
of quantum black holes but not of an AdS graviton gas. In other words, the Hawking-Page transition can be dynamically characterized as a chaotic-integrable transition.

\begin{figure}[t!]
	\centering
\resizebox{1.\textwidth}{!}{\includegraphics{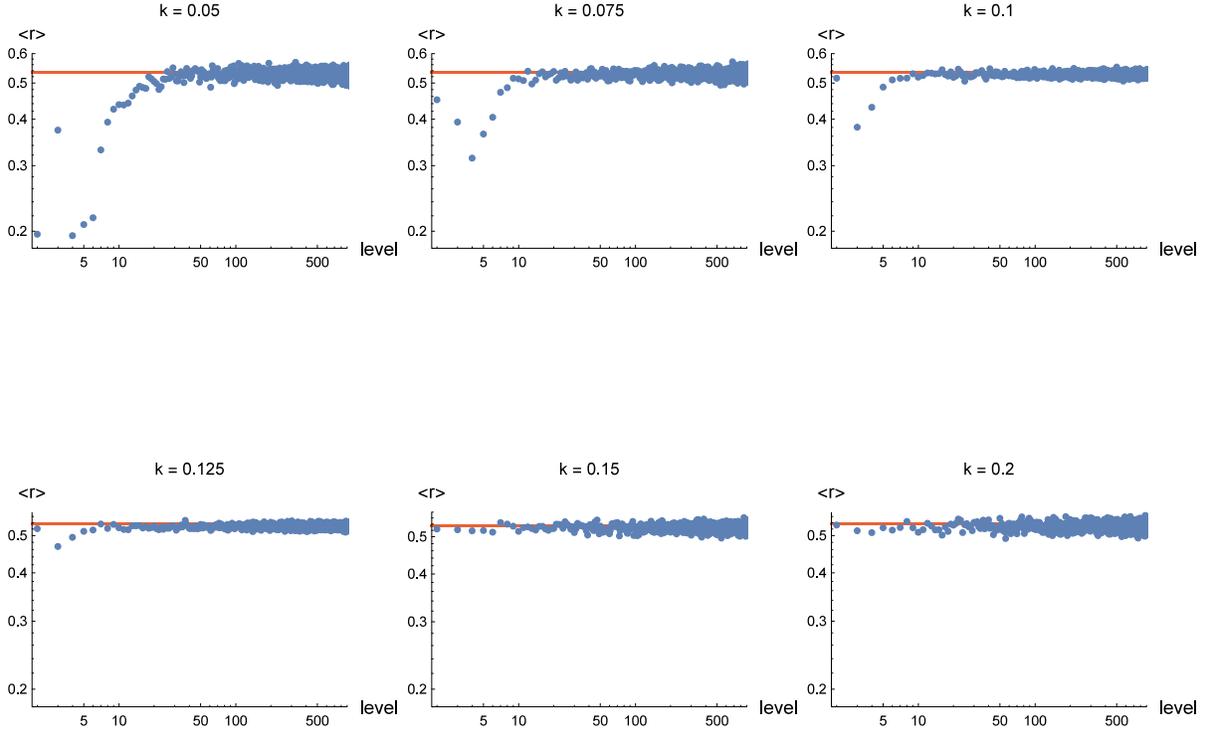}}  
	\vspace{-4mm}
	\caption{Adjacent gap ratio $\langle r \rangle$, Eq.~(\ref{eq:agr})
          for $N = 28$, ${\alpha} = 1$
%           for $N = 28$, $\gamma = 1$
          in Eq.(~\ref{Htotal}) and different values of the coupling parameter $k$. We employ only eigenvalues
          corresponding to the sector of lowest spin $S = -N/4$ that includes the ground state.
          Results are qualitatively similar to those of  different couplings. Deviations from GOE random
          matrix results (orange line) are observed for small $k$ which gradually disappear as $k$ approaches
          the critical point $k_c \sim 0.17$ where the Hawking-Page transition terminates.
        } 
	\label{rdario1}
\end{figure}
 
In order to test this hypothesis, we compare explicitly the number of eigenvalues with random matrix
like correlations with the spectral density (one point function) in that region.
For that purpose, we employ again the adjacent  gap ratio $\langle r \rangle$,
but with a larger $N = 34$, so that more eigenvalues show the anticipated anomalous behavior.
Regarding the spectral density, for which the prediction of gravitational excitations of
Ref.~\cite{maldacena2018} applies, we employ the local spectral density and
$\langle d \rangle \equiv \langle E_{i+1} - E_i \rangle$. Results, depicted in Fig.~\ref{fig2a}, clearly show
that the number of eigenvalues with random matrix correlations is in good agreement with the number
of eigenvalues whose $\langle d \rangle$ shows a pronounced dependence on $k$ for sufficiently
large $k$. 
Although further research is required to confirm this point, this quantitative agreement is encouraging. It strongly suggests that full quantum ergodicity is typical only of black holes, while other geometries like quantum wormholes may be closer to Poisson statistics typical of integrable dynamics. We now move to study level statistics for $\alpha =1$ in order to confirm that these findings are not particular to the value of $\alpha \sim 1$ used in previous sections.

\begin{figure}[t!]
	\centering
	\resizebox{0.80\textwidth}{!}{\includegraphics{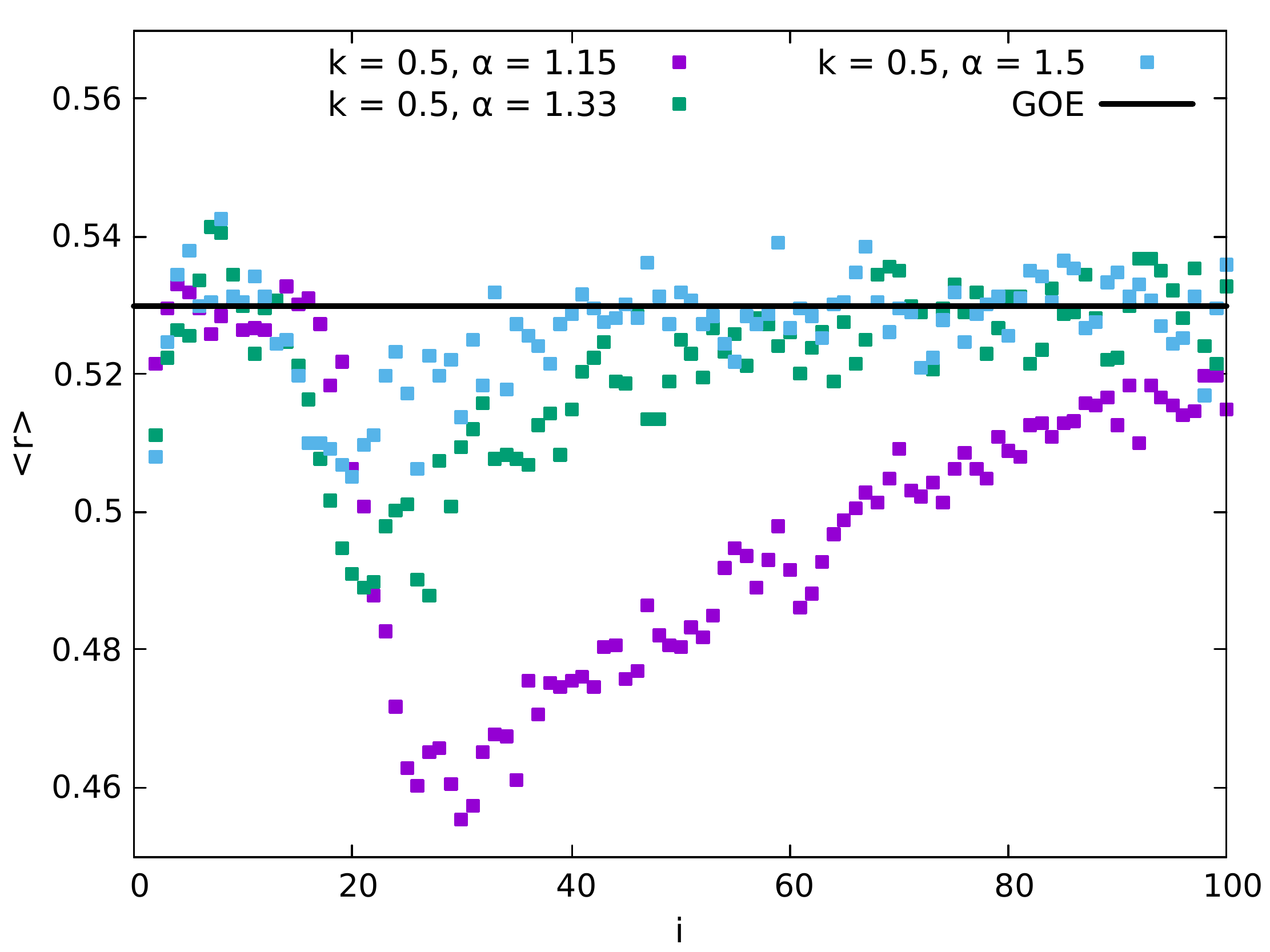}}  
	\vspace{-4mm}
	\caption{Adjacent gap ratio $\langle r \rangle$, Eq.~(\ref{eq:agr}), for $N = 26$ and
          different values of the asymmetry parameter. We observe that as the asymmetry increases
          the separation between the low lying eigenvalues gradually disappears.} 
	\label{fig3}
\end{figure}
\subsection{Dependence of Level correlations on $\alpha =1$}
To study level correlations in this case, the additional global spin symmetry for equal couplings makes it necessary to only consider eigenvalues with the same spin mod $4$ eigenvalue, in this case the block containing $S = -N/4$ which includes the ground state.
This requires an explicit calculation of eigenvectors which further limits
the maximum $N$ which we can reach numerically to $N = 28$ also taking into account that we need at least
$500$ disorder realizations to have relatively good statistics.
Results for $\langle r \rangle$, shown in Fig.~\ref{rdario1}, clearly show similar features as
in the slightly asymmetric case: 
deviation from GOE for small $k$ followed by the sudden appearance of the lowest eigenvalues correlated
according to random matrix theory at around the same value $k_c \sim 0.15$.
However, for $N = 28$, the crossover region is very narrow, only the two lowest eigenvalues become GOE correlated before the full spectrum becomes GOE correlated,
and the distinction between the tail and the bulk of the spectrum becomes more difficult to identify.

The spin symmetry is broken by the coupling asymmetry.
However, as mentioned previously, a small asymmetry does not change the physics of the model but makes
it technically easier to study the level statistics. 
In order to confirm this point more explicitly, we investigate the dependence of level statistics
 on the asymmetry parameter
$\alpha$ in Eq.~(\ref{Htotal}). In Fig.~\ref{fig3}, we depict results for the adjacent gap ratio
and different values of ${\alpha}$. As $\alpha$ increases, the differences between the tail and bulk of the spectrum becomes
increasingly small. Eventually, the full spectrum becomes correlated according to random matrix theory
even with a comparatively weak coupling. Physically this is an indication that the wormhole phase
related to the  graviton gas disappears if the asymmetry between left and right is sufficiently strong.
This is consistent with the results of Ref.~\cite{maldacena2018} where it was found that both the ground state energy gap and the left-right correlation decreases as the asymmetry increases. However further research
is required to determine whether there is a sharp  maximum value of the asymmetry for the wormhole phase
to occur.

\section{Outlook and Conclusions}

We have studied a coupled two-site SYK model whose gravity dual is conjectured to be \cite{maldacena2018} a
traversable wormhole geometry in the low temperature, small coupling, limit and dual to
a black hole geometry for sufficiently strong coupling or high temperature.
 We have analyzed the
spectral and thermodynamic properties of this model  and have found that it has a previously unknown
discrete spin symmetry which among other things is important for the analysis of level correlations. 

In Ref.~\cite{maldacena2018}, it was shown that the phase transition of this model
is first order
for weak coupling and seems to terminate for stronger coupling. 
An important point in the analysis of \cite{maldacena2018} is that the ground state in the wormhole phase is well approximated by a TFD state. In this paper we have found
that except in the limit of strong coupling between the two SYK models, TFD is never a good approximation of the ground state for the number of Majoranas ($N \leq 34$) that we can investigate by exact diagonalization.
For weaker coupling, coefficients of the expansion of the ground state in a tensor product
of the two separate SYK models have a slower power law decay, in contrast to the fast exponential decay with the energy expected in a TFD state.
Physically, this suggests a non-thermal state with entanglement properties qualitatively different from those of the TFD state. As the coupling $k$ increases, the overlap of the ground state with the TFD state decreases with a minimum at $k_c \sim 0.1$. For stronger coupling, the ground state becomes increasingly well approximated by the TFD state. In this large coupling limit, eigenstates of the associated spin operator with the
lowest eigenvalues are both close to the ground state of the system and to a TFD state at infinite temperature.
 Note that the deviations from the TFD state at  weak coupling mentioned above do not implicate that the overlap between
 the ground state and the TFD state is small. The reason is that for small coupling the strength of the
 wave function in a tensor basis of $H_{SYK} \otimes H_{SYK}$ is concentrated in only one or a few components of the eigenvectors which determine the overlap with the TFD state independently of the distribution in energy of the components.
 
We have also studied thermodynamic properties and level statistics that provide valuable information on the long time dynamics of the model. 
Regarding thermodynamic properties, we have fully confirmed \cite{maldacena2018} that the transition is
of first order for weak coupling and it terminates abruptly at $k_c \sim  0.2$. A careful analysis of
the free energy and specific heat shows that for stronger coupling, at fixed temperature, there is no second
order transition but just a sharp crossover. To a good approximation, this is also the value of the
coupling where the $k$ dependence of the energy gap changes from a power law to a
linear dependence which is
typical of a non-interacting system. In the strong coupling limit,
the spectral density develops well separated blobs centered around the eigenvalues of
the spin operator.

Level statistics are also affected by the coupling strength. For $k < k_c$, we observe
in the tail of the spectrum important deviations from the random matrix prediction, that we speculate,
is an indication that gravitational bound states related to the wormhole geometry are not quantum chaotic.
Around the transition we have observed that, suddenly, the lowest eigenvalues become correlated according
to random matrix theory. This quantum chaotic behavior is an expected  feature \cite{garcia2016} of
quantum black hole backgrounds. For larger $k$, the number of eigenvalues in the tail of the spectrum,
that is well described by random matrix theory, increases. Eventually, for $k \sim 1$, the full spectrum
is correlated according to the random matrix theory prediction. Physically, random matrix spectral correlations
are an indication of quantum chaotic features and full quantum ergodicity which is believed
to be a distinctive feature of quantum black holes. Therefore, the sudden appearance of random matrix correlations
deep in the infrared may be an indication of the instability of the wormhole phase towards the black hole phase. On the (quantum) gravity side, the exact meaning of a sharp crossover between the two geometries
is unclear.

In the opposite limit of very small $k$, well below the transition $k = k_c$,
we observe Poisson statistics for very small coupling strength $k_c \lessapprox 0.03$
(once the degeneracies are removed) which suggests that, at least in the range of finite $N$ we can explore, a minimum coupling is necessary to access the wormhole phase.
For larger coupling, the bulk of the spectrum becomes quickly correlated according to random matrix theory while deviations in the tail, very likely related to the wormhole phase, are still present until the transition occurs.
We did not manage to quantitatively understand  these deviations. The limited range of sizes we can explore numerically prevent us to perform a finite size scaling analysis. Additional research is required to clarify whether this limit of very small $k$ or the region of sudden appearance of random matrix correlations mentioned above broadens the phase diagram of the system.  

We note that the coupling term is one-body and therefore integrable. This leads us
to speculate that for a fixed temperature, the effective interactions among left and right Majoranas decrease as the coupling $k$ increases. It is tempting to speculate that the full phenomenology of the model
is consistent with a strong-weak coupling transition as $k$ increases such as the Gross-Witten
transition \cite{gross1980}  or the BEC-BCS crossover \cite{bourdel2004} observed in cold atom physics.

Interesting problems for further research include, the universality of these results for non-minimal
couplings between the two sites or to clarify whether similar physics is observed in supersymmetric analogues of this problem.
Another still unresolved problem is the gravity dual interpretation, if any, of the sharp crossover that
we observed around the transition. 
More specifically, it would be interesting to clarify whether
both geometries can somehow coexist in this region or simply the system ceases
to have a gravity dual interpretation. 
Another point which would be nice to investigate further is the study of the level statistics for the case of
equal couplings: in particular, it would be important to push the numerical studies to larger $N$ to see if
the distinction between the bulk and the tail of the spectrum becomes more prominent. To reach this goal,
we believe that the method of \cite{Gur-Ari:2018okm} could be useful.
The non TFD nature of the ground state for weak coupling, characterized by a power law decay of the coefficients in a tensor product basis, such as entanglement properties, deserves further investigation as well.
Does this power law dependence have a gravity dual interpretation? 
We plan to address some of these problems in the near future. 
\acknowledgements
DR wants to thank Frank Ferrari for discussions and Shanghai Jiao Tong University for hospitality during the completion of this work.
TN and DR thank Korea Institute for Advanced Study for providing computing resources (KIAS Center for Advanced Computation Abacus System) for this work.
JV was partially
supported by U.S.\ DOE Grant No.\ DE-FAG-88FR40388
and  performed part of this work at the Aspen Center for Physics, which is supported by National Science Foundation grant PHY-1607611.

\appendix

\section{Notations for Gamma matrices}
\label{Gammaapp1}

The Gamma matrices in \eqref{eq:gammapsi} are defined as usual
\begin{align}
\{\Gamma_i,\Gamma_j\}=2\delta_{ij}
\end{align}
For each SYK we have a chiral operator  
\be
\Gamma_5^R =i^{-N/4} \prod_{k=1}^{N/2} \Gamma_k, \qquad \Gamma_5^L =i^{-N/4} \prod_{k=1}^{N/2} \Gamma_{k+N/2},
\ee
the charge conjugation symmetry operators are defined as
\be
C^R = \prod_{k=1}^{N/4} \Gamma_{2k} K, \qquad C^L = \prod_{k=1}^{N/4} \Gamma_{2k+N/2} K,
\ee
%\jvc{We also use C as the charge conjugation matrix.}
with $K$ the complex conjugation operator.
They satisfy the commutation relations
\be
[H_R,\Gamma_5^R]&=&0, \qquad  [H_L,\Gamma_5^L]=0,\\
{[ H_R,\Gamma_5^L ]}&=&0, \qquad  [H_L,\Gamma_5^R]=0,\\
\{S,\Gamma_5^L\}&=&0, \qquad  \{S,\Gamma_5^R\}=0,\\
{[H_R,C^R]}&=&0, \qquad  [H_L,C^L]=0,
\ee
We also define
\be
\Gamma_5 = \Gamma_5^R \Gamma_5^L,
\ee
which is a symmetry of the total Hamiltonian.
The chiral projectors are given by
\be
P^R =\frac 12( 1 + \Gamma_5^R),\qquad
P^L=\frac 12( 1 + \Gamma_5^L),\qquad
\ee
If $N/4$ is even, the charge conjugation operator commutes with
the corresponding projection
operator, but this is not the case when $N/4$ is odd.
Then $\{\Gamma_5^R, C^R\} = 0$ and if we have an eigenstate of $H_R$,
\be
H_R|n\rangle= \lambda|n\rangle,\quad  {\rm with}\quad \Gamma_5|n\rangle= g_5|n\rangle
\ee
then also
\be
H_RC_R |n\rangle= \lambda C_R|n\rangle,
\ee
but now
\be
\Gamma_5^R C^R|n\rangle= -g_5C^R|n\rangle.
\ee

\section{Alternative derivation of $S$ mod 4 symmetry}
\label{Gammaapp13} 
The coupled model has an additional spin symmetry
\be
[e^{i\frac \pi 2S}, H_R+H_L + k S]=0,
\label{smod4}
\ee
with
\be
e^{\frac{\pi i}2 S} =\prod_{k=1}^{N/2} \frac 1{\sqrt{2}} ( 1 + \Gamma_{k+N/2}\Gamma_k).
\ee
To see that a product of the four factors
\be
1 + \Gamma_{k+N/2}\Gamma_k
\label{factors}
\ee
also commutes with terms  from the Hamiltonian that contain the index $k$, i.e.
terms of the from
\be
\Gamma_{k+ N/2}\Gamma_{l+ N/2}\Gamma_{m+ N/2}\Gamma_{n+ N2}+
\Gamma_k\Gamma_l\Gamma_m\Gamma_n
\ee
it is simplest to use the decomposition
\be
 \Gamma_{k+N/2}\Gamma_{l+N/2}\Gamma_{m+N/2}\Gamma_{n+N/2}+
\Gamma_k\Gamma_l\Gamma_m\Gamma_n
 =\frac 12(P_{kl}P_{mn}+Q_{kl}Q_{mn})
\ee
with
\be
P_{kl} &=& \Gamma_k\Gamma_l+\Gamma_{k+ N/2} \Gamma_{l+N/2},\\
Q_{kl} &=& \Gamma_k\Gamma_l-\Gamma_{k+ N/2} \Gamma_{l+ N/2}.
\ee
It is clear that each of the factors in \eref{factors} commutes with
the $P_{kl}$. They do not commute separately or even pairwise with
the $Q_{kl}$ term. The commutator can  be rewritten as 
\be
[F_{kl}F_{mn},Q_{kl}Q_{mn}] =[Q_{kl},F_{kl}]F_{mn}Q_{mn}+Q_{kl}F_{kl}[Q_{mn},F_{mn}]
\ee
with $F_{kl} =-(1+\Gamma_{k+N/2}\Gamma_k)(1+\Gamma_{l+N/2}\Gamma_l)/2$.
We have that
\be
[Q_{kl},F_{kl}]&=&2(\Gamma_k\Gamma_{l+\frac N2}-\Gamma_l\Gamma_{k+\frac N2}),\nn \\
F_{kl}Q_{kl}&=&\Gamma_l\Gamma_{k+\frac N2}-\Gamma_k\Gamma_{l+\frac N2},\nn\\
Q_{kl}F_{kl}&=&\Gamma_k\Gamma_{l+\frac N2}-\Gamma_l\Gamma_{k+\frac N2},\nn\\
\ee
so that the commutator vanishes. This proves the $S$ mod 4 symmetry.

\section{Ground state of the spin operator for $N/2 \mod 8 =2$}
\label{sec:tfdN20}
In this appendix we derive that the ground state of the spin operator
$N/2 \mod 8 =2$ 
is given by
\begin{align}
|S=-N/4\rangle=\frac{1}{\sqrt{L(N/2)}}\sum_{n,\pm}|n_\pm\rangle_L\otimes |n_\mp\rangle_R
\label{Slowest}
\end{align}
for an appropriate choice of  the phases of the states $|n_\pm\rangle_{L,R}$.

Let us consider the case $N=20,28,\cdots$ ($N/2 \mod 8 =2$).
In this case the charge conjugation matrix anti-commutes with the chiral $\gamma$
matrix
\be
\{ \gamma_c, C\}
\label{antiC}
\ee
with $\gamma_c=i^{-\frac{N}{4}}\gamma_1\gamma_2\cdots\gamma_{{N}/{2}}$.
Both $\gamma_c$ and $CK$ commute with the Hamiltonion
of a single SYK system
($K$ is the complex conjugation
operator)
\be
[H_{\rm SYK},\gamma_c ]=0, \qquad [H_{\rm SYK},CK ]=0
\ee
In a chiral basis 
with eigenvectors ${\widetilde v}_{n,\pm}$ of $H_{\rm SYK}$
we thus have that  $CK {\widetilde v}_{n,\pm}$ is also an eigenvector of
$H_{\rm SYK}$ but with opposite chirality because of the anti-commutation relations
\eref{antiC}.

We now define the left/right states above as
\begin{align}
|n_+\rangle_L={\widetilde v}_{n,+},\quad
|n_-\rangle_L=C({\widetilde v}_{n,+})^*,\quad
|n_+\rangle_R=e^{-\frac{\pi i}{4}}{\widetilde v}_{n,+},\quad
|n_-\rangle_R=e^{\frac{\pi i}{4}}C({\widetilde v}_{n,+})^*.
\label{fixphases}
\end{align}
With this convention we obtain
\begin{align}
\langle m_\pm |_R\gamma_i|n_{\mp}\rangle_R=\pm i\langle m_{\pm}|_L\gamma_i|n_{\mp}\rangle_L,\quad
\langle n_{\mp}|_L\gamma_i|m_{\pm}\rangle_L=(\langle n_{\pm}|_L\gamma_i|m_{\mp}\rangle_L)^*,
\end{align}
where the first equation holds due to the prefactor $\exp {\mp \frac{\pi i}{4}}$ in $|n_\pm\rangle_R$, and the second equation is a consequence of the relation
${\widetilde v}_{n,-}=C({\widetilde v}_{n,+})^*$.
Combining these two equations we find
\begin{align}
\langle m_\pm |_R\gamma_i|n_{\mp}\rangle_R=\pm i\langle n_{\pm}|_L\gamma_i|m_{\mp}\rangle_L.
\label{tip1}
\end{align}

To show that the state \eref{Slowest} is the ground state of the spin operator
we calculate the expectation value
\be
\langle S=- N/4 |S |S=- N/4 \rangle &=&\frac{i}{2^{N/4+1}}
\sum_{i=1}^{N/2}
\sum_{m,n,\pm,\pm'}\langle m_\pm|_L\otimes \langle m_\mp|_R\Gamma_i\Gamma_{{N}/{2}+i}|n_{\pm'}\rangle_L\otimes |n_{\mp'}\rangle_R\nonumber \\
&=&\frac{i}{2^{N/4+1}}\sum_{i=1}^{N/2}\sum_{m,n,\pm,\pm'}\langle m_\pm|_L\gamma_i\gamma_c|n_{\pm'}\rangle_L
\langle m_\mp|_R\gamma_i|n_{\mp'}\rangle_R.
\ee
Since $\gamma_i$ flips the chirality, only terms with $\pm'=\mp$ contributes:
\be
  \langle S=- N/4 |S |S=-N/4 \rangle &=&
  \frac{i}{2^{N/4+1}}\sum_{i=1}^{N/2}\sum_{m,n,\pm}(\langle m_\pm|_L\gamma_i\gamma_c|n_{\mp}\rangle_L)(\langle m_\mp|_R\gamma_i|n_{\pm}\rangle_R)\nonumber \\
&=&-\frac{1}{2^{N/4+1}}\sum_{i=1}^{N/2}\sum_{m,n,\pm}(\langle m_\pm|_L\gamma_i|n_{\mp}\rangle_L)(\langle n_\mp|_L\gamma_i|m_{\pm}\rangle_L),
\ee
where we have used $\gamma_c|n_\pm\rangle=\pm |n_\pm\rangle$ and \eqref{tip1}.
Now again noticing the fact that $\gamma_i$ flips chirality, we can safely replace $\sum_n|n_\mp\rangle_L\langle n_\mp|_L\rightarrow\sum_n(|n_+\rangle_L\langle n_+|_L+|n_-\rangle_L\langle n_+|_L)=1$ in the last line
\be
\langle S=- N/4 |S |S=- N/4 \rangle
&=&-\frac{1}{2^{N/4+1}}\sum_{i=1}^{N/2}\sum_{m,\pm}\langle m_\pm|_L\gamma_i\gamma_i|m_{\pm}\rangle_L\nonumber \\
&=&-\frac N4,
\ee
which is the lowest eigenvalue of the spin operator.

From Eq. \eref{fixphases} we can derive
\be
|n_-\rangle_R &=& e^{\frac{\pi i}4 \gamma_c} CK |n_+\rangle_L,\\
|n_+\rangle_R &=& e^{\frac{\pi i}4\gamma_c } CK |n_-\rangle_L,
\ee
  which allows us to rewrite the TFD state as
  \be
|S=-N/4\rangle=\frac{1}{\sqrt{L(N/2)}}\sum_{n,\pm}|n_\pm\rangle_L\otimes e^{\frac{\pi i}4 \gamma_c} CK |n_\mp\rangle_L.
\label{Slowest}
\ee
This is exactly the structure that was used in the general proof \eref{tfd-gen}.

\section{Lowest spin state and $|\text{TFD}\rangle_{\beta=0}$ for $N/2\mod 8 =4$}
\label{Gammaapp14}

We now construct explicitly the lowest eigenstate of the spin operator  for $N/2\mod 8 =4$.
In this case, the spectrum of $H_{SYK}$ with $N/2$ fermions is two-fold degenerate in both blocks,
while there are no degeneracies between the upper block and the lower block.

In this case, $H_{\rm SYK}$, the charge conjugation operator $CK$ and the chiral matrix $\gamma_c$ are
a set of commuting operators that can be diagonalized simultaneously. However, we have that
\be
(CK)^2=-1,
\ee
so that a state $\phi$ and a state $C\phi$ are linearly independent, which explains the two-fold degeneracy
of the states (which is the Kramers degeneracy).
Let us introduce an extra index $a=1,2$ for this degeneracy and denote the eigenvectors of $H_{\rm SYK}$ as
${\widetilde v}_{n,\pm,a}$. They satisfy the orthogonality relations
\begin{align}
({\widetilde v}_{m,\pm,a})^\dagger{\widetilde v}_{n,\pm',b}=\delta_{mn}\delta_{\pm,\pm'}\delta_{ab}.
\label{orthonormal}
\end{align}

To define the TFD state
 we choose the left/right states $|n_{\pm,a}\rangle_L$, $|n_{\pm,a}\rangle_R$ as
\be
|n_{\pm,1}\rangle_L&=&{\widetilde v}_{n,\pm,1},\\
|n_{\pm,2}\rangle_L&=&C({\widetilde v}_{n,\pm,1})^*,\\
|n_{\pm,1}\rangle_R&=&e^{\pm\frac{\pi i}{4}} |n_{\pm,1}\rangle_L
=-e^{\frac{\pi i}{4}\gamma_c} CK|n_{\pm,2}\rangle_L,  \label{d5}       \\
|n_{\pm,2}\rangle_R&=&-e^{\pm\frac{\pi i}{4}} |n_{\pm,2}\rangle_L=-e^{\frac{\pi i}{4}\gamma_c} CK|n_{\pm,1}\rangle_L .\label{d6}
\label{choiceofstates82440}
\ee
Then we can show 
\begin{align}
\phi=\frac{1}{\sqrt{L(N/2)}}\sum_{n,\pm}(|n_{\pm,1}\rangle_L\otimes |n_{\pm,2}\rangle_R+|n_{\pm,2}\rangle_L\otimes|n_{\pm,1}\rangle_R)
\end{align}
is the ground state of the spin operator.
This follows by using the Eqs. \eref{d5} and \eref{d6} to rewrite $\phi$ as
\begin{align}
\phi=-\frac{1}{\sqrt{L(N/2)}}\sum_{n,\pm,a}(|n_{\pm,a}\rangle_L\otimes  e^{\frac{\pi i}{4}\gamma_c} CK|n_{\pm,a}\rangle_L.
\end{align}
This is exactly the structure of the state \eref{tfd-gen}, so that the proof applies that
it is the ground state of the spin operator.

\section{Ground state of spin operator for $N/2 \mod 8=0$ }

\label{app:mod0}
In this case $(KC)^2 =1$ and it is always possible to choose a basis for which
\be
CK|n\rangle_{R.L} = |n\rangle_{R,L}.
\ee
The phases of the right states will be chosen as
\be
|n_\pm\rangle_R = e^{\frac{\pi i}4 \gamma_c} |n_\pm\rangle_L.
\ee
Then the TFD state is given by
\be
|\phi\rangle = \sum_{n,\pm} |n_\pm \rangle_L e^{\frac{\pi i}4 \gamma_c}  |n_\pm \rangle_L.
\ee
The expectation values of the spin operator is given by
\be
\langle \phi | S| \phi \rangle  &=&
\frac i2 2^{-N/4}\sum_{k=1}^{N/2} \sum_{mn\pm \pm'} \langle n_\pm|_L \gamma_k\gamma_c|m_{\pm'} \rangle_L
\langle n_\pm|_R \gamma_k|m_{\pm'} \rangle_R\nn\\
&=&
\frac i2 2^{-N/4} \sum_{k=1}^{N/2}\sum_{mn\pm } \langle n_\pm|_L \gamma_k\gamma_c|m_{\mp} \rangle_L
\langle n_\pm|_Le^{-\frac{\pi i}4 \gamma_c} \gamma_k e^{\frac{\pi i}4 \gamma_c}|m_{\mp} \rangle_L\nn\\
&=&
\frac i2 2^{-N/4} \sum_{k=1}^{N/2}\sum_{mn\pm }e^{\mp\frac{\pi i}2} (\mp)
\langle n_\pm|_L \gamma_k|m_{\mp} \rangle_L
\langle n_\pm|_L \gamma_k  |m_{\mp} \rangle_L\nn\\
&=&
-\frac 12 2^{-N/4} \sum_{k=1}^{N/2}\sum_{mn\pm }\langle n_\pm|_L \gamma_k|m_{\mp} \rangle_L
\langle m_\mp|_L \gamma_k  |n_{\pm} \rangle_L\nn\\
  &=& -\frac N4.
\ee
For the last equality we have used that that the $\gamma$-matrices for $N/2 \mod 8 =0$
can be chosen real symmetric.

\bibliography{library2_190302_Nosaka}

\begin{thebibliography}{63}%
\makeatletter
\providecommand \@ifxundefined [1]{%
 \@ifx{#1\undefined}
}%
\providecommand \@ifnum [1]{%
 \ifnum #1\expandafter \@firstoftwo
 \else \expandafter \@secondoftwo
 \fi
}%
\providecommand \@ifx [1]{%
 \ifx #1\expandafter \@firstoftwo
 \else \expandafter \@secondoftwo
 \fi
}%
\providecommand \natexlab [1]{#1}%
\providecommand \enquote  [1]{``#1''}%
\providecommand \bibnamefont  [1]{#1}%
\providecommand \bibfnamefont [1]{#1}%
\providecommand \citenamefont [1]{#1}%
\providecommand \href@noop [0]{\@secondoftwo}%
\providecommand \href [0]{\begingroup \@sanitize@url \@href}%
\providecommand \@href[1]{\@@startlink{#1}\@@href}%
\providecommand \@@href[1]{\endgroup#1\@@endlink}%
\providecommand \@sanitize@url [0]{\catcode `\\12\catcode `\$12\catcode
  `\&12\catcode `\#12\catcode `\^12\catcode `\_12\catcode `\%12\relax}%
\providecommand \@@startlink[1]{}%
\providecommand \@@endlink[0]{}%
\providecommand \url  [0]{\begingroup\@sanitize@url \@url }%
\providecommand \@url [1]{\endgroup\@href {#1}{\urlprefix }}%
\providecommand \urlprefix  [0]{URL }%
\providecommand \Eprint [0]{\href }%
\providecommand \doibase [0]{http://dx.doi.org/}%
\providecommand \selectlanguage [0]{\@gobble}%
\providecommand \bibinfo  [0]{\@secondoftwo}%
\providecommand \bibfield  [0]{\@secondoftwo}%
\providecommand \translation [1]{[#1]}%
\providecommand \BibitemOpen [0]{}%
\providecommand \bibitemStop [0]{}%
\providecommand \bibitemNoStop [0]{.\EOS\space}%
\providecommand \EOS [0]{\spacefactor3000\relax}%
\providecommand \BibitemShut  [1]{\csname bibitem#1\endcsname}%
\let\auto@bib@innerbib\@empty
%</preamble>
\bibitem [{\citenamefont {Wigner}(1951)}]{wigner1951}%
  \BibitemOpen
  \bibfield  {author} {\bibinfo {author} {\bibfnamefont {E.}~\bibnamefont
  {Wigner}},\ }\href {\doibase 10.1017/S0305004100027237} {\bibfield  {journal}
  {\bibinfo  {journal} {Math. Proc. Cam. Phil. Soc.}\ }\textbf {\bibinfo
  {volume} {49}},\ \bibinfo {pages} {790} (\bibinfo {year} {1951})}\BibitemShut
  {NoStop}%
\bibitem [{\citenamefont {Bertrand}\ and\ \citenamefont
  {Garc\'{\i}a-Garc\'{\i}a}(2016)}]{bertrand2016}%
  \BibitemOpen
  \bibfield  {author} {\bibinfo {author} {\bibfnamefont {C.~L.}\ \bibnamefont
  {Bertrand}}\ and\ \bibinfo {author} {\bibfnamefont {A.~M.}\ \bibnamefont
  {Garc\'{\i}a-Garc\'{\i}a}},\ }\href {\doibase 10.1103/PhysRevB.94.144201}
  {\bibfield  {journal} {\bibinfo  {journal} {Phys. Rev. B}\ }\textbf {\bibinfo
  {volume} {94}},\ \bibinfo {pages} {144201} (\bibinfo {year}
  {2016})}\BibitemShut {NoStop}%
\bibitem [{\citenamefont {Efetov}(1983)}]{efetov1983}%
  \BibitemOpen
  \bibfield  {author} {\bibinfo {author} {\bibfnamefont {K.}~\bibnamefont
  {Efetov}},\ }\href {\doibase 10.1080/00018738300101531} {\bibfield  {journal}
  {\bibinfo  {journal} {Advances in Physics}\ }\textbf {\bibinfo {volume}
  {32}},\ \bibinfo {pages} {53} (\bibinfo {year} {1983})}\BibitemShut {NoStop}%
\bibitem [{\citenamefont {Bohigas}\ \emph {et~al.}(1984)\citenamefont
  {Bohigas}, \citenamefont {Giannoni},\ and\ \citenamefont
  {Schmit}}]{bohigas1984}%
  \BibitemOpen
  \bibfield  {author} {\bibinfo {author} {\bibfnamefont {O.}~\bibnamefont
  {Bohigas}}, \bibinfo {author} {\bibfnamefont {M.~J.}\ \bibnamefont
  {Giannoni}}, \ and\ \bibinfo {author} {\bibfnamefont {C.}~\bibnamefont
  {Schmit}},\ }\href {\doibase 10.1103/PhysRevLett.52.1} {\bibfield  {journal}
  {\bibinfo  {journal} {Phys. Rev. Lett.}\ }\textbf {\bibinfo {volume} {52}},\
  \bibinfo {pages} {1} (\bibinfo {year} {1984})}\BibitemShut {NoStop}%
\bibitem [{\citenamefont {Verbaarschot}\ and\ \citenamefont
  {Zahed}(1993)}]{verbaarschot1993a}%
  \BibitemOpen
  \bibfield  {author} {\bibinfo {author} {\bibfnamefont {J.~J.~M.}\
  \bibnamefont {Verbaarschot}}\ and\ \bibinfo {author} {\bibfnamefont
  {I.}~\bibnamefont {Zahed}},\ }\href {\doibase 10.1103/PhysRevLett.70.3852}
  {\bibfield  {journal} {\bibinfo  {journal} {Phys. Rev. Lett.}\ }\textbf
  {\bibinfo {volume} {70}},\ \bibinfo {pages} {3852} (\bibinfo {year}
  {1993})}\BibitemShut {NoStop}%
\bibitem [{\citenamefont {Shuryak}\ and\ \citenamefont
  {Verbaarschot}(1993)}]{shuryak1993a}%
  \BibitemOpen
  \bibfield  {author} {\bibinfo {author} {\bibfnamefont {E.}~\bibnamefont
  {Shuryak}}\ and\ \bibinfo {author} {\bibfnamefont {J.}~\bibnamefont
  {Verbaarschot}},\ }\href {\doibase
  https://doi.org/10.1016/0375-9474(93)90098-I} {\bibfield  {journal} {\bibinfo
   {journal} {Nuclear Physics A}\ }\textbf {\bibinfo {volume} {560}},\ \bibinfo
  {pages} {306 } (\bibinfo {year} {1993})}\BibitemShut {NoStop}%
\bibitem [{\citenamefont {Maldacena}\ \emph
  {et~al.}(2016{\natexlab{a}})\citenamefont {Maldacena}, \citenamefont
  {Shenker},\ and\ \citenamefont {Stanford}}]{maldacena2015}%
  \BibitemOpen
  \bibfield  {author} {\bibinfo {author} {\bibfnamefont {J.}~\bibnamefont
  {Maldacena}}, \bibinfo {author} {\bibfnamefont {S.~H.}\ \bibnamefont
  {Shenker}}, \ and\ \bibinfo {author} {\bibfnamefont {D.}~\bibnamefont
  {Stanford}},\ }\href {\doibase 10.1007/JHEP08(2016)106} {\bibfield  {journal}
  {\bibinfo  {journal} {Journal of High Energy Physics}\ }\textbf {\bibinfo
  {volume} {08}},\ \bibinfo {pages} {106} (\bibinfo {year}
  {2016}{\natexlab{a}})}\BibitemShut {NoStop}%
\bibitem [{\citenamefont {Larkin}\ and\ \citenamefont
  {Ovchinnikov}(1969)}]{larkin1969}%
  \BibitemOpen
  \bibfield  {author} {\bibinfo {author} {\bibfnamefont {A.}~\bibnamefont
  {Larkin}}\ and\ \bibinfo {author} {\bibfnamefont {Y.~N.}\ \bibnamefont
  {Ovchinnikov}},\ }\href@noop {} {\bibfield  {journal} {\bibinfo  {journal}
  {Sov Phys JETP}\ }\textbf {\bibinfo {volume} {28}},\ \bibinfo {pages} {1200}
  (\bibinfo {year} {1969})}\BibitemShut {NoStop}%
\bibitem [{\citenamefont {Kitaev}()}]{kitaev2015}%
  \BibitemOpen
  \bibfield  {author} {\bibinfo {author} {\bibfnamefont {A.}~\bibnamefont
  {Kitaev}},\ }\href@noop {} {\enquote {\bibinfo {title} {A simple model of
  quantum holography},}\ }\bibinfo {note} {KITP strings seminar and
  Entanglement 2015 program, 12 February, 7 April and 27 May 2015,
  http://online.kitp.ucsb.edu/online/entangled15/}\BibitemShut {NoStop}%
\bibitem [{\citenamefont {Maldacena}\ and\ \citenamefont
  {Stanford}(2016)}]{maldacena2016}%
  \BibitemOpen
  \bibfield  {author} {\bibinfo {author} {\bibfnamefont {J.}~\bibnamefont
  {Maldacena}}\ and\ \bibinfo {author} {\bibfnamefont {D.}~\bibnamefont
  {Stanford}},\ }\href {\doibase 10.1103/PhysRevD.94.106002} {\bibfield
  {journal} {\bibinfo  {journal} {Phys. Rev. D}\ }\textbf {\bibinfo {volume}
  {94}},\ \bibinfo {pages} {106002} (\bibinfo {year} {2016})}\BibitemShut
  {NoStop}%
\bibitem [{\citenamefont {Bohigas}\ and\ \citenamefont
  {Flores}(1971{\natexlab{a}})}]{bohigas1971}%
  \BibitemOpen
  \bibfield  {author} {\bibinfo {author} {\bibfnamefont {O.}~\bibnamefont
  {Bohigas}}\ and\ \bibinfo {author} {\bibfnamefont {J.}~\bibnamefont
  {Flores}},\ }\href {\doibase http://dx.doi.org/10.1016/0370-2693(71)90598-3}
  {\bibfield  {journal} {\bibinfo  {journal} {Physics Letters B}\ }\textbf
  {\bibinfo {volume} {34}},\ \bibinfo {pages} {261 } (\bibinfo {year}
  {1971}{\natexlab{a}})}\BibitemShut {NoStop}%
\bibitem [{\citenamefont {Bohigas}\ and\ \citenamefont
  {Flores}(1971{\natexlab{b}})}]{bohigas1971a}%
  \BibitemOpen
  \bibfield  {author} {\bibinfo {author} {\bibfnamefont {O.}~\bibnamefont
  {Bohigas}}\ and\ \bibinfo {author} {\bibfnamefont {J.}~\bibnamefont
  {Flores}},\ }\href {\doibase http://dx.doi.org/10.1016/0370-2693(71)90399-6}
  {\bibfield  {journal} {\bibinfo  {journal} {Physics Letters B}\ }\textbf
  {\bibinfo {volume} {35}},\ \bibinfo {pages} {383 } (\bibinfo {year}
  {1971}{\natexlab{b}})}\BibitemShut {NoStop}%
\bibitem [{\citenamefont {French}\ and\ \citenamefont
  {Wong}(1970)}]{french1970}%
  \BibitemOpen
  \bibfield  {author} {\bibinfo {author} {\bibfnamefont {J.}~\bibnamefont
  {French}}\ and\ \bibinfo {author} {\bibfnamefont {S.}~\bibnamefont {Wong}},\
  }\href {\doibase http://dx.doi.org/10.1016/0370-2693(70)90213-3} {\bibfield
  {journal} {\bibinfo  {journal} {Physics Letters B}\ }\textbf {\bibinfo
  {volume} {33}},\ \bibinfo {pages} {449 } (\bibinfo {year}
  {1970})}\BibitemShut {NoStop}%
\bibitem [{\citenamefont {French}\ and\ \citenamefont
  {Wong}(1971)}]{french1971}%
  \BibitemOpen
  \bibfield  {author} {\bibinfo {author} {\bibfnamefont {J.}~\bibnamefont
  {French}}\ and\ \bibinfo {author} {\bibfnamefont {S.}~\bibnamefont {Wong}},\
  }\href {\doibase http://dx.doi.org/10.1016/0370-2693(71)90424-2} {\bibfield
  {journal} {\bibinfo  {journal} {Physics Letters B}\ }\textbf {\bibinfo
  {volume} {35}},\ \bibinfo {pages} {5 } (\bibinfo {year} {1971})}\BibitemShut
  {NoStop}%
\bibitem [{\citenamefont {Mon}\ and\ \citenamefont {French}(1975)}]{mon1975}%
  \BibitemOpen
  \bibfield  {author} {\bibinfo {author} {\bibfnamefont {K.}~\bibnamefont
  {Mon}}\ and\ \bibinfo {author} {\bibfnamefont {J.}~\bibnamefont {French}},\
  }\href {\doibase http://dx.doi.org/10.1016/0003-4916(75)90045-7} {\bibfield
  {journal} {\bibinfo  {journal} {Annals of Physics}\ }\textbf {\bibinfo
  {volume} {95}},\ \bibinfo {pages} {90 } (\bibinfo {year} {1975})}\BibitemShut
  {NoStop}%
\bibitem [{\citenamefont {Benet}\ and\ \citenamefont
  {Weidenm{\"u}ller}(2003)}]{benet2003}%
  \BibitemOpen
  \bibfield  {author} {\bibinfo {author} {\bibfnamefont {L.}~\bibnamefont
  {Benet}}\ and\ \bibinfo {author} {\bibfnamefont {H.~A.}\ \bibnamefont
  {Weidenm{\"u}ller}},\ }\href {http://stacks.iop.org/0305-4470/36/i=12/a=340}
  {\bibfield  {journal} {\bibinfo  {journal} {Journal of Physics A:
  Mathematical and General}\ }\textbf {\bibinfo {volume} {36}},\ \bibinfo
  {pages} {3569} (\bibinfo {year} {2003})}\BibitemShut {NoStop}%
\bibitem [{\citenamefont {Kota}(2014)}]{kota2014}%
  \BibitemOpen
  \bibfield  {author} {\bibinfo {author} {\bibfnamefont {V.~K.~B.}\
  \bibnamefont {Kota}},\ }\href@noop {} {\emph {\bibinfo {title} {Embedded
  random matrix ensembles in quantum physics}}},\ Vol.\ \bibinfo {volume}
  {884}\ (\bibinfo  {publisher} {Springer},\ \bibinfo {year}
  {2014})\BibitemShut {NoStop}%
\bibitem [{\citenamefont {Sachdev}\ and\ \citenamefont
  {Ye}(1993)}]{sachdev1993}%
  \BibitemOpen
  \bibfield  {author} {\bibinfo {author} {\bibfnamefont {S.}~\bibnamefont
  {Sachdev}}\ and\ \bibinfo {author} {\bibfnamefont {J.}~\bibnamefont {Ye}},\
  }\href {\doibase 10.1103/PhysRevLett.70.3339} {\bibfield  {journal} {\bibinfo
   {journal} {Phys. Rev. Lett.}\ }\textbf {\bibinfo {volume} {70}},\ \bibinfo
  {pages} {3339} (\bibinfo {year} {1993})}\BibitemShut {NoStop}%
\bibitem [{\citenamefont {Sachdev}(2010)}]{sachdev2010}%
  \BibitemOpen
  \bibfield  {author} {\bibinfo {author} {\bibfnamefont {S.}~\bibnamefont
  {Sachdev}},\ }\href {\doibase 10.1103/PhysRevLett.105.151602} {\bibfield
  {journal} {\bibinfo  {journal} {Phys. Rev. Lett.}\ }\textbf {\bibinfo
  {volume} {105}},\ \bibinfo {pages} {151602} (\bibinfo {year}
  {2010})}\BibitemShut {NoStop}%
\bibitem [{\citenamefont {Fu}\ \emph {et~al.}(2017)\citenamefont {Fu},
  \citenamefont {Gaiotto}, \citenamefont {Maldacena},\ and\ \citenamefont
  {Sachdev}}]{fu2018}%
  \BibitemOpen
  \bibfield  {author} {\bibinfo {author} {\bibfnamefont {W.}~\bibnamefont
  {Fu}}, \bibinfo {author} {\bibfnamefont {D.}~\bibnamefont {Gaiotto}},
  \bibinfo {author} {\bibfnamefont {J.}~\bibnamefont {Maldacena}}, \ and\
  \bibinfo {author} {\bibfnamefont {S.}~\bibnamefont {Sachdev}},\ }\href
  {\doibase 10.1103/PhysRevD.95.026009} {\bibfield  {journal} {\bibinfo
  {journal} {Phys. Rev. D}\ }\textbf {\bibinfo {volume} {95}},\ \bibinfo
  {pages} {026009} (\bibinfo {year} {2017})}\BibitemShut {NoStop}%
\bibitem [{\citenamefont {Garc\'{\i}a-Garc\'{\i}a}\ and\ \citenamefont
  {Verbaarschot}(2017)}]{garcia2017}%
  \BibitemOpen
  \bibfield  {author} {\bibinfo {author} {\bibfnamefont {A.~M.}\ \bibnamefont
  {Garc\'{\i}a-Garc\'{\i}a}}\ and\ \bibinfo {author} {\bibfnamefont {J.~J.~M.}\
  \bibnamefont {Verbaarschot}},\ }\href {\doibase 10.1103/PhysRevD.96.066012}
  {\bibfield  {journal} {\bibinfo  {journal} {Phys. Rev. D}\ }\textbf {\bibinfo
  {volume} {96}},\ \bibinfo {pages} {066012} (\bibinfo {year}
  {2017})}\BibitemShut {NoStop}%
\bibitem [{\citenamefont {Stanford}\ and\ \citenamefont
  {Witten}(2017)}]{stanford2017}%
  \BibitemOpen
  \bibfield  {author} {\bibinfo {author} {\bibfnamefont {D.}~\bibnamefont
  {Stanford}}\ and\ \bibinfo {author} {\bibfnamefont {E.}~\bibnamefont
  {Witten}},\ }\href {\doibase 10.1007/JHEP10(2017)008} {\bibfield  {journal}
  {\bibinfo  {journal} {JHEP}\ }\textbf {\bibinfo {volume} {10}},\ \bibinfo
  {pages} {008} (\bibinfo {year} {2017})},\ \Eprint
  {http://arxiv.org/abs/1703.04612} {arXiv:1703.04612 [hep-th]} \BibitemShut
  {NoStop}%
%%CITATION = ARXIV:1703.04612;%%
\bibitem [{\citenamefont {Mertens}\ \emph {et~al.}(2017)\citenamefont
  {Mertens}, \citenamefont {Turiaci},\ and\ \citenamefont
  {Verlinde}}]{mertens2017}%
  \BibitemOpen
  \bibfield  {author} {\bibinfo {author} {\bibfnamefont {T.~G.}\ \bibnamefont
  {Mertens}}, \bibinfo {author} {\bibfnamefont {G.~J.}\ \bibnamefont
  {Turiaci}}, \ and\ \bibinfo {author} {\bibfnamefont {H.~L.}\ \bibnamefont
  {Verlinde}},\ }\href {\doibase 10.1007/JHEP08(2017)136} {\bibfield  {journal}
  {\bibinfo  {journal} {JHEP}\ }\textbf {\bibinfo {volume} {08}},\ \bibinfo
  {pages} {136} (\bibinfo {year} {2017})},\ \Eprint
  {http://arxiv.org/abs/1705.08408} {arXiv:1705.08408 [hep-th]} \BibitemShut
  {NoStop}%
%%CITATION = ARXIV:1705.08408;%%
\bibitem [{\citenamefont {Belokurov}\ and\ \citenamefont
  {Shavgulidze}(2017)}]{belokurov2017}%
  \BibitemOpen
  \bibfield  {author} {\bibinfo {author} {\bibfnamefont {V.~V.}\ \bibnamefont
  {Belokurov}}\ and\ \bibinfo {author} {\bibfnamefont {E.~T.}\ \bibnamefont
  {Shavgulidze}},\ }\href {\doibase 10.1103/PhysRevD.96.101701} {\bibfield
  {journal} {\bibinfo  {journal} {Phys. Rev. D}\ }\textbf {\bibinfo {volume}
  {96}},\ \bibinfo {pages} {101701} (\bibinfo {year} {2017})}\BibitemShut
  {NoStop}%
\bibitem [{\citenamefont {Bagrets}\ \emph {et~al.}(2016)\citenamefont
  {Bagrets}, \citenamefont {Altland},\ and\ \citenamefont
  {Kamenev}}]{bagrets2016}%
  \BibitemOpen
  \bibfield  {author} {\bibinfo {author} {\bibfnamefont {D.}~\bibnamefont
  {Bagrets}}, \bibinfo {author} {\bibfnamefont {A.}~\bibnamefont {Altland}}, \
  and\ \bibinfo {author} {\bibfnamefont {A.}~\bibnamefont {Kamenev}},\
  }\href@noop {} {\bibfield  {journal} {\bibinfo  {journal} {Nuclear Physics
  B}\ }\textbf {\bibinfo {volume} {911}},\ \bibinfo {pages} {191} (\bibinfo
  {year} {2016})}\BibitemShut {NoStop}%
\bibitem [{\citenamefont {Bagrets}\ \emph {et~al.}(2017)\citenamefont
  {Bagrets}, \citenamefont {Altland},\ and\ \citenamefont
  {Kamenev}}]{bagrets2017}%
  \BibitemOpen
  \bibfield  {author} {\bibinfo {author} {\bibfnamefont {D.}~\bibnamefont
  {Bagrets}}, \bibinfo {author} {\bibfnamefont {A.}~\bibnamefont {Altland}}, \
  and\ \bibinfo {author} {\bibfnamefont {A.}~\bibnamefont {Kamenev}},\ }\href
  {\doibase https://doi.org/10.1016/j.nuclphysb.2017.06.012} {\bibfield
  {journal} {\bibinfo  {journal} {Nuclear Physics B}\ }\textbf {\bibinfo
  {volume} {921}},\ \bibinfo {pages} {727 } (\bibinfo {year}
  {2017})}\BibitemShut {NoStop}%
\bibitem [{\citenamefont {Jevicki}\ \emph {et~al.}(2016)\citenamefont
  {Jevicki}, \citenamefont {Suzuki},\ and\ \citenamefont {Yoon}}]{jevicki2016}%
  \BibitemOpen
  \bibfield  {author} {\bibinfo {author} {\bibfnamefont {A.}~\bibnamefont
  {Jevicki}}, \bibinfo {author} {\bibfnamefont {K.}~\bibnamefont {Suzuki}}, \
  and\ \bibinfo {author} {\bibfnamefont {J.}~\bibnamefont {Yoon}},\ }\href
  {\doibase 10.1007/JHEP07(2016)007} {\bibfield  {journal} {\bibinfo  {journal}
  {Journal of High Energy Physics}\ }\textbf {\bibinfo {volume} {07}},\
  \bibinfo {pages} {1} (\bibinfo {year} {2016})}\BibitemShut {NoStop}%
\bibitem [{\citenamefont {Aref~'eva}\ \emph {et~al.}(2018)\citenamefont
  {Aref~'eva}, \citenamefont {Khramtsov}, \citenamefont {Tikhanovskaya},\ and\
  \citenamefont {Volovich}}]{arefeva2018}%
  \BibitemOpen
  \bibfield  {author} {\bibinfo {author} {\bibfnamefont {I.}~\bibnamefont
  {Aref~'eva}}, \bibinfo {author} {\bibfnamefont {M.}~\bibnamefont
  {Khramtsov}}, \bibinfo {author} {\bibfnamefont {M.}~\bibnamefont
  {Tikhanovskaya}}, \ and\ \bibinfo {author} {\bibfnamefont {I.}~\bibnamefont
  {Volovich}},\ }\href@noop {} {\  (\bibinfo {year} {2018})},\ \Eprint
  {http://arxiv.org/abs/1811.04831} {arXiv:1811.04831 [hep-th]} \BibitemShut
  {NoStop}%
%%CITATION = ARXIV:1811.04831;%%
\bibitem [{\citenamefont {Altland}\ and\ \citenamefont
  {Bagrets}(2018)}]{Altland:2017eao}%
  \BibitemOpen
  \bibfield  {author} {\bibinfo {author} {\bibfnamefont {A.}~\bibnamefont
  {Altland}}\ and\ \bibinfo {author} {\bibfnamefont {D.}~\bibnamefont
  {Bagrets}},\ }\href {\doibase 10.1016/j.nuclphysb.2018.02.015} {\bibfield
  {journal} {\bibinfo  {journal} {Nucl. Phys.}\ }\textbf {\bibinfo {volume}
  {B930}},\ \bibinfo {pages} {45} (\bibinfo {year} {2018})},\ \Eprint
  {http://arxiv.org/abs/1712.05073} {arXiv:1712.05073 [cond-mat.str-el]}
  \BibitemShut {NoStop}%
%%CITATION = ARXIV:1712.05073;%%
\bibitem [{\citenamefont {Jensen}(2016)}]{jensen2016}%
  \BibitemOpen
  \bibfield  {author} {\bibinfo {author} {\bibfnamefont {K.}~\bibnamefont
  {Jensen}},\ }\href {\doibase 10.1103/PhysRevLett.117.111601} {\bibfield
  {journal} {\bibinfo  {journal} {Phys. Rev. Lett.}\ }\textbf {\bibinfo
  {volume} {117}},\ \bibinfo {pages} {111601} (\bibinfo {year}
  {2016})}\BibitemShut {NoStop}%
\bibitem [{\citenamefont {Cotler}\ \emph {et~al.}(2017)\citenamefont {Cotler},
  \citenamefont {Gur-Ari}, \citenamefont {Hanada}, \citenamefont {Polchinski},
  \citenamefont {Saad}, \citenamefont {Shenker}, \citenamefont {Stanford},
  \citenamefont {Streicher},\ and\ \citenamefont {Tezuka}}]{cotler2016}%
  \BibitemOpen
  \bibfield  {author} {\bibinfo {author} {\bibfnamefont {J.~S.}\ \bibnamefont
  {Cotler}}, \bibinfo {author} {\bibfnamefont {G.}~\bibnamefont {Gur-Ari}},
  \bibinfo {author} {\bibfnamefont {M.}~\bibnamefont {Hanada}}, \bibinfo
  {author} {\bibfnamefont {J.}~\bibnamefont {Polchinski}}, \bibinfo {author}
  {\bibfnamefont {P.}~\bibnamefont {Saad}}, \bibinfo {author} {\bibfnamefont
  {S.~H.}\ \bibnamefont {Shenker}}, \bibinfo {author} {\bibfnamefont
  {D.}~\bibnamefont {Stanford}}, \bibinfo {author} {\bibfnamefont
  {A.}~\bibnamefont {Streicher}}, \ and\ \bibinfo {author} {\bibfnamefont
  {M.}~\bibnamefont {Tezuka}},\ }\href {\doibase 10.1007/JHEP05(2017)118}
  {\bibfield  {journal} {\bibinfo  {journal} {Journal of High Energy Physics}\
  }\textbf {\bibinfo {volume} {05}},\ \bibinfo {pages} {118} (\bibinfo {year}
  {2017})}\BibitemShut {NoStop}%
\bibitem [{\citenamefont {Garc\'{\i}a-Garc\'{\i}a}\ and\ \citenamefont
  {Verbaarschot}(2016)}]{garcia2016}%
  \BibitemOpen
  \bibfield  {author} {\bibinfo {author} {\bibfnamefont {A.~M.}\ \bibnamefont
  {Garc\'{\i}a-Garc\'{\i}a}}\ and\ \bibinfo {author} {\bibfnamefont {J.~J.~M.}\
  \bibnamefont {Verbaarschot}},\ }\href {\doibase 10.1103/PhysRevD.94.126010}
  {\bibfield  {journal} {\bibinfo  {journal} {Phys. Rev. D}\ }\textbf {\bibinfo
  {volume} {94}},\ \bibinfo {pages} {126010} (\bibinfo {year}
  {2016})}\BibitemShut {NoStop}%
\bibitem [{\citenamefont {Garc\'{\i}a-Garc\'{\i}a}\ \emph
  {et~al.}(2018)\citenamefont {Garc\'{\i}a-Garc\'{\i}a}, \citenamefont {Jia},\
  and\ \citenamefont {Verbaarschot}}]{garcia2018a}%
  \BibitemOpen
  \bibfield  {author} {\bibinfo {author} {\bibfnamefont {A.~M.}\ \bibnamefont
  {Garc\'{\i}a-Garc\'{\i}a}}, \bibinfo {author} {\bibfnamefont
  {Y.}~\bibnamefont {Jia}}, \ and\ \bibinfo {author} {\bibfnamefont {J.~J.~M.}\
  \bibnamefont {Verbaarschot}},\ }\href {\doibase 10.1103/PhysRevD.97.106003}
  {\bibfield  {journal} {\bibinfo  {journal} {Phys. Rev. D}\ }\textbf {\bibinfo
  {volume} {97}},\ \bibinfo {pages} {106003} (\bibinfo {year}
  {2018})}\BibitemShut {NoStop}%
\bibitem [{\citenamefont {Maldacena}\ \emph
  {et~al.}(2016{\natexlab{b}})\citenamefont {Maldacena}, \citenamefont
  {Stanford},\ and\ \citenamefont {Yang}}]{maldacena2016a}%
  \BibitemOpen
  \bibfield  {author} {\bibinfo {author} {\bibfnamefont {J.}~\bibnamefont
  {Maldacena}}, \bibinfo {author} {\bibfnamefont {D.}~\bibnamefont {Stanford}},
  \ and\ \bibinfo {author} {\bibfnamefont {Z.}~\bibnamefont {Yang}},\ }\href
  {\doibase 10.1093/ptep/ptw124} {\bibfield  {journal} {\bibinfo  {journal}
  {Progress of Theoretical and Experimental Physics}\ }\textbf {\bibinfo
  {volume} {2016}},\ \bibinfo {pages} {12C104} (\bibinfo {year}
  {2016}{\natexlab{b}})}\BibitemShut {NoStop}%
\bibitem [{\citenamefont {Almheiri}\ and\ \citenamefont
  {Polchinski}(2015)}]{almheiri2015}%
  \BibitemOpen
  \bibfield  {author} {\bibinfo {author} {\bibfnamefont {A.}~\bibnamefont
  {Almheiri}}\ and\ \bibinfo {author} {\bibfnamefont {J.}~\bibnamefont
  {Polchinski}},\ }\href {\doibase 10.1007/JHEP11(2015)014} {\bibfield
  {journal} {\bibinfo  {journal} {Journal of High Energy Physics}\ }\textbf
  {\bibinfo {volume} {11}},\ \bibinfo {pages} {1} (\bibinfo {year}
  {2015})}\BibitemShut {NoStop}%
\bibitem [{\citenamefont {Jackiw}(1985)}]{jackiw1985}%
  \BibitemOpen
  \bibfield  {author} {\bibinfo {author} {\bibfnamefont {R.}~\bibnamefont
  {Jackiw}},\ }\href {\doibase https://doi.org/10.1016/0550-3213(85)90448-1}
  {\bibfield  {journal} {\bibinfo  {journal} {Nuclear Physics B}\ }\textbf
  {\bibinfo {volume} {252}},\ \bibinfo {pages} {343 } (\bibinfo {year}
  {1985})}\BibitemShut {NoStop}%
\bibitem [{\citenamefont {Teitelboim}(1983)}]{teitelboim1983}%
  \BibitemOpen
  \bibfield  {author} {\bibinfo {author} {\bibfnamefont {C.}~\bibnamefont
  {Teitelboim}},\ }\href {\doibase
  https://doi.org/10.1016/0370-2693(83)90012-6} {\bibfield  {journal} {\bibinfo
   {journal} {Physics Letters B}\ }\textbf {\bibinfo {volume} {126}},\ \bibinfo
  {pages} {41 } (\bibinfo {year} {1983})}\BibitemShut {NoStop}%
\bibitem [{\citenamefont {Förste}\ and\ \citenamefont
  {Golla}(2017)}]{forste2017}%
  \BibitemOpen
  \bibfield  {author} {\bibinfo {author} {\bibfnamefont {S.}~\bibnamefont
  {Förste}}\ and\ \bibinfo {author} {\bibfnamefont {I.}~\bibnamefont
  {Golla}},\ }\href {\doibase https://doi.org/10.1016/j.physletb.2017.05.039}
  {\bibfield  {journal} {\bibinfo  {journal} {Physics Letters B}\ }\textbf
  {\bibinfo {volume} {771}},\ \bibinfo {pages} {157 } (\bibinfo {year}
  {2017})}\BibitemShut {NoStop}%
\bibitem [{\citenamefont {Maldacena}\ and\ \citenamefont
  {Qi}(2018)}]{maldacena2018}%
  \BibitemOpen
  \bibfield  {author} {\bibinfo {author} {\bibfnamefont {J.}~\bibnamefont
  {Maldacena}}\ and\ \bibinfo {author} {\bibfnamefont {X.-L.}\ \bibnamefont
  {Qi}},\ }\href@noop {} {\  (\bibinfo {year} {2018})},\ \Eprint
  {http://arxiv.org/abs/1804.00491} {arXiv:1804.00491 [hep-th]} \BibitemShut
  {NoStop}%
%%CITATION = ARXIV:1804.00491;%%
\bibitem [{\citenamefont {Bak}\ \emph {et~al.}(2018)\citenamefont {Bak},
  \citenamefont {Kim},\ and\ \citenamefont {Yi}}]{bak2018}%
  \BibitemOpen
  \bibfield  {author} {\bibinfo {author} {\bibfnamefont {D.}~\bibnamefont
  {Bak}}, \bibinfo {author} {\bibfnamefont {C.}~\bibnamefont {Kim}}, \ and\
  \bibinfo {author} {\bibfnamefont {S.-H.}\ \bibnamefont {Yi}},\ }\href
  {\doibase 10.1007/JHEP08(2018)140} {\bibfield  {journal} {\bibinfo  {journal}
  {JHEP}\ }\textbf {\bibinfo {volume} {08}},\ \bibinfo {pages} {140} (\bibinfo
  {year} {2018})},\ \Eprint {http://arxiv.org/abs/1805.12349} {arXiv:1805.12349
  [hep-th]} \BibitemShut {NoStop}%
%%CITATION = ARXIV:1805.12349;%%
\bibitem [{\citenamefont {Kim}\ \emph {et~al.}(2019)\citenamefont {Kim},
  \citenamefont {Klebanov}, \citenamefont {Tarnopolsky},\ and\ \citenamefont
  {Zhao}}]{kim2019}%
  \BibitemOpen
  \bibfield  {author} {\bibinfo {author} {\bibfnamefont {J.}~\bibnamefont
  {Kim}}, \bibinfo {author} {\bibfnamefont {I.~R.}\ \bibnamefont {Klebanov}},
  \bibinfo {author} {\bibfnamefont {G.}~\bibnamefont {Tarnopolsky}}, \ and\
  \bibinfo {author} {\bibfnamefont {W.}~\bibnamefont {Zhao}},\ }\href@noop {}
  {\  (\bibinfo {year} {2019})},\ \Eprint {http://arxiv.org/abs/1902.02287}
  {arXiv:1902.02287 [hep-th]} \BibitemShut {NoStop}%
%%CITATION = ARXIV:1902.02287;%%
\bibitem [{\citenamefont {Gao}\ \emph {et~al.}(2017)\citenamefont {Gao},
  \citenamefont {Jafferis},\ and\ \citenamefont {Wall}}]{gao2016}%
  \BibitemOpen
  \bibfield  {author} {\bibinfo {author} {\bibfnamefont {P.}~\bibnamefont
  {Gao}}, \bibinfo {author} {\bibfnamefont {D.~L.}\ \bibnamefont {Jafferis}}, \
  and\ \bibinfo {author} {\bibfnamefont {A.}~\bibnamefont {Wall}},\ }\href
  {\doibase 10.1007/JHEP12(2017)151} {\bibfield  {journal} {\bibinfo  {journal}
  {JHEP}\ }\textbf {\bibinfo {volume} {12}},\ \bibinfo {pages} {151} (\bibinfo
  {year} {2017})},\ \Eprint {http://arxiv.org/abs/1608.05687} {arXiv:1608.05687
  [hep-th]} \BibitemShut {NoStop}%
%%CITATION = ARXIV:1608.05687;%%
\bibitem [{\citenamefont {Maldacena}\ \emph {et~al.}(2017)\citenamefont
  {Maldacena}, \citenamefont {Stanford},\ and\ \citenamefont
  {Yang}}]{maldacena2017}%
  \BibitemOpen
  \bibfield  {author} {\bibinfo {author} {\bibfnamefont {J.}~\bibnamefont
  {Maldacena}}, \bibinfo {author} {\bibfnamefont {D.}~\bibnamefont {Stanford}},
  \ and\ \bibinfo {author} {\bibfnamefont {Z.}~\bibnamefont {Yang}},\ }\href
  {\doibase 10.1002/prop.201700034} {\bibfield  {journal} {\bibinfo  {journal}
  {Fortsch. Phys.}\ }\textbf {\bibinfo {volume} {65}},\ \bibinfo {pages}
  {1700034} (\bibinfo {year} {2017})},\ \Eprint
  {http://arxiv.org/abs/1704.05333} {arXiv:1704.05333 [hep-th]} \BibitemShut
  {NoStop}%
%%CITATION = ARXIV:1704.05333;%%
\bibitem [{\citenamefont {Fu}\ \emph {et~al.}(2018)\citenamefont {Fu},
  \citenamefont {Grado-White},\ and\ \citenamefont {Marolf}}]{marlof2018}%
  \BibitemOpen
  \bibfield  {author} {\bibinfo {author} {\bibfnamefont {Z.}~\bibnamefont
  {Fu}}, \bibinfo {author} {\bibfnamefont {B.}~\bibnamefont {Grado-White}}, \
  and\ \bibinfo {author} {\bibfnamefont {D.}~\bibnamefont {Marolf}},\
  }\href@noop {} {\  (\bibinfo {year} {2018})},\ \Eprint
  {http://arxiv.org/abs/1807.07917} {arXiv:1807.07917 [hep-th]} \BibitemShut
  {NoStop}%
%%CITATION = ARXIV:1807.07917;%%
\bibitem [{\citenamefont {Tumurtushaa}\ and\ \citenamefont
  {Yeom}(2018)}]{tumurtushaa2018}%
  \BibitemOpen
  \bibfield  {author} {\bibinfo {author} {\bibfnamefont {G.}~\bibnamefont
  {Tumurtushaa}}\ and\ \bibinfo {author} {\bibfnamefont {D.-H.}\ \bibnamefont
  {Yeom}},\ }\href@noop {} {\  (\bibinfo {year} {2018})},\ \Eprint
  {http://arxiv.org/abs/1808.01103} {arXiv:1808.01103 [hep-th]} \BibitemShut
  {NoStop}%
%%CITATION = ARXIV:1808.01103;%%
\bibitem [{\citenamefont {Maldacena}\ \emph {et~al.}(2018)\citenamefont
  {Maldacena}, \citenamefont {Milekhin},\ and\ \citenamefont
  {Popov}}]{maldacena2018b}%
  \BibitemOpen
  \bibfield  {author} {\bibinfo {author} {\bibfnamefont {J.}~\bibnamefont
  {Maldacena}}, \bibinfo {author} {\bibfnamefont {A.}~\bibnamefont {Milekhin}},
  \ and\ \bibinfo {author} {\bibfnamefont {F.}~\bibnamefont {Popov}},\
  }\href@noop {} {\  (\bibinfo {year} {2018})},\ \Eprint
  {http://arxiv.org/abs/1807.04726} {arXiv:1807.04726 [hep-th]} \BibitemShut
  {NoStop}%
%%CITATION = ARXIV:1807.04726;%%
\bibitem [{\citenamefont {Anabalón}\ and\ \citenamefont
  {Oliva}(2018)}]{anabalon2018}%
  \BibitemOpen
  \bibfield  {author} {\bibinfo {author} {\bibfnamefont {A.}~\bibnamefont
  {Anabalón}}\ and\ \bibinfo {author} {\bibfnamefont {J.}~\bibnamefont
  {Oliva}},\ }\href@noop {} {\  (\bibinfo {year} {2018})},\ \Eprint
  {http://arxiv.org/abs/1811.03497} {arXiv:1811.03497 [hep-th]} \BibitemShut
  {NoStop}%
%%CITATION = ARXIV:1811.03497;%%
\bibitem [{\citenamefont {Gao}\ and\ \citenamefont {Liu}(2018)}]{gao2018}%
  \BibitemOpen
  \bibfield  {author} {\bibinfo {author} {\bibfnamefont {P.}~\bibnamefont
  {Gao}}\ and\ \bibinfo {author} {\bibfnamefont {H.}~\bibnamefont {Liu}},\
  }\href@noop {} {\  (\bibinfo {year} {2018})},\ \Eprint
  {http://arxiv.org/abs/1810.01444} {arXiv:1810.01444 [hep-th]} \BibitemShut
  {NoStop}%
%%CITATION = ARXIV:1810.01444;%%
\bibitem [{\citenamefont {Bao}\ \emph {et~al.}(2018)\citenamefont {Bao},
  \citenamefont {Chatwin-Davies}, \citenamefont {Pollack},\ and\ \citenamefont
  {Remmen}}]{bao2018}%
  \BibitemOpen
  \bibfield  {author} {\bibinfo {author} {\bibfnamefont {N.}~\bibnamefont
  {Bao}}, \bibinfo {author} {\bibfnamefont {A.}~\bibnamefont {Chatwin-Davies}},
  \bibinfo {author} {\bibfnamefont {J.}~\bibnamefont {Pollack}}, \ and\
  \bibinfo {author} {\bibfnamefont {G.~N.}\ \bibnamefont {Remmen}},\ }\href
  {\doibase 10.1007/JHEP11(2018)071} {\bibfield  {journal} {\bibinfo  {journal}
  {JHEP}\ }\textbf {\bibinfo {volume} {11}},\ \bibinfo {pages} {071} (\bibinfo
  {year} {2018})},\ \Eprint {http://arxiv.org/abs/1808.05963} {arXiv:1808.05963
  [hep-th]} \BibitemShut {NoStop}%
%%CITATION = ARXIV:1808.05963;%%
\bibitem [{\citenamefont {Azeyanagi}\ \emph {et~al.}(2018)\citenamefont
  {Azeyanagi}, \citenamefont {Ferrari},\ and\ \citenamefont
  {Massolo}}]{ferrari2017}%
  \BibitemOpen
  \bibfield  {author} {\bibinfo {author} {\bibfnamefont {T.}~\bibnamefont
  {Azeyanagi}}, \bibinfo {author} {\bibfnamefont {F.}~\bibnamefont {Ferrari}},
  \ and\ \bibinfo {author} {\bibfnamefont {F.~I.~S.}\ \bibnamefont {Massolo}},\
  }\href {\doibase 10.1103/PhysRevLett.120.061602} {\bibfield  {journal}
  {\bibinfo  {journal} {Phys. Rev. Lett.}\ }\textbf {\bibinfo {volume} {120}},\
  \bibinfo {pages} {061602} (\bibinfo {year} {2018})}\BibitemShut {NoStop}%
\bibitem [{\citenamefont {Gross}\ and\ \citenamefont
  {Witten}(1980{\natexlab{a}})}]{Gross:1980he}%
  \BibitemOpen
  \bibfield  {author} {\bibinfo {author} {\bibfnamefont {D.~J.}\ \bibnamefont
  {Gross}}\ and\ \bibinfo {author} {\bibfnamefont {E.}~\bibnamefont {Witten}},\
  }\href {\doibase 10.1103/PhysRevD.21.446} {\bibfield  {journal} {\bibinfo
  {journal} {Phys. Rev.}\ }\textbf {\bibinfo {volume} {D21}},\ \bibinfo {pages}
  {446} (\bibinfo {year} {1980}{\natexlab{a}})}\BibitemShut {NoStop}%
%%CITATION = PHRVA,D21,446;%%
\bibitem [{\citenamefont {Verbaarschot}\ and\ \citenamefont
  {Brussaard}(1979)}]{Verbaarschot:1979kts}%
  \BibitemOpen
  \bibfield  {author} {\bibinfo {author} {\bibfnamefont {J.~J.~M.}\
  \bibnamefont {Verbaarschot}}\ and\ \bibinfo {author} {\bibfnamefont {P.~J.}\
  \bibnamefont {Brussaard}},\ }\href {\doibase 10.1016/0370-2693(79)90953-5}
  {\bibfield  {journal} {\bibinfo  {journal} {Phys. Lett.}\ }\textbf {\bibinfo
  {volume} {87B}},\ \bibinfo {pages} {155} (\bibinfo {year}
  {1979})}\BibitemShut {NoStop}%
%%CITATION = PHLTA,87B,155;%%
\bibitem [{\citenamefont {Cottrell}\ \emph {et~al.}(2018)\citenamefont
  {Cottrell}, \citenamefont {Freivogel}, \citenamefont {Hofman},\ and\
  \citenamefont {Lokhande}}]{Cottrell2018}%
  \BibitemOpen
  \bibfield  {author} {\bibinfo {author} {\bibfnamefont {W.}~\bibnamefont
  {Cottrell}}, \bibinfo {author} {\bibfnamefont {B.}~\bibnamefont {Freivogel}},
  \bibinfo {author} {\bibfnamefont {D.~M.}\ \bibnamefont {Hofman}}, \ and\
  \bibinfo {author} {\bibfnamefont {S.~F.}\ \bibnamefont {Lokhande}},\
  }\href@noop {} {\  (\bibinfo {year} {2018})},\ \Eprint
  {http://arxiv.org/abs/1811.11528} {arXiv:1811.11528 [hep-th]} \BibitemShut
  {NoStop}%
%%CITATION = ARXIV:1811.11528;%%
\bibitem [{\citenamefont {Gross}\ and\ \citenamefont
  {Witten}(1980{\natexlab{b}})}]{gross1980}%
  \BibitemOpen
  \bibfield  {author} {\bibinfo {author} {\bibfnamefont {D.~J.}\ \bibnamefont
  {Gross}}\ and\ \bibinfo {author} {\bibfnamefont {E.}~\bibnamefont {Witten}},\
  }\href {\doibase 10.1103/PhysRevD.21.446} {\bibfield  {journal} {\bibinfo
  {journal} {Phys. Rev. D}\ }\textbf {\bibinfo {volume} {21}},\ \bibinfo
  {pages} {446} (\bibinfo {year} {1980}{\natexlab{b}})}\BibitemShut {NoStop}%
\bibitem [{\citenamefont {Bourdel}\ \emph {et~al.}(2004)\citenamefont
  {Bourdel}, \citenamefont {Khaykovich}, \citenamefont {Cubizolles},
  \citenamefont {Zhang}, \citenamefont {Chevy}, \citenamefont {Teichmann},
  \citenamefont {Tarruell}, \citenamefont {Kokkelmans},\ and\ \citenamefont
  {Salomon}}]{bourdel2004}%
  \BibitemOpen
  \bibfield  {author} {\bibinfo {author} {\bibfnamefont {T.}~\bibnamefont
  {Bourdel}}, \bibinfo {author} {\bibfnamefont {L.}~\bibnamefont {Khaykovich}},
  \bibinfo {author} {\bibfnamefont {J.}~\bibnamefont {Cubizolles}}, \bibinfo
  {author} {\bibfnamefont {J.}~\bibnamefont {Zhang}}, \bibinfo {author}
  {\bibfnamefont {F.}~\bibnamefont {Chevy}}, \bibinfo {author} {\bibfnamefont
  {M.}~\bibnamefont {Teichmann}}, \bibinfo {author} {\bibfnamefont
  {L.}~\bibnamefont {Tarruell}}, \bibinfo {author} {\bibfnamefont {S.~J. J.
  M.~F.}\ \bibnamefont {Kokkelmans}}, \ and\ \bibinfo {author} {\bibfnamefont
  {C.}~\bibnamefont {Salomon}},\ }\href {\doibase
  10.1103/PhysRevLett.93.050401} {\bibfield  {journal} {\bibinfo  {journal}
  {Phys. Rev. Lett.}\ }\textbf {\bibinfo {volume} {93}},\ \bibinfo {pages}
  {050401} (\bibinfo {year} {2004})}\BibitemShut {NoStop}%
\bibitem [{\citenamefont {Mehta}(2004)}]{mehta2004}%
  \BibitemOpen
  \bibfield  {author} {\bibinfo {author} {\bibfnamefont {M.~L.}\ \bibnamefont
  {Mehta}},\ }\href@noop {} {\emph {\bibinfo {title} {Random matrices}}}\
  (\bibinfo  {publisher} {Academic press},\ \bibinfo {year} {2004})\BibitemShut
  {NoStop}%
\bibitem [{\citenamefont {Guhr}\ \emph {et~al.}(1998)\citenamefont {Guhr},
  \citenamefont {Mueller-Groeling},\ and\ \citenamefont
  {Weidenmueller}}]{guhr1998}%
  \BibitemOpen
  \bibfield  {author} {\bibinfo {author} {\bibfnamefont {T.}~\bibnamefont
  {Guhr}}, \bibinfo {author} {\bibfnamefont {A.}~\bibnamefont
  {Mueller-Groeling}}, \ and\ \bibinfo {author} {\bibfnamefont {H.~A.}\
  \bibnamefont {Weidenmueller}},\ }\href {\doibase
  http://dx.doi.org/10.1016/S0370-1573(97)00088-4} {\bibfield  {journal}
  {\bibinfo  {journal} {Physics Reports}\ }\textbf {\bibinfo {volume} {299}},\
  \bibinfo {pages} {189 } (\bibinfo {year} {1998})}\BibitemShut {NoStop}%
\bibitem [{\citenamefont {Luitz}\ \emph {et~al.}(2015)\citenamefont {Luitz},
  \citenamefont {Laflorencie},\ and\ \citenamefont {Alet}}]{luitz2015}%
  \BibitemOpen
  \bibfield  {author} {\bibinfo {author} {\bibfnamefont {D.~J.}\ \bibnamefont
  {Luitz}}, \bibinfo {author} {\bibfnamefont {N.}~\bibnamefont {Laflorencie}},
  \ and\ \bibinfo {author} {\bibfnamefont {F.}~\bibnamefont {Alet}},\ }\href
  {\doibase 10.1103/PhysRevB.91.081103} {\bibfield  {journal} {\bibinfo
  {journal} {Phys. Rev. B}\ }\textbf {\bibinfo {volume} {91}},\ \bibinfo
  {pages} {081103} (\bibinfo {year} {2015})}\BibitemShut {NoStop}%
\bibitem [{\citenamefont {Oganesyan}\ and\ \citenamefont
  {Huse}(2007)}]{oganesyan2007}%
  \BibitemOpen
  \bibfield  {author} {\bibinfo {author} {\bibfnamefont {V.}~\bibnamefont
  {Oganesyan}}\ and\ \bibinfo {author} {\bibfnamefont {D.~A.}\ \bibnamefont
  {Huse}},\ }\href {\doibase 10.1103/PhysRevB.75.155111} {\bibfield  {journal}
  {\bibinfo  {journal} {Phys. Rev. B}\ }\textbf {\bibinfo {volume} {75}},\
  \bibinfo {pages} {155111} (\bibinfo {year} {2007})}\BibitemShut {NoStop}%
\bibitem [{\citenamefont {Atas}\ \emph {et~al.}(2013)\citenamefont {Atas},
  \citenamefont {Bogomolny}, \citenamefont {Giraud},\ and\ \citenamefont
  {Roux}}]{atas2016}%
  \BibitemOpen
  \bibfield  {author} {\bibinfo {author} {\bibfnamefont {Y.~Y.}\ \bibnamefont
  {Atas}}, \bibinfo {author} {\bibfnamefont {E.}~\bibnamefont {Bogomolny}},
  \bibinfo {author} {\bibfnamefont {O.}~\bibnamefont {Giraud}}, \ and\ \bibinfo
  {author} {\bibfnamefont {G.}~\bibnamefont {Roux}},\ }\href {\doibase
  10.1103/PhysRevLett.110.084101} {\bibfield  {journal} {\bibinfo  {journal}
  {Phys. Rev. Lett.}\ }\textbf {\bibinfo {volume} {110}},\ \bibinfo {pages}
  {084101} (\bibinfo {year} {2013})}\BibitemShut {NoStop}%
\bibitem [{\citenamefont {Numasawa}(2019)}]{numasawa2019}%
  \BibitemOpen
  \bibfield  {author} {\bibinfo {author} {\bibfnamefont {T.}~\bibnamefont
  {Numasawa}},\ }\href@noop {} {\  (\bibinfo {year} {2019})},\ \Eprint
  {http://arxiv.org/abs/1901.02025} {arXiv:1901.02025 [hep-th]} \BibitemShut
  {NoStop}%
%%CITATION = ARXIV:1901.02025;%%
\bibitem [{\citenamefont {Kourkoulou}\ and\ \citenamefont
  {Maldacena}(2017)}]{Kourkoulou:2017zaj}%
  \BibitemOpen
  \bibfield  {author} {\bibinfo {author} {\bibfnamefont {I.}~\bibnamefont
  {Kourkoulou}}\ and\ \bibinfo {author} {\bibfnamefont {J.}~\bibnamefont
  {Maldacena}},\ }\href@noop {} {\  (\bibinfo {year} {2017})},\ \Eprint
  {http://arxiv.org/abs/1707.02325} {arXiv:1707.02325 [hep-th]} \BibitemShut
  {NoStop}%
%%CITATION = ARXIV:1707.02325;%%
\bibitem [{\citenamefont {Gur-Ari}\ \emph {et~al.}(2018)\citenamefont
  {Gur-Ari}, \citenamefont {Mahajan},\ and\ \citenamefont
  {Vaezi}}]{Gur-Ari:2018okm}%
  \BibitemOpen
  \bibfield  {author} {\bibinfo {author} {\bibfnamefont {G.}~\bibnamefont
  {Gur-Ari}}, \bibinfo {author} {\bibfnamefont {R.}~\bibnamefont {Mahajan}}, \
  and\ \bibinfo {author} {\bibfnamefont {A.}~\bibnamefont {Vaezi}},\ }\href
  {\doibase 10.1007/JHEP11(2018)070} {\bibfield  {journal} {\bibinfo  {journal}
  {JHEP}\ }\textbf {\bibinfo {volume} {11}},\ \bibinfo {pages} {070} (\bibinfo
  {year} {2018})},\ \Eprint {http://arxiv.org/abs/1806.10145} {arXiv:1806.10145
  [hep-th]} \BibitemShut {NoStop}%
%%CITATION = ARXIV:1806.10145;%%
\end{thebibliography}%
% \bibliography{library2_190227_Antonio}
 
\end{document}